\documentclass[12pt,preprint]{aastex}

\slugcomment{Accepted for publication in ApJ}

\shorttitle{The First \ion{He}{1}$\lamba 10830$ BALQSO}
\shortauthors{Leighly, Dietrich \& Barber}

\begin{document}


\title{The Discovery of the First \ion{He}{1}$\lambda 10830$
  Broad-Absorption-Line Quasar} 


\author{Karen M. Leighly\altaffilmark{1}}
\affil{Homer L. Dodge Department of Physics and Astronomy, The
  University of Oklahoma, 440 W.\ Brooks St., Norman, OK 73019}
\email{leighly@nhn.ou.edu}

\author{Matthias Dietrich \altaffilmark{1}}
\affil{Department of Astronomy, The Ohio State University, 4055
  McPherson Lab., 140 W.\ 18th Ave., Columbus, OH 43210}

\and

\author{Sara Barber}
\affil{Homer L. Dodge Department of Physics and Astronomy, The
  University of Oklahoma, 440 W.\ Brooks St., Norman, OK 73019}

\altaffiltext{1}{Visiting Astronomer at the Infrared Telescope
  Facility, which is operated by the University of Hawaii under
  Cooperative Agreement no. NCC 5-538 with the National Aeronautics
  and Space Administration, Science Mission Directorate, Planetary
  Astronomy Program.}


\begin{abstract}
We report the discovery of the first \ion{He}{1}*$\lambda 10830$ broad
absorption line quasar FBQS~J1151$+$3822.  Using new infrared and
optical spectra, as well as the SDSS spectrum, we extracted the
apparent optical depth profiles as a function of velocity of the
3889\AA\/ and 10830\AA\/ \ion{He}{1}* absorption lines. Since these
lines have the same lower levels, inhomogeneous absorption models
could be used to extract the average true \ion{He}{1}* column density;
the log of that number was 14.9.  The total hydrogen column density
was obtained using {\it Cloudy} models.  A range of ionization
parameters and densities were allowed, with the lower limit on the
ionization parameter of $\log U=-1.4$ determined by the requirement
that there be sufficient \ion{He}{1}*, and the upper limit on the
density of $\log n = 8$ determined by the lack of Balmer absorption.
Simulated UV spectra showed that the ionization parameter could be
further constrained in principle using a combination of low and high
ionization lines (such as \ion{Mg}{2} and \ion{P}{5}), but the only
density-sensitive line predicted to be observable and not
significantly blended was \ion{C}{3}$\lambda 1176$.  We estimated the
outflow rate and kinetic energy, finding them to be consistent but on
the high side compared with analysis of other objects.  Assuming that
radiative line driving is the responsible acceleration mechanism, a
force multipler model was constructed.  A dynamical argument using the
model results strongly constrained the density to be $\log n \geq \sim
7$.  Consequently, the log hydrogen column density is constrained to
be between 21.7 and 22.9, the mass outflow rate to be between 11 and
56 solar masses per year, the ratio of the mass outflow rate to the
accretion rate to be between 1.2 and 5.8, and the kinetic energy to be
between 1 and $5\times 10^{44}\rm \, erg\, s^{-1}$.  We discuss the
advantages of using \ion{He}{1}* to detect high-column-density BALQSOs
and and measure their properties. We find that the large $\lambda
f_{ik}$ ratio of 23.3 between the 10830\AA\/ and 3889\AA\/ components
makes \ion{He}{1}* analysis sensitive to a large range of high column
densities.  We discuss the prospects for finding other
\ion{He}{1}*$\lambda 10830$ BALQSOs and examine the advantages of
studying the properties of a sample identified using \ion{He}{1}*.

\end{abstract}



\section{Introduction}

Active galactic nuclei (AGN) are powered by mass accretion onto a
supermassive black hole.  But while most of the gas is accreted
by the black hole, some fraction of the gas is blown out of the
central engine in powerful winds.  These outflows are seen as the
blue-shifted absorption lines primarily in the rest-frame UV spectra
for about $\sim 50$\,\%\ of AGN as narrow absorption lines
\citep{crenshaw03} and for about $\sim 10$ to 40\,\%\ as broad
blue-shifted absorption troughs \citep[e.g.,][]{weymann91, gibson09,
  dai08}.  Outflows are an essential part of the AGN phenomenon 
because they can carry away angular momentum and thus facilitate
accretion through a disk. Winds are important probes of the chemical
abundances in AGN, which appear to be elevated \citep{hf99}. They can
distribute chemically-enriched gas through the intergalactic medium
\citep{cavaliere02}. They may carry kinetic energy to the host galaxy,
influencing its evolution, and contributing to the coevolution of
black holes and galaxies \citep[e.g.,][]{so04}.

Physical properties, including the mass and origin of AGN outflows, remain 
largely mysterious.   It is particularly important to understand
the acceleration mechanism.   Resonance-line absorption (i.e.,
continuum photons are absorbed as permitted transitions in ions)  is a
promising mechanism, and there is compelling evidence that it is
present in a few objects \citep[e.g.][]{arav95,north06}.  Other
mechanisms include hydromagnetic acceleration \citep{everett05}  and
acceleration due to dust \citep{sn95}.  As discussed by \citet{ge07},
different 
acceleration mechanisms may dominate in different parts of the
outflow. These models could in principle be distinguished by measuring
the
properties of the outflow.  So, for example, while resonance-line
absorption is a  compelling mechanism (since we see the troughs
created  by absorption), there is evidence in some cases that the
mass-outflow rates may be too high for it to be feasible (e.g., Hamann
et al.\  1997; Leighly et al.\ 2009).  Thus, kinetic energy is a 
sensible discriminant for these models; e.g., we ask, is there 
sufficient energy available to accelerate wind of a particular column
density to a particular velocity?  

Determining the kinetic energy deposited into the wind requires
measurement of the velocity, which can be obtained directly from the
absorption line profile, assuming that the flow is radial, and the
column density, which is much more difficult to  measure.  The UV
lines have apparent optical depths generally less than one, so it
appears that column densities could be determined by simply
integrating over the apparent optical depth profile. However, it is
now known that the lines are generally saturated, 
although not black, implying that the absorbing material only
partially covers the source \citep[e.g.,][]{sabra05}.   The column
density and covering fraction can be solved for in special cases when
two or more lines from the same lower level can be measured
\citep[e.g.,][]{hamann97, arav05}.  For example, atomic physics requires
that the 1548\,\AA\ component of the \ion{C}{4} doublet have twice
the optical depth of the 1551\,\AA\ component. Partial covering gives
an apparent optical depth ratio of less than two. The measurements of
the two lines can be used to solve for the two unknowns, the optical
depth and covering fraction. 

This method, while powerful, has limitations.  Blending can be a
problem when the lines are broad.  For example, the \ion{C}{4} doublet
at 1548 and 1551\AA\/ is a suitable pair of lines for this method,
since these two transitions have the same lower level.  However,
their separation is only 2.6\AA\/, corresponding to only $500\rm \,
km\, s^{-1}$.  While these lines may be resolved in narrow-line
objects, they will be profoundly blended in objects with broad lines.   

In addition, this method fails if the lines are saturated.  Lines from
permitted transitions in ions from relatively abundant elements
saturate easily at relatively low column densities. Examples of such
lines are those that are most easily recognized in BALQSOs, such as
\ion{C}{4}.  \ion{P}{5}$\lambda\lambda 1118, 1128$ has been recognized
as a valuable probe of higher column densities \citep{hamann98}
because of its low elemental abundance, only $9.3 \times 10^{-4}$ that
of carbon 
\citep{grevesse07}.  So, from the  observation of \ion{P}{5} in a
quasar, we can generally infer that \ion{C}{4} and similar lines are
saturated, even though they may not be black.  

However, \ion{P}{5} has the problem in that it falls in the far UV
part of the spectrum.  It is therefore, in practice, accessible only
by space-based observatories, except in BALQSOs with redshifts greater
than $\sim 2.5$.  In high-redshift objects, however, Ly$\alpha$ forest
lines ablate the spectrum in the far UV, and \ion{P}{5} may be
difficult to measure.  

In this paper, we report  the discovery of the first
\ion{He}{1}*$\lambda 10830$ broad absorption line quasar
FBQS~J1151$+$3822.  Absorption has been seen in the
\ion{He}{1}*$\lambda 3889$ transition (a transition to 3p from the
metastable 2s level in the \ion{He}{1} triplet) in several quasars 
and Seyfert galaxies \citep[e.g., Mrk~231;][]{boksenberg77}, but it has
never before been reported in the 10830\AA\/ transition (a transition
to 2p from the metastable state).   

We also discuss  the use of \ion{He}{1}* absorption lines at
3889 and 10830\AA\/ for detecting low-redshift, high-column-density
BALQSOs.  \ion{He}{1}* offers a number of advantages.  The metastable
state is populated by recombination from He$^+$, so it is a
high-ionization line, produced in the same gas as the usual resonance
lines such as \ion{C}{4}$\lambda 1549$.   The metastable
state  has a low abundance, comparable to that of $P^{+4}$.  The two
\ion{He}{1}* transitions are widely separated and located in a region of the
spectrum where there are few other absorption lines, so blending is
not a problem.  They are located in the optical and infrared, so
observation is possible from the ground.  Observation is then limited
to lower-redshift quasars, but turns out to be a valuable property
because there are now only a handful of low-redshift BALQSOs known,
many discovered serendipitously in space-UV spectra taken for other
reasons. Finally, these two lines have a high ratio of $\lambda
f_{ik}=23.3$, compared with $\sim 2$ for the usual resonance lines.
This means that the two lines 
together are useful over a larger range of high column densities.
Although the \ion{He}{1}*$\lambda 10830$ line has so far not been used 
extensively in the study of quasars, its value has been recognized for
the study of other objects, including winds in young stars
\citep[e.g.,][]{edwards03}.

How metastable is the \ion{He}{1}* 2s triplet state?  The calculation
of the rate of decay for this and for \ion{H}{1}* 2s has a
fascinating history.      Electric dipole transitions are  strictly
forbidden.   Electric quadrupole and magnetic dipole transitions are
forbidden.  However, decay via a two-photon process is possible
\citep{bs57}.  The rate for the two-photon process was computed for
\ion{H}{1} and discussed for \ion{He}{1} by \citet{bt40}.  They found that
while the lifetime of the singlet state of \ion{He}{1} should be
similar to that of hydrogen, at $0.11$--$0.15$ seconds, the lifetime
of the triplet state should be about $10^{-6}$ times smaller, or $\sim
1.5$ days.  The lifetime was then calculated by \citet{m57} to be
about 0.5 days.   But \citet{dd68} showed that the analysis performed
by \citet{bt40} was not applicable to the triplet state, because while
the singlet state decay has net $\Delta m=0$, the triplet state decay
transports one quantum of angular momentum.   In \citet{dvd69} they
show that the lifetime of the metastable state for the two-photon
decay should be  7.9 years.  But then features from transitions from the
triplet state to the ground from helium-like ions were observed in the
laboratory and the sun, implying that there must be a direct
transition from the metastable state to the ground \citep{gj69}. The
key lay apparently in the relativistic corrections to the magnetic
dipole transition.  \citet{griem69} derived rates commensurate with
the observation using an approximate Dirac theory.  Finally,
\citet{drake71} calculated the rate using higher order terms yielding
the current best estimate of then lifetime of 2.2 hours.  

This paper is organized as follows.  In \S 2, we describe the
infrared and optical observations of FBQS~J1151$+$3822, the
extraction of the apparent optical depth profiles, and application of
partial covering models to extract the true column density and
covering fraction.  In \S 3, we use {\it Cloudy} modeling to extract
column density information from the results, and to compute synthetic
UV spectra.  \S 4 provides a discussion of the kinetic luminosity, the
viability of acceleration mechanisms, the column densities and
covering fractions that can be reliably measured using the
\ion{He}{1}* lines, and the prospects for using \ion{He}{1}* to find
low-redshift quasars.  A brief appendix explores the dependence of our
total hydrogen column density estimates on the spectral energy
distribution, the metallicity, and turbulence.  We use cosmological
parameters $\Omega_\lambda=0.73$, $\Omega_{M}=0.27$,$H_0=71\rm\,
km\,s^{-1} Mpc^{-1}$ unless otherwise specified.

\section{The Observations and Analysis}
 
FBQS~J1151$+$3822 ($z=0.3344$, $M_B=-26.2$) was identified in the First
Bright Quasar Study \citep{white00}, although it is not a radio-loud
quasar ($\log R* = 0.52 < 1$).  

\subsection{Infrared Observations}

\begin{figure}
\epsscale{1.0}
\plotone{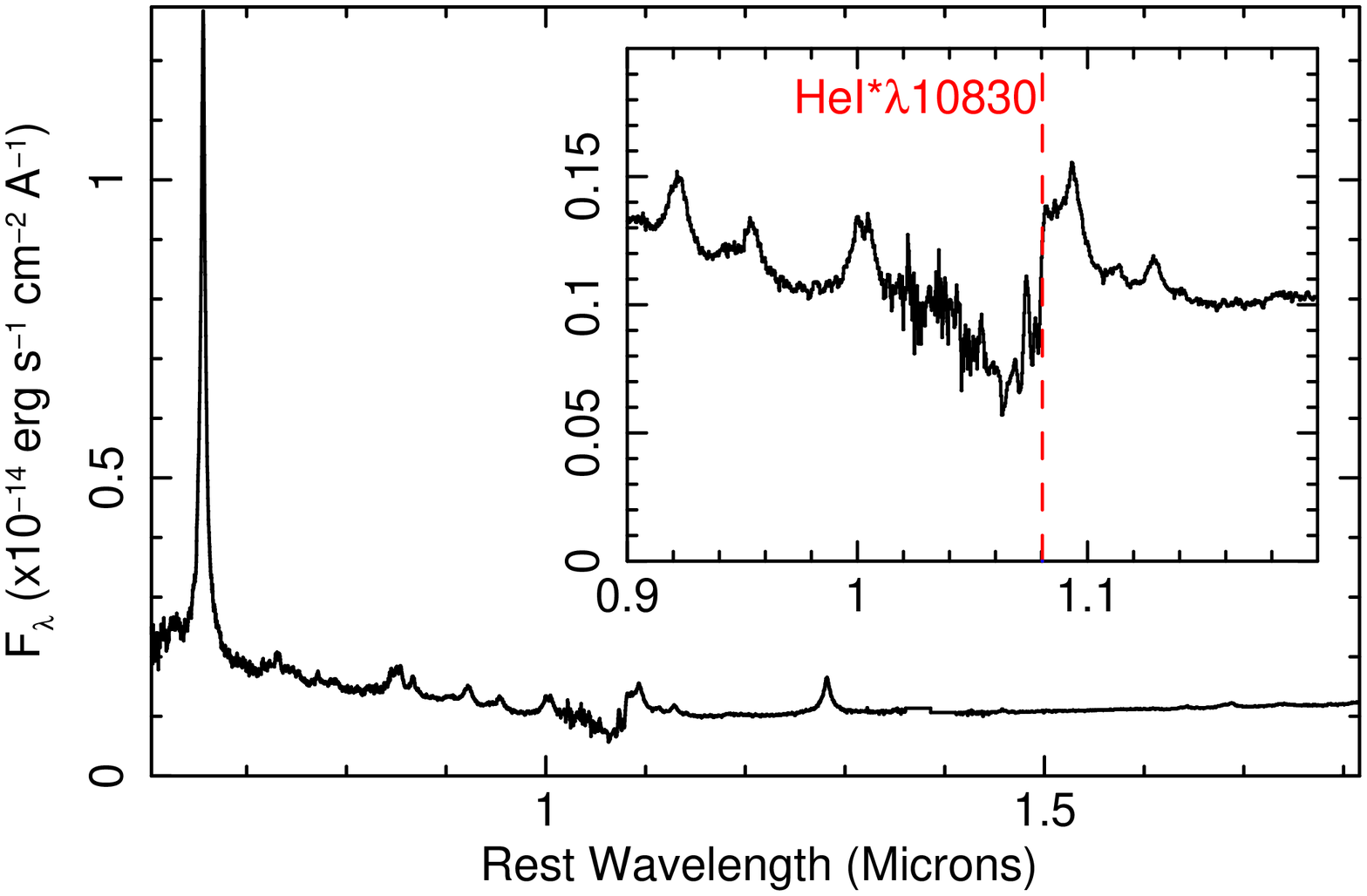}
\caption{IRTF SpeX spectrum of FBQS J1151$+$3822.  The inset shows the
  area of the \ion{He}{1}*$\lambda 10830$ absorption line, where the
  vertical line shows the rest wavelength of \ion{He}{1}*$\lambda
  10830$.  \label{fig1}}
\end{figure}

The infrared spectroscopic observations were made at the NASA Infrared
Telescope Facility 
(IRTF) using the SpeX spectrograph \citep{rayner03} in the short
cross-dispersed mode (SXD) with an effective bandpass of
0.8--2.4$\rm\, \mu m$ as part of another project.  The observations 
were made 2008 March 28--30 and 2008 August 22--24.  Conditions were
non-photometric.  As 
noted above, the scope of this paper is limited to the \ion{He}{1}
absorption, and so we will defer the detailed description of the
observations and reduction to Barber et al.\ (in prep.) which
will include analysis of the entire dataset.  For this paper, we use
the spectra from FBQS~J1151$+$3822, PG~1543$+$489, PHL~1811, and
FBQS~J1702$+$3247; these were all observed with a 0.8'' slit for
effective exposures of 2, 1.7, 2.9 and 1 hour(s) respectively.
  Flat-field lamps and line lamps
were observed along with each object.  A nearby early A-type star was
also observed along with each object.  The spectra were reduced using the
IDL software {\tt   SpexTool} \citep{cushing04}. The flux calibration
and correction for telluric features was performed using the A star
spectrum and employing distributed IDL software that use the methods
described in \citet{vacca03}.    In the vicinity of the \ion{He}{1}
absorption feature (at approximate 1.4$\rm\, \mu m$ in the observed
frame), the FWHM resolution measured from the line-lamp spectra ranged
from $\sim350$ to $\sim400 \rm \, km\,s^{-1}$ at the short wavelength end
of the 4th spectral order to $\sim 300\rm\,  km\,s^{-1}$ at the long 
wavelength end of the 5th spectral order.  The spectrum is shown in
Fig.~\ref{fig1}.

The short wavelength end of the absorption feature 
occurs at the long wavelength end of the 5th spectral order.  There is
a prominent telluric absorption band between $\sim 1.35$ and $1.45\rm\,
\mu m$.  In addition, the effective area is dropping off at the end of
the order.  Unfortunately, our feature is partially compromised by
these issues.  Is it possible that the absorption feature is somehow
an artifact of the data reduction?  This is not likely, primarily
because the long wavelength end of the absorption feature is found on
both the long wavelength end of the 5th spectral order, and on the
short wavelength end of the 4th spectral order, and both have the
precisely the same shape.  In addition, FBQS~J1151$+$3822 is a
relatively bright quasar for the {\it IRTF}, and the telluric
correction method is very reliable \citep{vacca03}.

The spectra were corrected for Galactic reddening using $E(B-V)=0.023$
\citep[derived from the infrared cirrus,][]{schlegel98} using the CCM
reddening curve \citep{cardelli89}.  The redshifts were derived from the
{\it IRTF} spectra assuming that H$\alpha$, Paschen $\beta$ and
Paschen $\alpha$ are observed at the rest wavelengths.  For
FBQS~J1151$+$3822, PG~1543$+$489, PHL~1811, and FBQS~J1702$+$3247, the
redshifts were found to be 0.3354, 0.4104, 0.1928, and 0.1644,
respectively.  These are insignificantly different from the values
listed in NED.  The spectra were shifted into the rest frame using these
redshift values.   Finally, the spectra were modestly
smoothed\footnote{The smoothing function is
  $0.2(f(i-1)+3f(i)+f(i+1))$.}.

\subsection{Derivation of the \ion{He}{1}$\lambda 10830$ Apparent
  Optical Depth}

As discussed in e.g., \citet{ss91}, the apparent optical depth of
an absorption feature is given by
$\tau(\lambda)=\ln[I_0(\lambda)/I_{obs}(\lambda)]$, where
$I_{obs}(\lambda)$ is the observed spectral intensity, and
$I_0(\lambda)$ is the unabsorbed spectral intensity.  The {IRTF}
spectrum of FBQS~J1151$+$3822 yields $I_{obs}(\lambda)$, of course.
In order to obtain the apparent optical depth, we need to obtain the
intrinsic continuum 
$I_0$.  The lines in the near IR spectra of quasars are relatively
unblended, compared with the optical and near UV bands; 
however, the continuum in this region of the spectrum transitions from
the AGN power law at longer wavelengths to the dust-reprocessing
feature at shorter wavelengths (Fig.~\ref{fig1}).  In
FBQS~J1151$+$3822, it appears that this change in the continuum
occurs very near the absorption feature.  

To obtain $I_0$, we used the following procedure.  First, we
determined that of the 12 objects that we observed, PG~1543$+$489,
PHL~1811, and FBQS~J1702$+$3247 had spectra most similar to that of
FBQS~J1151$+$3822 outside of the absorption region.  In addition, we
downloaded the  spectra presented by \citet{landt08} and found two
additional objects that resemble FBQS~J1151$+$3822: PDS~456 and
HE~1228$+$013.  The spectra were similar to that of FBQS J1151$+$3822
in terms of emission lines, but different in their continuum slopes.
We used line-free regions in FBQS~J1151$+$3822 and each of the
comparison objects to determine power laws that the comparison-object
spectra could be divided by so that their continua would match that of
FBQS~J1151$+$3822.  We then performed this continuum transformation.  

\begin{figure}
\epsscale{1.0}
\plotone{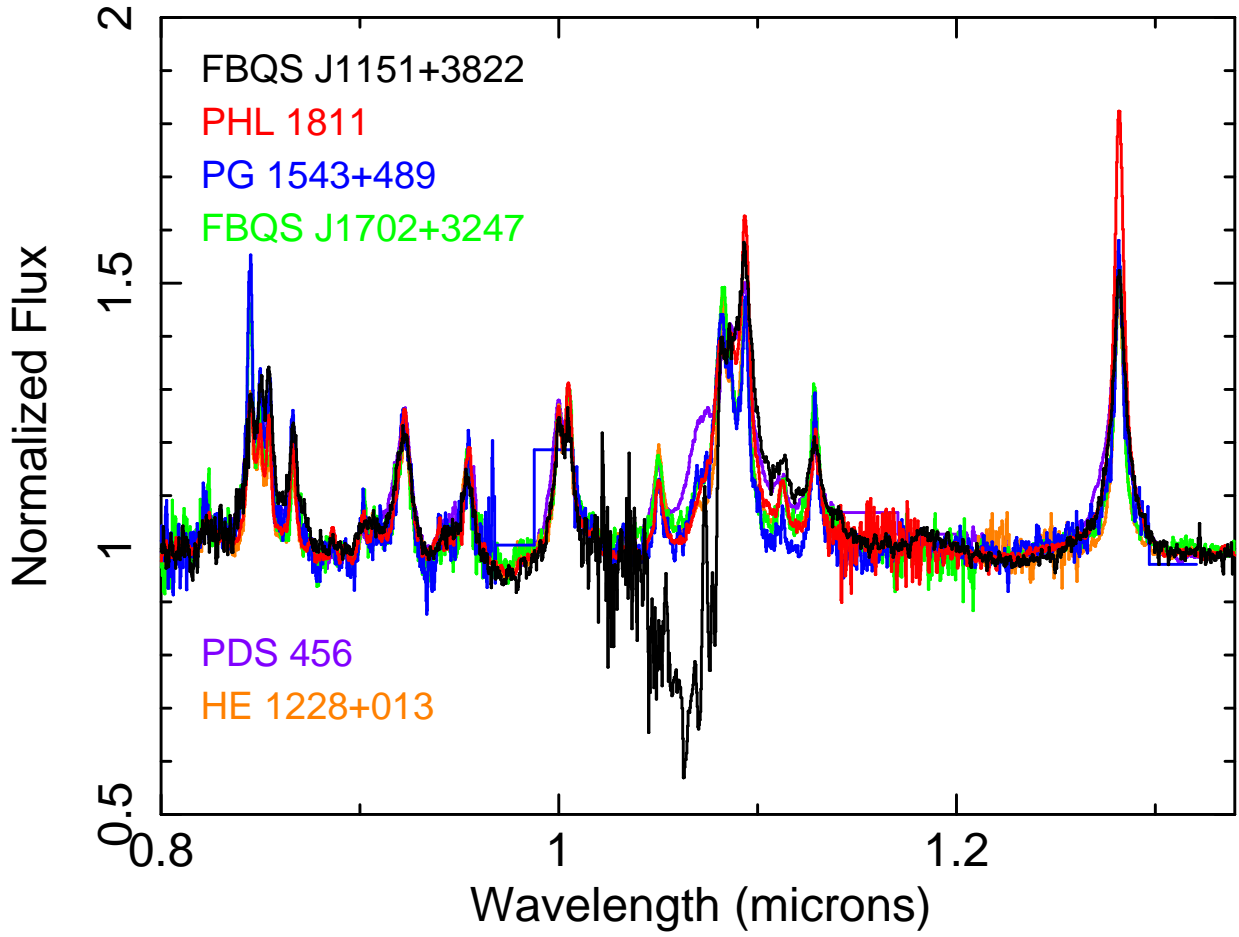}
\caption{SpeX spectrum of FBQS J1151$+$3822 along with the comparison
  object spectra that  have been normalized and scaled (see text for
  details).   \label{fig2}} 
\end{figure}

The procedure above works quite well for the continuum correction.
However, several of the objects have significantly higher equivalent
widths than FBQS~J1151$+$3822.  To scale the lines, we need to divide
each spectrum by its continuum.  We focus on a limited wavelength
range between 0.79 and 1.4$\rm\, \mu m$ and fit the line-free
continuum bands of each object with a 4th order polynomial.  Each
spectrum was then divided by this polynomial.  Finally, the spectra
were suitably scaled.  The following scale factors were used:
PG~1543$+$489: 0.73; PHL~1811: 1.1; FBQS~J1702$+$3247: 0.75; PDS~456:
1.0; HE~1228$+$013: 0.61.  The results are shown in Fig.~\ref{fig2}. 

The figure shows that all objects are rather similar over most of the
range shown except for the long-wavelength side of the 
\ion{He}{1}/Pa$\gamma$ feature.  In particular, in the region of the
absorption feature which is seen to stretch between 1.044 and
1.082$\rm\, \mu m$, most objects are quite similar except for PDS~456
which is seen to exhibit some strange feature near 1.07 microns
\citep{landt08}. Since PHL~1811 has by far the best signal-to-noise
ratio of all of the spectra presented, we  use the scaled version
shown in Fig.~\ref{fig2} as the intrinsic spectral intensity $I_0$.
Except for PDS~456, the other 
spectra match that of PHL~1811 within about 5\%, and we take that
value as the systematic uncertainty in our estimate of $I_0$.  

The resolution of the IRTF spectra in the region of the feature is
about $350 \rm\, km\,s^{-1}$.  Binning the FBQS~J1151$+$3822
spectrum by a factor of two yields slightly more than 
two bins per resolution element.  The PHL 1811 spectrum was resampled
on the same wavelengths as the FBQS~J1151$+$3822 spectrum.  

The IRTF spectra were were extracted using the optimal extraction
technique \citep{cushing04}.  This method produces statistical errors
that appear to be too small. Specifically, the standard deviation in
the data in relatively flat and  smooth regions of the continua are
systematically several factors larger than the assigned errors.  To
account for this trend, we increase the size of the error bars by a
factor of 6 for the PHL 1811 spectrum, and a factor of 4 for the
FBQS~J1151$+$3822 spectrum. Then, between 1.0 and 1.1 microns, in the
region of the absorption feature, the signal to noise ratios range
between about 16 and 130 for the FBQS~J1151$+$3822 spectrum and are
about 200 uniformly for the PHL 1811 spectrum.  

\begin{figure}
\epsscale{.80}
\plotone{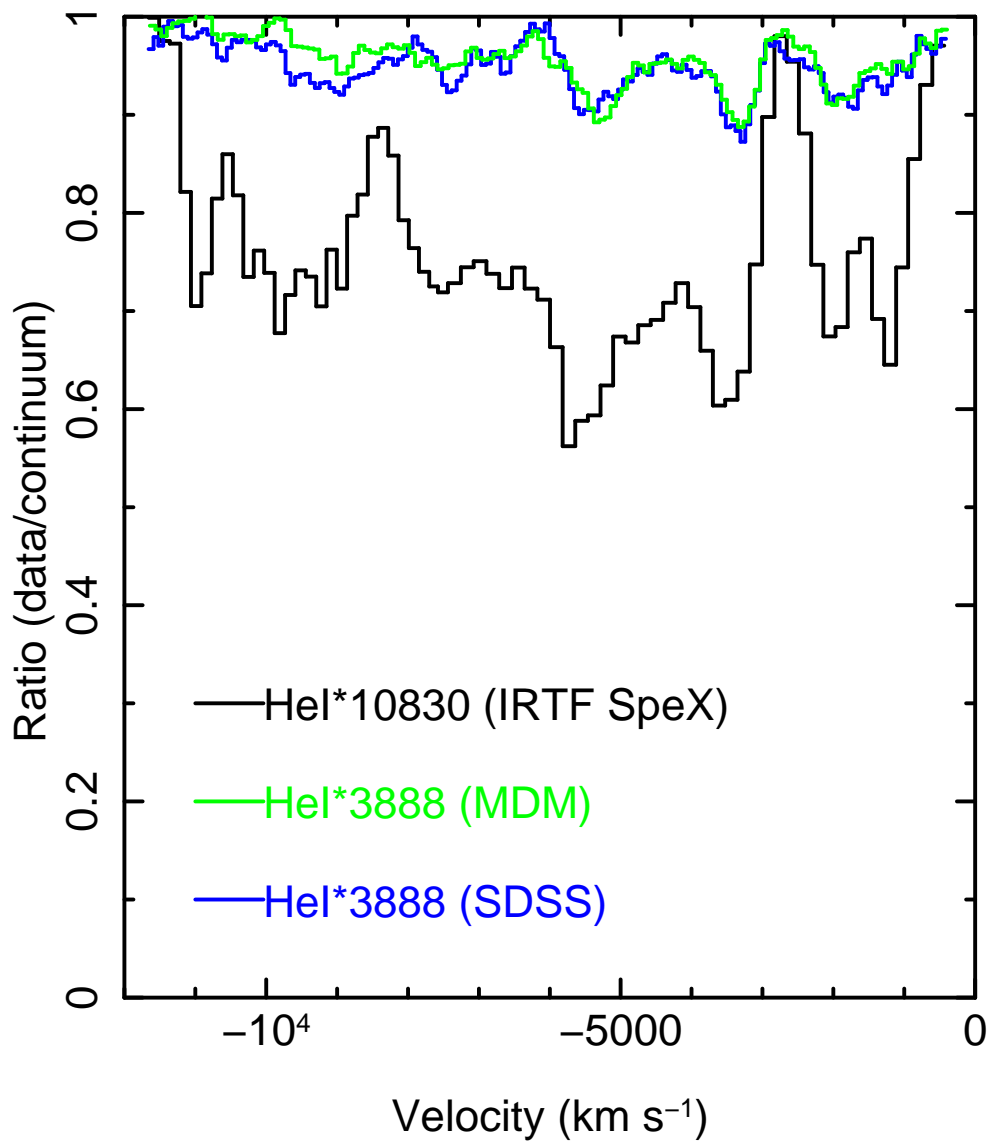}
\caption{The ratio of the data to the continuum ($R$) is shown for
  the \ion{He}{1}*$\lambda 10830$ and the \ion{He}{1}*$\lambda 3889$
  lines.  The apparent optical depth is $\tau=-\ln R$.  The ratios are
  similar for the two lines, but not
  identical, implying the presence of velocity-dependent optical
  depth and covering fraction.  On the linear part of the curve of
  growth, the optical depth of  \ion{He}{1}*$\lambda 3889$ should be
  23.3 times lower than that of \ion{He}{1}*$\lambda 10830$.  It is
  clearly higher than that, implying the presence of partial
  covering.  \label{fig3}}  
\end{figure}

Using these spectra, the ratio and apparent optical depth were
computed as a function of velocity.  The resulting ratio is shown in
Fig.~\ref{fig3}.  The maximum velocity $v_{max}$ is $\sim 11,000 \rm
\, km\, s^{-1}$, showing that FBQS~J1151$+$3822 is truely a broad
absorption line quasar.

\subsection{Optical Observations}

We analyzed two optical spectra of FBQS J1151$+$3822.  The Sloan
Digital Sky Survey observed the object 2005 March 11 for 4600.4
seconds.   The spectrum was extracted from the archive, rewritten in
text format, and a moderate amount of smoothing was
applied as mentioned in \S 2.1.  The SDSS redshift was $0.3351 \pm
0.002$.  We remeasured the redshift using the Balmer lines and
estimated it to be 0.3341.  The spectrum was corrected for redshift
and reddening as above. 

The object was also observed at the MDM observatory 2.4 meter Hilter
telescope and the CCDS spectrograph on January 3, 2009
for 1 hour.  We used a grating with $350 \, \rm mm^{-1}$ and a 1''
slit yielding observed-frame spectra between $\sim 4300$\AA\/ and
$\sim 5800$\AA\/.  Conditions were non-photometric.  The spectra were
reduced in a standard way independently using
MIDAS\footnote{Trade-mark of the European Southern Observatory} (MD)
and IRAF\footnote{IRAF is distributed by the National Optical
  Astronomy Observatories, which are operated by the Association of
  Universities for Research in Astronomy, Inc., under cooperative
  agreement with the National Science Foundation.}
(KML, SB); the results were consistent.  The signal-to-noise ratio of
the FBQS~J1151$+$3822 was estimated to be approximately 70.  The FWHM
resolution in the vicinity of the  \ion{He}{1}$\lambda 3889$ line was
found to be about $175\rm\, km\,s^{-1}$ from the line lamps.  

\subsection{Derivation of the \ion{He}{1}*$\lambda 3889$ Apparent
  Optical Depth} 

\begin{figure}
\epsscale{1.0}
\plotone{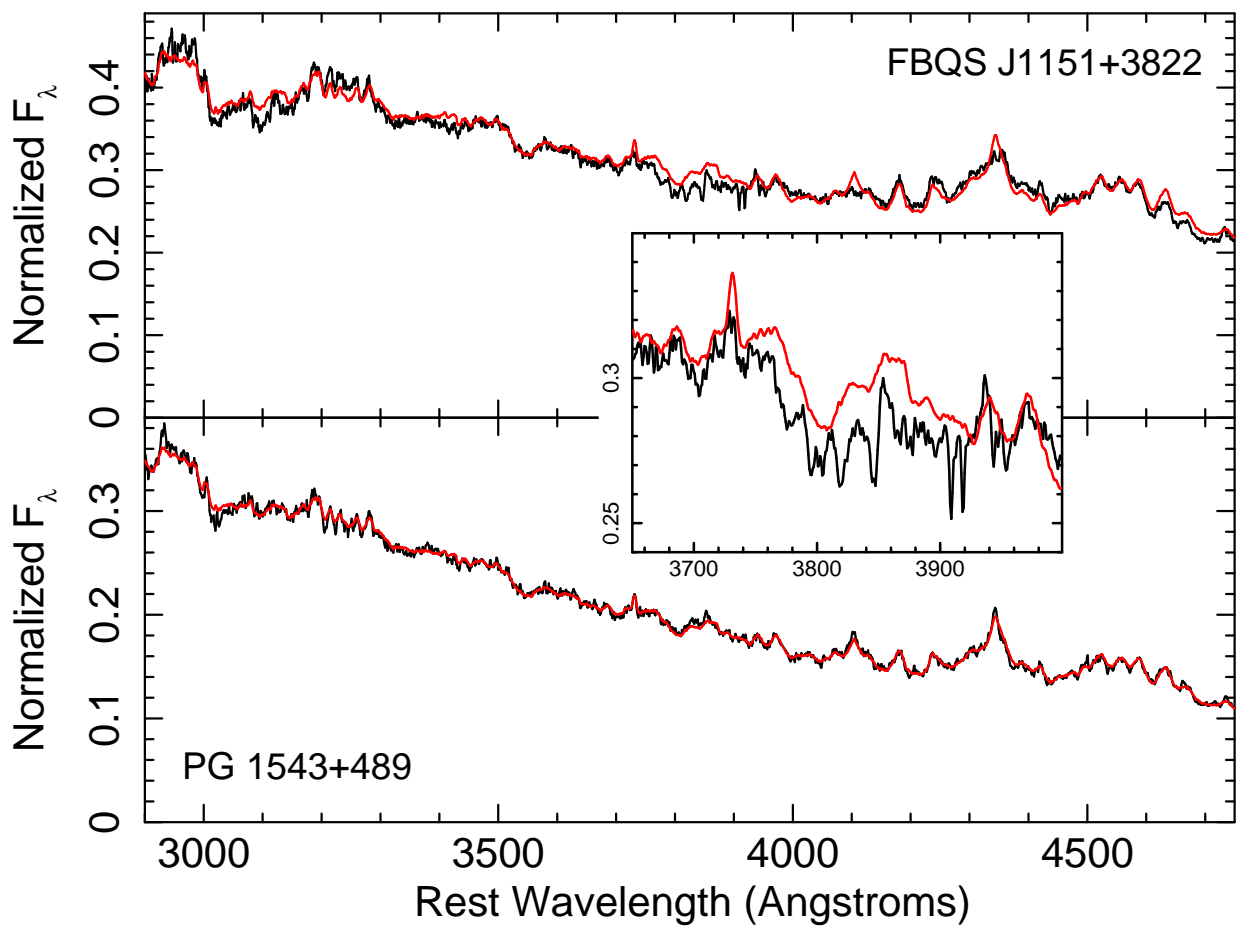}
\caption{Fitting the optical spectra to the template model.  {\it Top:} the
  SDSS   FBQS~J1151$+$3822 spectrum fitted with the template created from
  28 SDSS spectra excluding the region of the absorption lines.  The
  MDM spectrum yields a similar result but with a smaller wavelength
  range (3200--4400\AA\/).  
  {\it Bottom:} The SDSS spectrum from PG~1543$+$489 fit with the same
  model for comparison.  {\it Inset:} the region of the
  FBQS~J1151$+$3822  spectrum in the vicinity of the absorption
  lines.  \label{fig4}}   
\end{figure}

Extraction of the apparent optical depth of the \ion{He}{1}*$\lambda 3889$
feature from either the SDSS or MDM spectrum is challenging.  
First, as can be seen in Fig.~\ref{fig4}, the
apparent optical depth of the 3889 \AA\/ feature is quite
small. This is expected, since the ratio of $f_{ik}\lambda$ between
the 10830 and 3889\AA\/ components is 23.3, which means that on the
linear part of the curve of growth, the true optical depths of these
lines should have the a ratio of 23.3 \citep{ss91}.  Second, the
continuum is not a power law; rather, FBQS~J1151$+$3822 is a strong
\ion{Fe}{2} emitter and the continuum includes a forest of \ion{Fe}{2}
lines.  In fact, to the casual observer, the only clearly apparent
absorption lines are the \ion{Ca}{2} H\&K lines.  Without knowing that
there is a 10830\AA\/ absorption trough in this object, the
\ion{He}{1} lines are easily confused with the \ion{Fe}{2} continuum.
Thus, FBQS~J1151$+$3822 had never before been recognized to be a
BALQSO.  

We need a suitable estimation of the \ion{Fe}{2} continuum to extract
the optical depths.  The usual I~Zw~1 template \citep{bg92} is
inadequate because it does not cover the shorter wavelengths around
\ion{He}{1}*$\lambda 3889$ that we need, so we developed a
template from SDSS spectra.  We selected a 
sample of 28 bright low-redshift quasars that have sufficiently narrow
lines that the \ion{Fe}{2} complexes near H$\beta$ are resolved.  From
these, we constructed an average line spectrum as follows.  The
spectra, de-reddened and transformed to the rest frame, were resampled
onto a common wavelength range between 2869\AA\/ (the lower wavelength
limit of the SDSS spectrum of FBQS~J1151$+$3822) and 4750\AA\/
(considered the upper limit to avoid H$\beta$, as that line can be
quite variable between objects). The spectra were normalized by
dividing by the average flux obtained from a 21-element bin located in
the middle of the chosen wavelength range.  The power law slope of the
continuum was estimated 
by fitting a power law to the flux in line-free bands, and all spectra
were normalized using this estimated value to slope of 1.96 (i.e,
where $F(\lambda) \propto \lambda^{-1.96}$). At each
wavelength, the mean spectrum was computed, and finally, the power law
estimated from the line-free bands was subtracted yielding an average
line spectrum.

The SDSS and MDM spectra were then fitted using this \ion{Fe}{2} line
model excluding the regions where absorption lines, both originating in
\ion{He}{1} and \ion{Ca}{2}, may be present.  For the
SDSS spectrum the excluded regions were 3015--3188.7\AA\/ and
3678--3958 \AA\/.  For the MDM spectrum, with its more limited
wavelength coverage, only the second region was excluded.  Four
parameters were fit: the reddening of the input spectrum (i.e., the
observed spectrum was dereddened by an input amount), the power law
index, the power law normalization, and the line normalization. A
figure of merit was computed at each grid point, and the minimum of
the grid identified the best fit.  The figure of merit
used was the absolute value of the difference between the unreddened
input spectrum, and the power law plus line flux model, divided by the
unreddened uncertainty in the case of the SDSS spectrum (the MDM
spectrum, produced by IRAF, had no formal uncertainties).  

The SDSS spectrum includes a notable feature between $\sim
3200$--3300\AA\/ probably originating in \ion{Fe}{2} multiplets 
M6 and M7 \citep[][e.g.,]{phillips78}.  The
shape of the iron continuum varies from object to object in
the UV \citep[e.g.,][]{leighly07} and so it is possible that it also
varies in the optical.  We therefore create an alternative line
spectrum from seven of the 28 objects which shows the largest excess
variance\footnote{Excess variance is defined as sum of the difference
  between the variance in the mean-normalized flux and the square of
  the mean-normalized uncertainty.} between 3170--3300 \AA\/.  A large
value of excess variance chooses objects which both have prominent
features and good signal-to-noise ratios.   We then fit the MDM and
SDSS spectra with this alternative \ion{Fe}{2} continuum.  

It is clear that the uncertainty in the results produced by the  procedure
described above will be dominated by the systematic errors associated
with the fact that the \ion{Fe}{2} spectrum is not identical among
AGNs.  An estimate of the systematic error can be obtained from the standard
deviation of the mean spectrum from the sample used to create the
\ion{Fe}{2} continuum.   The standard deviation in the region
of the absorption feature has an offset due to normalization offsets
in the average spectra, plus fluctuations with scales of tens of
angstroms with an amplitude of about 2\%.  The offset is unimportant
since the fitting routine will remove uncertainty associated with
that.  However, the fluctuations in the standard deviation are
important since they measure the differences in the spectrum among
AGN.  Thus, we assign a systematic uncertainty of 1\% to the model
spectrum. The statistical uncertainty in the SDSS spectrum in the region of
interest is about 2\%.  Thus the ratio will have an uncertainty of
about 2.2\%, and the apparent optical depth should have an uncertainty
of about the same value.  Based on the number of counts in the MDM raw
spectrum, we estimate a statistical uncertainty of about 1.5\%.
Propagating our estimated 1\% uncertainty in the model spectrum yields
an uncertainty in the ratio for the MDM spectrum of about 1.8\%.  The
results of fitting the  28-SDSS-spectra model to the SDSS spectrum of
FBQS~J1151$+$3889 are shown in Fig.~\ref{fig4}; the other combinations
of spectra and \ion{Fe}{2} model were similar.  The resulting ratio
of data to model for the SDSS and MDM spectra are shown in
Fig.~\ref{fig3}. 

In principle, we should also observe \ion{He}{1}*$\lambda 3188$
absorption in the SDSS spectrum.  But $\lambda f_{ik}$ is 79.4 times
smaller for that component than that of  \ion{He}{1}*$\lambda 10830$, so it is expected
to be even weaker than \ion{He}{1}*$\lambda 3889$, where the
$\lambda f_{ik}$ ratio is 23.3.  Furthermore, the region of the spectrum
around 3188\AA\/ has exceptionally strong and complex \ion{Fe}{2}
emission in it.  In addition, the \ion{Fe}{2} emission in this region
seems to vary from object to object more profoundly than it does
around 3889\AA\/.  So while there is some suggestion of absorption in
this object (Fig.~\ref{fig4}), we cannot robustly extract an apparent
optical depth profile and therefore ignore this line for the rest of
the analysis.

\subsection{Balmer Absorption Optical Depth Upper Limit}

We do not see any significant Balmer line absorption in these
spectra.  However, it may be present at low equivalent widths.  We
fitted the SDSS spectrum  in the vicinity of the H$\alpha$ emission
line with a power law continuum, a Lorentzian profile for the Balmer
line, two \ion{Fe}{2} templates (which will be described in Leighly et
al.\ in prep.), and the optical depth profile obtained from the
\ion{He}{1}* $\lambda 10830$ absorption line.  The resulting component
was transformed to a ratio as a function of velocity, which could be
used to estimate the Balmer column density upper limit (\S 2.8).

The MDM spectral bandpass did not cover H$\alpha$.  The IRTF bandpass
does (Fig.~\ref{fig1}), but the spectrum had lower signal-to-noise
ratio in the vicinity of this line compared with the SDSS spectrum.
Therefore, we use the limit obtained from the SDSS spectrum
henceforth. 

\subsection{\ion{Ca}{2} H\& K Absorption}

\ion{Ca}{2} H\&K absorption lines are visible in the optical
spectra.  There are two relatively narrow ($\sim 300\rm \, km\,
s^{-1}$ FWHM) features that peak at $\sim -1200\rm \, km\, s^{-1}$ and
$\sim -2000\rm \, km\, s^{-1}$.  Both sets of features lie 
redward of the \ion{He}{1}* absorption.  

\begin{figure}
\epsscale{1.0}
\plotone{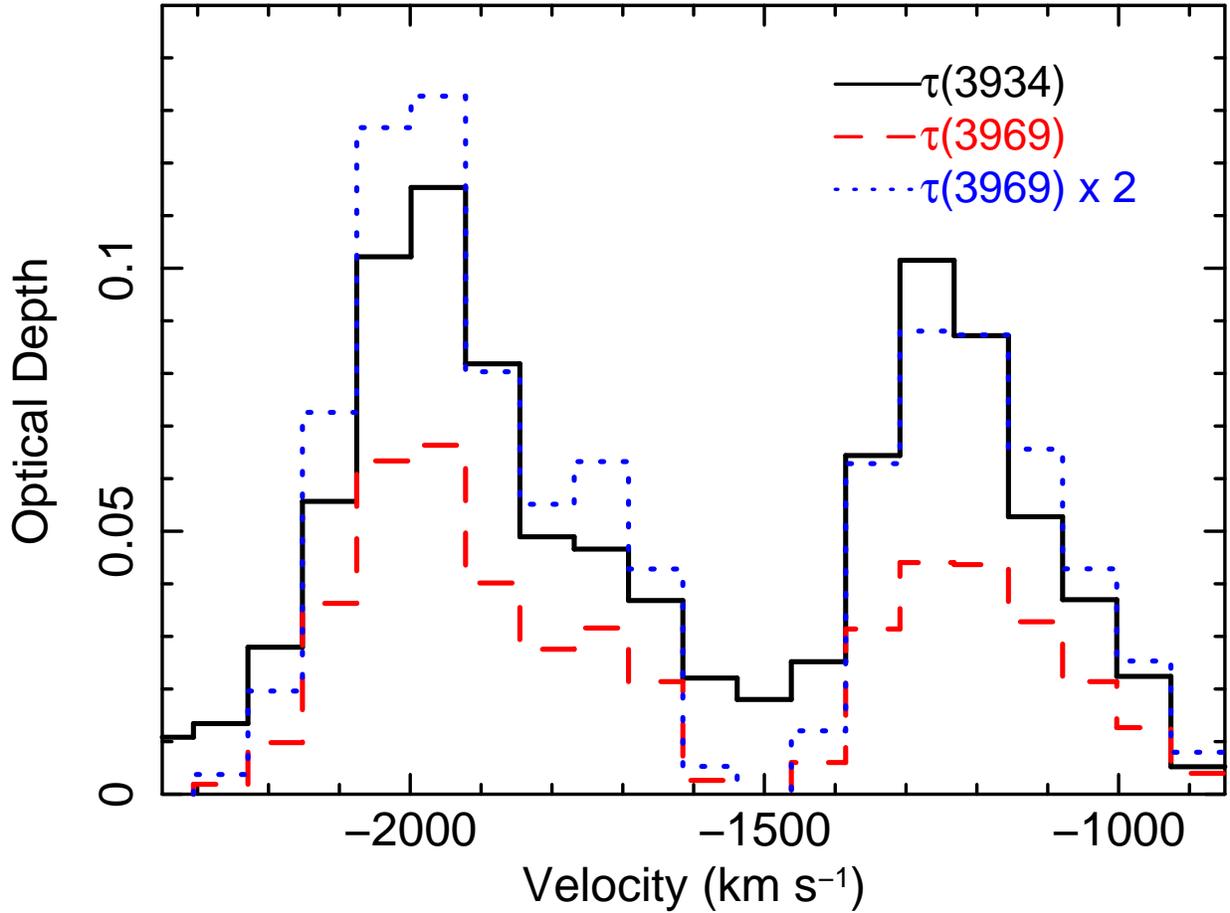}
\caption{The apparent optical depth of the \ion{Ca}{2} H\&K
  absorption feature.  The optical depth from the 3934\AA\/ component
  and the 3969\AA\/ component are shown.  The ratio of $\lambda
  f_{ik}$ for the two components is approximately 2.  We find that
  $2\times$ the apparent optical depth of the 3969\AA\/ component
  approximately matches that of the 3934\AA\/ component.  This implies
  that partial covering is not important and that the true optical
  depth is equal to the apparent optical depth.    \label{fig5}}  
\end{figure}

These two features have low apparent optical depth and are a challenge to
analyze.  We begin with the ratio of continuum to model derived in \S
2.4 for the MDM and SDSS spectra obtained using the two models for the
\ion{Fe}{2} emission discussed in \S 2.4.  The MDM and SDSS profiles
appear consistent, so we average the ratios for each \ion{Fe}{2}
model, leaving us with two spectra.  We extract the apparent optical
depths from both and display the results for the 28-spectrum
\ion{Fe}{2} model in Fig.~\ref{fig5}.  

\ion{Ca}{2} is a lithium-like ion and like other resonance transitions
in lithium-like atoms, the  $\lambda f_{ik}$ for the 3934\AA\/
component is about 2 times that of $\lambda f_{ik}$ for the 3969\AA\/
(actually equal to 2.05).  We found that the apparent optical depths
shown in Fig.~\ref{fig5} appear to have  a ratio of 2:1, and
indeed, if we double the apparent optical depth of the 3969\AA\/
component (dashed line in  Fig.~\ref{fig5}) we find a good
correspondence with the apparent optical depth of the 3934\AA\/
component (solid line).  This implies that partial covering is not
important for these lines and that they are not saturated.  Indeed, if
we use the apparent optical depth profiles to estimate the column
density of Ca$^+$ ions, we find that the measured column density from
the two lines among the two spectra ranges between $9.4\times
10^{12}\rm \, cm^{-2}$ and $1.07 \times 10^{13}\rm \, cm^{-2}$.  These
estimates are consistent, given the uncertainties on the ratio
spectra.  We take the mean of these estimates ($1.0\times 10^{13}\rm
\, cm^{-2}$) to be the column density of the Ca$^+$ ions. 

We believe that this absorption arises in  different gas than the
\ion{He}{1}* absorption. First, the \ion{He}{1}* absorption profile is
much broader; had that broad profile been present in the \ion{Ca}{2}
lines, the 3969\AA\/ component would have severely altered the 3934\AA\/
component, as these two lines are separated by only $2655\rm \, km\,
s^{-1}$.  Likewise, the 3934\AA\/ component would have altered the low
velocity region of the \ion{He}{1}* line, and strong absorption would
have been seen between the \ion{He}{1}*$\lambda 3889$ and the
\ion{Ca}{2}$\lambda 3934\AA\/$ lines (separated by $6100\rm \, km\,
s^{-1}$).  These features are not seen. In addition, the {\it Cloudy}
simulations discussed in \S 3.1 for the relatively high-ionization
\ion{He}{1}* lines do not produce significant Ca$^+$ column
density. The highest log column density 
predicted is $\log N_{Ca+}=12.3$, a factor of five too low; the
highest column densities are found at the lowest ionization
parameters.  Generally speaking, the ionization parameter is higher
and the column density is lower in our simulations than is generally
required for \ion{Ca}{2} absorption \citep[e.g.,][]{hall03}.  Our
interest in this paper is in the \ion{He}{1}* lines, and we do not
discuss the \ion{Ca}{2} absorption further.  

\subsection{Column Density Lower Limits}

As discussed by \citet{ss91} and others, the optical depth is related
to the ratio of the observed continuum $I(\lambda)$ to the true
continuum $I_0(\lambda)$ by
$$\tau(\lambda)=\ln [I(\lambda)/I_0(\lambda)].$$
If the absorption line is not saturated and the absorber fully covers
the source, the column density of the absorbing ion can be obtained
from the optical depth \citep[e.g.,][]{ss91}
$$N=\frac{m_e c}{\pi e^2 f \lambda} \int \tau(v) dv$$
where $m_e$ is the mass of the electron, $c$ is the speed of light,
$e$ is the charge on the electron, $f$ and $\lambda$ are the
oscillator strength and wavelength for the transition, respectively,
and $v$ is the velocity relative to the rest frame with respect to
$\lambda$.  

Using the ratio profiles discussed in \S 2.2, and 2.4, we find that the
log of the column density of HeI* is $\sim 14.3$ from the
\ion{He}{1}*$\lambda 10830$ line, and $\sim 14.75$--$14.95$ from the
various spectra for the \ion{He}{1}*$\lambda 3889$ line.  These two
estimates do not agree, although physically they have to be equal.
This means that the assumption that the absorber fully covers the
source is incorrect, and a partial covering analysis is appropriate
(\S 2.8). 

We discuss the Balmer absorption limit in the next section.

\subsection{Partial Covering Analysis}

HeI*$\lambda 10830$ and HeI*$\lambda 3889$ are transitions from the
same lower level.  Therefore, the ratio of optical depths is fixed by
atomic physics, provided that the lines are not saturated.  We can use
these ratios to solve for the parameters of a two-parameter inhomogeneous
absorption model as a function of velocity.  The simplest example is
the partial covering model, where  
$$R_{10830}=(1-C_f)+C_f e^{-\tau_{10830}}$$ 
$$R_{3889}=(1-C_f)+C_f e^{-\tau_{3889}}$$
where $R$ is the ratio of the observed spectrum to continuum, $C_f$ is
the covering 
fraction, and $\tau_{10830}$ and
$\tau_{3889}$ are the {\it true} optical depths.  All of these
parameters are functions of  velocity \citep[e.g.][]{sabra05}.  We can
solve these equations for 
$C_f$, $\tau_{10830}$ and $\tau_{3889}$, subject also to the
constraint that $\tau_{10830}=23.3 \tau_{3889}$.  

Partial covering models have been widely applied to quasar spectra.
One of the first applications was reported by
\citet{hamann97}, who analyzed metal resonance lines which have $\tau$ ratios 
very close to 2.  In that case, the two equations can be reduced to
one analytically.  For our $\tau$ ratio, an analytic expression cannot
be derived.   We
found that the following method worked most reliably.  First, we
computed a grid of simulated $R_{10830}$ and $R_{3889}$ for a
large range of input covering fractions and $\tau_{3889}$
(specifically, 5001 covering fraction points between 0 and 1, and 4001
logarithmically-spaced $\tau_{3889}$ points between 0.01 and 100).  
Then for each velocity, we determine which pair of $(C_f,\tau_{3889})$
best matches the data, where our figure of merit is $\chi^2$;
specifically,  $\sqrt
{(D_{3889}^2 + D_{10830}^2)}$ and the $D$ values are the difference
between the data points and the models divided by the uncertainty in the
data points.  At the point where the figure of merit is minimized,  we
evaluate an uncertainty on the model point 
where the confidence criterion is our best fitting $\chi^2$ plus 1.0,
corresponding to $1 \sigma$ for one parameter of interest, in the
direction along the parameter axis. Note that the best-fitting
$\chi^2$ values were all very small fractions of 1 since we have two
equations and two unknowns, essentially, and since our grid was finely
meshed.   Finally, we exclude points that are unphysical, where
$R_{3889} > R_{10830} > R_{3889}^{23.3}$ is not obeyed
\citep[e.g.,][]{hamann97}.  Generally, this amounts to only a few
points; i.e., of the 69 points in the velocity profile, 5--8 points
are rejected (Table 1).

\begin{figure}
\epsscale{0.6}
\plotone{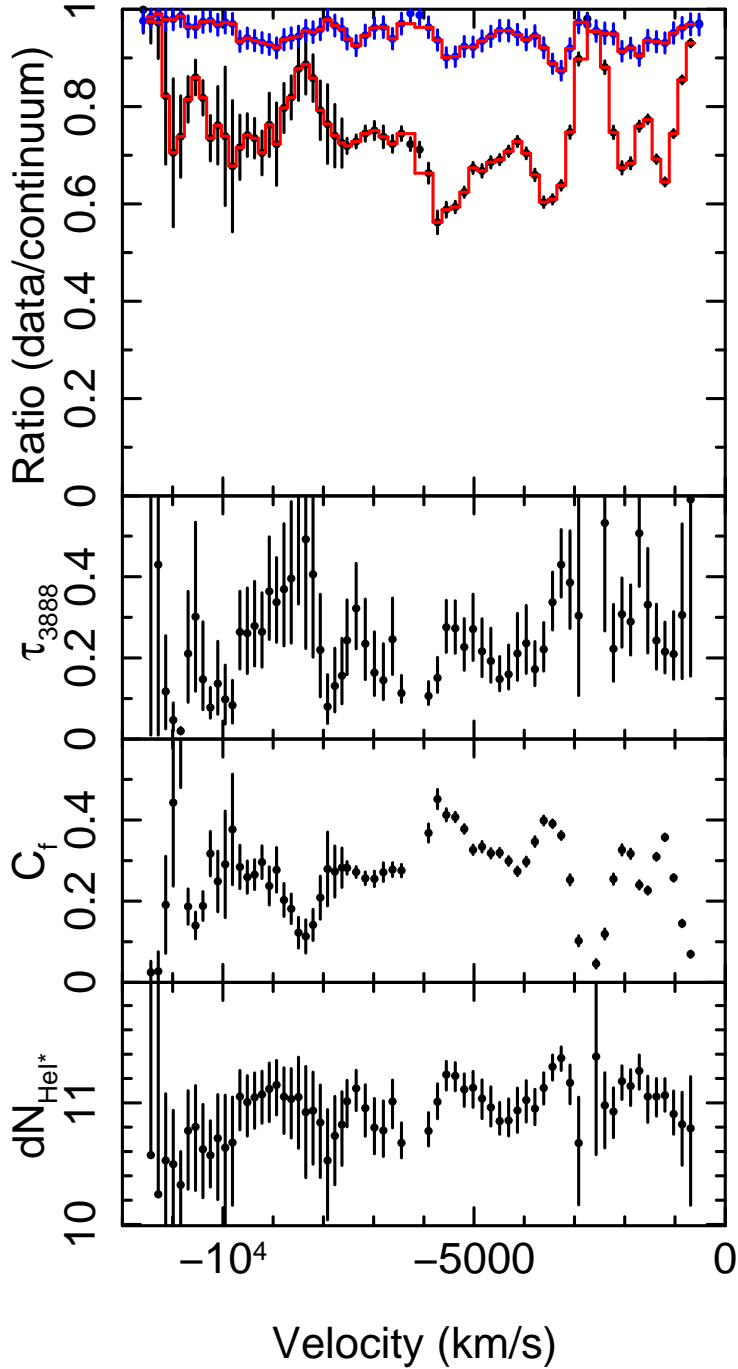}
\caption{Partial covering model of the IRTF ratio and the SDSS
  ratio extracted using the 28-spectrum \ion{Fe}{2} model. The results
  for the excess-variance \ion{Fe}{2} model and for the MDM ratio
  spectrum are similar.  The top 
  panel shows the ratios with the best-fitting model overlaid.  The second and
  third panels show the fitted optical depth and covering fraction.
  The bottom panel shows the incremental column density as a function
  of velocity computed from the average optical depth in each velocity
  bin. \label{fig6}}  
\end{figure}

Fig.~\ref{fig6} shows the results for the IRTF ratio for the
\ion{He}{1}*$\lambda 10830$ line and the SDSS spectrum fit  with
the 28--SDSS-spectrum \ion{Fe}{2} model for the \ion{He}{1}$\lambda
3889$ line; the others were similar.  The top panel shows the data and
the model 
fit.  The second panel shows the optical depth for the 3889\AA\/ line;
the value for the 10830\AA\/ line would be 23.3 times larger.  The
third panel shows the covering fraction as a function of velocity.
The fourth panel shows the incremental column density as a function of
velocity \citep[e.g.,][]{ss91} computed from the average $\tau$, i.e.,
the value integrated over the spatial dimension
\citep[e.g.][]{arav05}.  For the partial covering model,
$\bar{\tau}=C_f \tau$.  The uncertainties in the incremental column
density are conservatively estimated from the products of the limits
of the optical depth and covering fraction.   

\begin{figure}
\epsscale{0.6}
\plotone{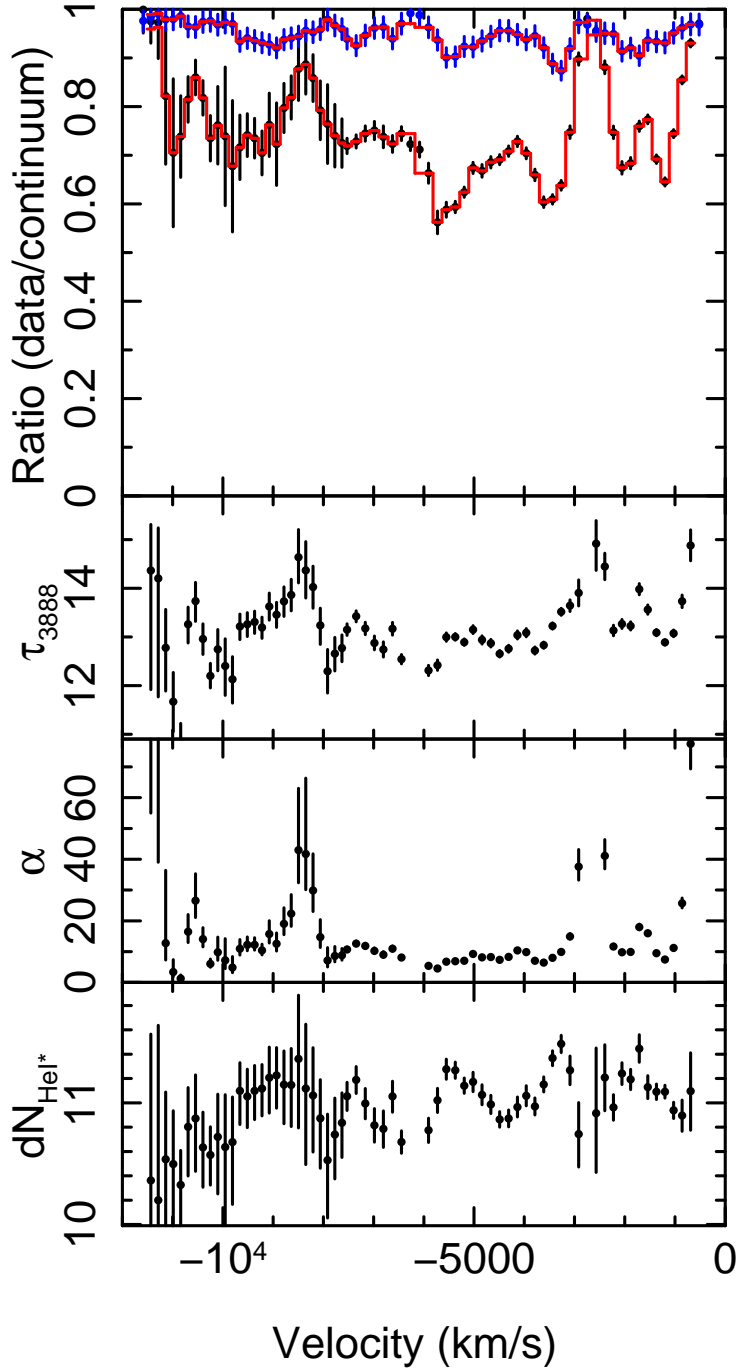}
\caption{Power law inhomogeneous absorber  model of the IRTF ratio and
  the SDSS   ratio extracted using the 28-spectrum \ion{Fe}{2} model.  The results
  for the excess-variance \ion{Fe}{2} model and for the MDM ratio
  spectrum are similar. 
  The top   panel shows the ratios with the model overlaid.  The
  second and   third panels show the fitted optical depth and power
  law index.  The bottom panel shows the incremental column density as a function
  of velocity computed from the average optical depth. \label{fig7}}  
\end{figure}

The simple partial covering model, above, posits a physical scenario
where $1-C_f$ of continuum source is seen directly, while $C_f$ of the
source is uniformly covered by gas with a single value of
$\tau$.  This scenario may be appropriate in some cases, for example,
when a fraction of the continuum is scattered into our line of sight
by electrons located in the vicinity of the symmetry axis; the
presence of these electrons is indicated by the large polarization
frequently seen in the absorption troughs   \citep[e.g.,][]{ogle98}.
But it is also possible that the 
optical depth is not uniform over the source.  These so-called
inhomogeneous absorbers have been discussed by
e.g., \citet{dekool02}, \citet{sabra05}, and \citet{arav05}.  We
choose investigate the power law inhomogeneous absorber model. We use a
power 
law function, i.e.,  $\tau(x)=\tau_{max} x^{a}$, where $\tau_{max}$
and $a$ are fit parameters.  We use the analytical formula for the
ratio of the data to model given in \citet{sabra05} (their equation
14), and follow the approach outlined above.  Our model grid
consisted of 3501 values of $a$ spaced logarithmically between 0.0316
and 100, and 4001 values of $\tau_{max}$ spaced logarithmically
between 0.01 and 100.  The resulting fit for the IRTF spectrum and the
SDSS spectrum fit with the 28-spectrum \ion{Fe}{2} model is shown in
Fig.~\ref{fig7}.  In this case, the average optical 
depth is $\tau_{max}/(a+1)$.  The upper limit on the column density
was conservatively estimated using the upper limit on $\tau_{max}$
divided by the lower limit on $a$; similarly, the lower limit was
estimated using the lower limit of $\tau_{max}$ and the upper limit of
$a$.  

For both models, the total \ion{He}{1}* column density is obtained by
integrating over the the average $\tau(v)$ profile \citep[e.g.,][]{ss91},
and the uncertainties obtained by integrating over the upper and lower
bounds. The results are given in Table 1.  The values for the partial
covering model and power law model are essentially identical.  In
addition, the values are very similar  among the four different
\ion{He}{1}*$\lambda 3889$ ratio profiles. The average of the log of
the eight column density estimates is 14.9.  This is the value that we use
for {\it Cloudy} simulations and the remainder of the paper.   

The derived value of \ion{He}{1}* column density is very similar to
the lower limit derived from the 3889\AA\/ component (\S 2.7; Table 1).
This is not an accident.  To obtain the lower limit, we assume
$$R_s \approx exp(-\tau_s)$$ 
for the weaker component, for example.  As noted above,  the partial
covering model is
$$R_s=(1-C_f)+C_f \exp(-\tau_s).$$ 
If the optical depth of the weaker component is sufficiently small,
the exponent can be Taylor expanded, as follows:
$$R_s \approx (1-C_f)+C_f(1-\tau_s)=1-C_f \tau_s.$$
Inverting the Taylor expansion, we find that 
$$R_s \approx exp(-C_f \tau_s)= exp(-\bar{\tau_s}).$$
Thus, for sufficiently low optical depth, the apparent optical depth is
approximately equal to the average optical depth.  

We use this fact to derive the upper limit on the column density of
the hydrogen $n=2$.  As discussed in \S 2.5, there is no apparent
Balmer line absorption in this object, and we treat the observed small
optical depth as a lower limit.  Needless to say,  the optical depth
is very small.  Thus, we can assume that the apparent optical depth is
approximately equal to the average optical depth.  Next, we need the
oscillator strength for the Balmer absorption line.  The sum of the
oscillator strengths for the 2s state is 0.44 and the sum for the 2p
state is 1.42.  If the $n=2$ Hydrogen is distributed among the 2s and
2p states according to their oscillator strengths, we obtain an
estimate of the log of the hydrogen $n=2$ column equal to 12.9.  But
there is no permitted transition to ground from the 2s state, so one
might expect that it be more highly populated than the 2p state.  If
all the absorption lines were from 2s only, we would obtain an estimate
of the log of the Hydrogen $n=2$ column equal to 13.5.  In the {\it
  Cloudy} modeling, we can separate hydrogen $n=2$ in the 2s state and
in the 2p state, and we will adjust the hydrogen $n=2$ column upper
limit according to the fraction predicted in these two states. We
discuss this in the next section.

\section{Cloudy Modeling}

\subsection{Ionization Parameter, Density and Hydrogen Column Density
  Constraints}  

Armed with our estimates of the \ion{He}{1}* column density and
hydrogen $n=2$ column density upper limit, we can proceed to use these
to constrain the  properties of the absorbing gas.  We use {\it Cloudy} 08.00,
last described by \citet{ferland98}.  Initially, we used the so-called
{\it Cloudy} AGN continuum with the same parameters as used by
\citet{korista97}\footnote{The {\it Cloudy} command for this
  continuum is ``AGN Kirk''.};
this is taken to be a typical AGN continuum.  We examine a wide range
of ionization parameters ($-2.0 \leq \log U \leq 1.2$) and densities
($3.0 \leq \log n \leq 9.0$).  We use the full
\ion{Fe}{2} model \citep[i.e., 371 levels;][]{verner99} for every run.
\ion{He}{1}* is populated essentially solely by recombination of
He$^+$, and therefore, we expect the metastable state of the
\ion{He}{1} ion to be coincident with 
He$^+$.  Therefore, initially, we set the column density equal to
$\log N = 24.5 + \log U$; this integrates through the hydrogen
ionization front and therefore certainly through the entire He$^+$
region.  

The column density of \ion{He}{1}* can be obtained from the
output in two ways.  First, the total column density can be output
using the {\tt punch some column densities} command\footnote{This
  command is described {\it
    Hazy} Volume 1 on page 142 for the 08 release.}  Second, the
fraction of helium in different states  in every calculational zone
can be output if the {\tt print every} command is used.  The 
fraction can be converted to a density using the helium abundance; it
can then be integrated over depth.  We process our {\it Cloudy}
results by first obtaining the log of the total \ion{He}{1}* column
density and 
determining whether or not it is greater than or equal to 14.9.  If it
is not, that combination of ionization parameter and density are
rejected. If it is, we interpolate to determine the thickness of the
gas required to reach a log \ion{He}{1}* column density of 14.9.
Using the gas density and this thickness, we obtain the total hydrogen
column density. We then run a second set of {\it Cloudy} models as a
function of ionization parameter and density in which the total column
density is set to the value estimate from the depth and density.

The density as a function of depth of hydrogen in various levels such
as $n=2$ s and p can be output using the {\tt punch hydrogen
  populations} command.   The output of the second set of runs is
examined and the total column density of hydrogen in the the 2s and
2p states is computed.  The upper limit constraint is computed based
on the fraction of the $n=2$ hydrogen in the two states.  If the
amount exceeds the limit, that combination of ionization parameter and
density is rejected.  It is interesting to note that we can't combine
this step with the previous one because the density of hydrogen in
$n=2$ differs whether it is in the middle or back side of the cloud,
as a consequence of radiative transfer.  

The results are shown in Fig.~\ref{fig8}.  Solutions at low
ionization parameter are rejected because they fail to produced
sufficient \ion{He}{1}*.  Basically, for solutions with the lowest
allowed ionization parameter, we are integrating through the entire
\ion{He}{1}* zone.  For higher ionization parameter solutions, we
integrate only part of the way through the \ion{He}{1}* zone.  Solutions at
high density are rejected because they produce too much $n=2$
hydrogen. The inferred log hydrogen column density is larger than 21.6 and may
be as large as 23.8, or even larger, although it should be noted that
Thompson scattering becomes important for the higher column
densities.  The ionization parameter is greater or equal to $\log U =
-1.4$.    

The total hydrogen column density inferred, as well as the extent of
the allowed $\log U$--$\log n$ parameter space, will also depend on
the shape of the input spectral energy distribution, the metallicity
of the gas, and whether or not the gas has a differential velocity
field (i.e., turbulence).  The effect of these factors is discussed in
the Appendix.

How do these results compare with results from studies of
BALQSOs? \citet{dunn10} compile a collection of results from analyses
of 8 quasars  where partial covering has been taken into account.  The
range of column densities  in their Table 9 is 19.9--22.2; however,
our minimum value is larger than all but two.  \citet{dunn10} find a
range of $\log U$ between $-3.1$ and $-0.7$; our object has a larger
ionization parameter than all but three.  But these objects were not
chosen as high-column candidates.  \citet{hamann98} report analysis of
the \ion{P}{5} quasar PG~1254$+$047.   They find, for solar
metallicity, that $\log U > -0.6$ and $\log N_H > 22$.  These values
are larger than our minimums, but we note that if the ionization
parameter were $\log U=-0.6$, then the column density would be $\sim
22.3$, larger than \citet{hamann98}'s estimated lower limit.  Thus, it
appears that FBQS~J1151$+$3822 is a high-column BALQSO.  In addition,
as we will discuss later, these results suggest that \ion{He}{1}* is
useful to constrain high-column outflows. 

\subsection{Additional Constraints Possible with UV Spectra}

Our solutions encompass a fairly large range of ionization
parameter-density parameter space.  Is it possible to constrain the
solution better?  Certainly, we cannot with our current spectra,
limited, as they are, to the rest optical and infrared bandpass.
However, the rest UV offers numerous transitions that may produce
absorption in FBQS~J1151$+$3822.

Which ions could help us constrain $\log U$--$\log n$ parameter space?
Ideally, we need ions that have column 
densities large enough to make detectable absorption lines, but not so
large that the lines become saturated.
\citet{ss91} show that the optical depth as a function of velocity is
$\tau(v)=2.654 \times  10^{-15} f_{ik} \lambda N(v)$.  This equation
has to be integrated over velocity; in our case, the column-weighted
mean velocity is $\sim 5500\rm \, km\,s^{-1}$, so we multiply
$\tau(v)$ by that value to estimate the integrated column density.   Many
of the resonance lines of interest have total $f_{ik} \sim 0.3$--1.0,
and the lines are in the UV, so $\lambda$ is of the order of, say,
1500\AA\/.  Then, a ratio of observed to continuum of $\sim 0.5$
corresponds to a column density of $\sim 2 \times 10^{15}\rm \,
cm^{-2}$.  The natural log of the ratio is correlated with the column
density.  So a 10-times higher column density would yield a ratio of
$\sim 0.001$, and the line would be saturated.  Thus, for the commonly
observed resonance lines (e.g., \ion{C}{4}),  an ionic column around
$10^{16} \rm \, cm^{-2}$ would be 
likely to be saturated.  Likewise, a column density 10 times lower
would produce a ratio of 0.93, and as we saw with \ion{He}{1}*$\lambda
3889$, such a weak absorption line would be somewhat difficult to detect.
So we are looking for ions with log column densities in the vicinity
of $14.7$ to $16$.    This analysis of course breaks down for
transitions with lower oscillator strengths; such lines could be
detected at higher column densities.

We use {\it Cloudy} models to estimate the ratio of spectrum to continuum by
making use of the fact that $I/I_0\approx exp(-\bar\tau)$ and
$\tau(v)=2.654 \times 10^{-15} f_{ik} \lambda N(v)$ \citep{ss91} where
$N(v)$ is the ionic column density as a function of velocity in the
units of $\rm cm^{-2}$ per $\rm km s^{-1}$.  The total ionic column
densities are obtained from the {\it Cloudy} models. To estimate  $N(v)$, we
make the approximation that the total column density is equally
distributed in velocity space over the $11,000\,\rm km\,s^{-1}$
absorption line.   We examined most of the lines predicted to be
bright in quasar absorption systems by \citet{verner94} for wavelengths
longer than 900\AA\/.  \ion{He}{1}* is essentially a high-ionization
line, and many low-ionization species\footnote{These include
  \ion{H}{1}, \ion{C}{1}, \ion{N}{1}, \ion{N}{2}, \ion{O}{1},
  \ion{Na}{1}, \ion{Mg}{1}, \ion{Al}{1}, \ion{Al}{2}, \ion{Si}{1},
  \ion{P}{1}, \ion{S}{1}, \ion{Ca}{1}, \ion{Ca}{2}, \ion{Fe}{2},
  \ion{Fe}{3}} have low column densities and are predicted to  produce
essentially no absorption   lines.  A few, including \ion{C}{2},
\ion{Al}{3}, and \ion{Si}{2}, are predicted to produce weak lines for
the lower ionization parameter range, but none at the higher
ionization parameters.   Others, such as the resonance lines of
\ion{C}{3}, \ion{C}{4}, \ion{N}{5}, \ion{O}{6}, and \ion{Si}{4}, are 
expected to be saturated.  But some lines fall mid-way between these
two extremes, showing strong evolution in ratio as a function of
ionization parameter, and in some cases, also dependence on density.
The most important of these (plus \ion{C}{2} as 
an example of a line expected to be present only at lowest ionization
parameters) are shown in Fig.~\ref{fig11}.   

\subsubsection{Ionization Parameter and Density Discrimination}

Fig.~\ref{fig11} has several intriguing features.  First, several of the
lines, including \ion{Mg}{2}$\lambda 2800$, \ion{C}{2}$\lambda 1335$,
\ion{N}{3}, \ion{Si}{3}$\lambda 1206$, and \ion{S}{3}$\lambda 1195$
decrease with increasing ionization parameter, while
\ion{P}{5}$\lambda 1121$ increases with increasing  ionization
parameter.  This suggests that observing \ion{P}{5} with  along with
e.g., \ion{N}{3}$\lambda 990$, would allow us to constrain 
 the ionization parameter.  The ionization parameter is degenerate
 with the total column density (Fig.~\ref{fig8}), so this would be an
 important constraint for this outflow.

Several lines in this bandpass are expected to have density
dependence, and the {\it Cloudy} simulations verify that expectation.
 \ion{S}{4}$\lambda 1063$ was observed in WPVS~007 and the properties
 of this line were discussed in \citet{leighly09}.  As noted in that
 paper, \ion{S}{4} is composed of three components.  The ground state
 transition has a wavelength of 1062.7\AA\/.  The other two
 transitions from this configuration arise from an excited state with
 $E_i=951.43 \,\rm cm^{-1}$, with wavelengths of 1072.96 and
 1073.51\AA\/.  Population of the upper level requires $n_e >
 A_{ki}/q_{ki}$ where $A_{ki}=7.70\times 10^{-3}$ is the Einstein A
 value for the transition from the excited state to the ground state,
 resulting in the \ion{S}{4} $10.51\rm\, \mu$ line, and $q_{21}$ is
 the collision deexcitation rate, which can be found from the
 collision strength $\Upsilon$ for this transition. For a nebular
 temperature of 15,000 K, \citet{ss99} give
 $\Upsilon($\ion{S}{4})$=8.48$.  The resulting critical density is
 $n_{crit}=4.7\times 10^4\,\rm cm^{3}$. At densities above critical,
 the ratio would approach $g_{ik}(excited)/g_{ik}(ground)=2.02$.  This 
suggests that observing \ion{S}{4} would allow us to pin down the
density of the gas.  Unfortunately, this line is predicted to be
saturated in this object.

The \ion{C}{3}*$\lambda 1176$  line is also predicted to show
density dependence.  This line arises from the metastable
configuration $1s^22s2p$ to $1s^22p^2$, where both the upper and lower
levels have J substructure.   The transition from the lower level to
the ground state produces \ion{C}{3}]$\lambda 1909$.  The critical
  density for this transition is   $\sim 5 \times 10^{9}\rm\,
  cm^{-3}$, so this line should show density dependence in the
  vicinity   of that density, and that behavior is seen in the {\it
      Cloudy} model results shown in Fig.~\ref{fig11}.  This
  line has been seen in absorption in e.g., NGC~4151 \citep{kriss92,
    kraemer01, kraemer06}.  It has also been seen in emission in a
  number of AGN \citep[e.g., I Zw 1,][]{laor97}.  As shown in
  Fig.~\ref{fig11}, this line is expected to be present for lower
  ionization parameters and high densities, but generally, it seems to
  be characterized by slightly too low ionization for the outflow
  predicted from the \ion{He}{1}*   properties in FBQS~J1151$+$3822.

\ion{Si}{3} is similar in structure to \ion{C}{3}, so in principle
there may be absorption from the metastable level producing an
absorption line near 1112\AA\/.  The transition from the metastable
level to ground produces \ion{Si}{3}]$\lambda 1892$.  This transition
has a critical density of $1.1  \times 10^{11}\rm \, cm^{-3}$.  The 
 \ion{Si}{3}*$\lambda 1112$ line was seen in emission in   PHL~1811
 \citep{leighly07}.  Unfortunately, that line is not   modeled in
 {\it Cloudy}.  But with a critical density of $1.1 \times 10^{11}\rm
 \, cm^{-3}$, the presence of this line would violate our upper limit
 on the density imposed by the lack of Balmer absorption lines.  

Fig.~\ref{fig11} shows that the {\it Cloudy} modeling predicts that
\ion{P}{4}$\lambda 951$ should be density dependent.  The P$^{+3}$ ion
has a similar structure as the C$^{+3}$ ion, where the
\ion{P}{4}$\lambda 951$ transition is analogous to the
\ion{C}{3}$\lambda 977$ transition, and there is a semiforbidden
transition \ion{P}{4}]$\lambda 1467$ analogous to \ion{C}{3}]$\lambda
1909$.  We were unable to find a critical density for the 
\ion{P}{4}]$\lambda 1467$  line in the literature.  At any rate, the
  \ion{P}{4}$\lambda 951$ transition is permitted, so why would it
  show any density dependence?  We believe this happens because at
  high densities, collisional de-excitation causes the ground state to
  have a higher electron population than it would at lower densities
  where the metastable state is populated.   Thus this line is
  stronger at higher densities.  

In principle, the density dependence shown by \ion{P}{4}$\lambda 951$
  should also be seen in \ion{C}{3}$\lambda 977$; however, that line is
  predicted to be highly   saturated in this spectrum.  Similarly,
  there may be absorption from the metastable state of \ion{P}{4}.
  There are six   transitions from   the metastable level (which has
  fine structure)   between 1025.6 and   1035.5 \AA\/ with a total
  $f_{ik}=1.44$, and   six transitions between   823.2 and 827.9 with
  a total   $f_{ik}=2.39$.  Either of these could   produce
  significant   absorption assuming that the metastable state   were
  significantly   populated.  Considering phosphorus has low
  abundance, that may be a difficult situation to realize.

\subsubsection{Blending and Simulated Spectra}

The discussion in the previous section indicates that there should be
several lines that are 
neither too weak nor saturated  that can help us discriminate among
densities and ionization parameters for the solutions shown in
Fig.~\ref{fig11}.  However, with $v_{min}\sim 500\rm \,
km\,s^{-1}$ and $v_{max} \sim 11,000 \rm \, km\, s^{-1}$, blending
will be a problem.  To explore that
issue, we use the results of the {\it Cloudy} simulations to simulate
spectra.  We make the assumption that the optical depth profile as a
function of velocity can be uniformly approximated by the apparent
optical depth profile from the \ion{He}{1}*$\lambda 10830$ line, so
that we do not take into account partial covering explicitly.  We also
assume that the integrated optical depth for each line corresponds to
the ionic column density predicted for the appropriate ion. One
hundred four transitions were considered.  The resulting ratio of
spectrum to continuum is shown in Fig.~\ref{fig12} for three
combinations of ionization parameter and density from Fig.~\ref{fig11}.   

Comparison of the top and middle panel (both simulations using a
relatively low ionization parameter) with the bottom panel shows that
we should be able to distinguish between ionization parameters using
UV spectra.  This is important because the total hydrogen column
density is degenerate with the ionization parameter
(Fig.~\ref{fig8}).  Several low-ionization lines  (\ion{Mg}{2},
\ion{C}{2})  are sufficiently isolated that they can clearly be
identified for the low-ionization parameter models.  \ion{P}{5} is
also sufficiently isolated to identify in high-ionization models.   

Almost all of the density discriminators discussed above are predicted
to be strongly blended.
\ion{P}{4}$\lambda 951$ is obliterated by a combination of \ion{C}{3},
\ion{N}{3}, and \ion{S}{6}.  As we noted previously, \ion{S}{4} is
predicted to be highly saturated. \ion{Si}{3}* near 1112 is not
modeled by {\it Cloudy} but may not be present in these data due to
the high critical density; recall that high densities are excluded for
these data due to the lack of Balmer absorption (\S 3.1). However,
information about \ion{C}{3}$\lambda 1176$ may be useful. It is
predicted to be present for low ionization parameters, but not for
high ionization parameters. It is partially blended with
\ion{S}{3}$\lambda 1195$, which is very strong at low ionization
parameters (Fig.~\ref{fig9}), but the higher velocity portion of the
trough should be unblended.   

As mentioned above, this simulated spectrum is constructed only
considering the absorption component responsible for the
\ion{He}{1}*.  There may be both lower and higher ionization
absorption components.  We know there is a lower-ionization component
that is responsible for the \ion{Ca}{2} absorption; the simulations
shown here, even for the lowest ionization parameters, do not predict
any \ion{Ca}{2} absorption.  But that feature is relatively narrow 
and has low velocity, so confusion caused by blending would be less of
an issue.

\subsubsection{Broadband Continuum}

If the simulated UV spectrum approximates the real one, then we expect
that FBQS J1151$+$3822 should be highly attenuated in the UV simply
due to the absorption lines.  For example, between 800 and 1600\AA\/,
the simulated spectra predict that the observed flux should be about
half of the intrinsic flux.  With the covering fraction taken into
account, the observed flux would be a somewhat higher fraction of the
intrinsic flux.

Fig.~\ref{fig13} shows the infrared--UV broad band continuum spectrum
constructed from the SDSS, 2MASS and GALEX photometry.  It is compared
with the SED constructed by \citet{richards06} scaled to roughly match
the flux at the one-micron dip.   It is also compared with the SDSS
spectrum of FBQS~J1151$+$3822 (reddening corrected; see below).  For a
redshift of 0.3344, the GALEX FUV band spans $\sim 1000$--1300\AA\/.
In this band, the flux from the simulated spectra is predicted to be
about about 45\% of the intrinsic flux; this is marked by the arrow on
Fig.~\ref{fig13}.  Partial covering would increase the flux; hence the
upward-pointing arrow.  The observed flux in the FUV {\it GALEX} filter  
(central rest wavelength of 1136\AA\/) is only about 5\% of the
observed spectrum.  So the observed flux is much fainter than can be
explained by the simulated absorption alone.  

The optical continuum from the SDSS spectrum and also from the
photometry (which may not be strongly affected by absorption lines
except possibly in the U band) is slightly flat.  This could be a
consequence of reddening, but is is somewhat difficult to estimate how
much reddening is present.  We assume that the spectrum would
have the same slope as the \citet{richards06} SED once the reddening
has been removed.  We deredden the SDSS spectrum using various values
of $E(B-V)$ and an SMC reddening law 
\citep{pei92}.  We find that the maximum reddening that the spectrum
can accommodate without becoming much bluer than the composite is
$E(B-V)=0.1$.  Fig.~\ref{fig11} shows the spectrum and photometry
points dereddened by this amount.  In this case, the  dereddened
observed flux in the FUV filter is about 0.23 dex below the
\citet{richards06} curve, corresponding to $\sim 18$\% of the 
intrinsic flux.  So, with reddening taken into account, the FUV flux
is still lower than expected from the simulated spectrum.  Partial
covering would increase the discrepancy.  It should be noted that
these observations were not simultaneous, and evidence for
variability is seen between the two {\it GALEX} observations separated
by eleven months.

Interestingly, the dereddened SDSS spectrum matches the
\citep{richards06} photometry quite well.  
We use the scaled \citep{richards06} photometry to obtain the
intrinsic flux at 2500\AA\/, measuring that to be $8.95 \times
10^{-15}\rm\, erg \, cm^{-2}\, s^{-1}$\AA\/$^{-1}$.  This corresponds
to a luminosity density of $6.8 \times 10^{30}\rm\, erg\, s^{-1}\,
Hz^{-1}$ at 2500\AA\/.

\section{Discussion}

In this paper, we report the IRTF and MDM observations of
FBQS~J1151$+$3822, the analysis of those data, and the analysis of the
SDSS spectrum.   In particular, we analyze the absorption lines from
\ion{He}{1}* located at 3889\AA\/ and 10830\AA\/.  We extract the
apparent optical depth profile as a function of velocity from the
spectra, finding that FBQS~J1151$+$3822 is a true BALQSO, with
$v_{max} \approx 11,000\rm \, km\, s^{-1}$.  We integrate over these
optical depth profiles to find the \ion{He}{1}* column density.
The results for these two lines do not agree, indicating that
inhomogeneous covering is  present.  We also extract an upper limit on the
column density of hydrogen in $n=2$.  We perform an inhomogeneous
absorber analysis of the \ion{He}{1}* lines, using partial covering
and power law models.  We find that the average log column density of
the \ion{He}{1}* is 14.9.  {\it Cloudy} modeling, using this mean
column density and the upper limit on the hydrogen $n=2$ shows that
the ionization parameter must be greater than $\log U=-1.4$, the
log density must be less than $\log n=8$, and the log hydrogen column
density must be greater than $N_H=21.6$.  Using the {\it Cloudy}
models, we produced simulated UV spectra to try to identify lines the
could help us better constrain the ionization parameter and density.
We found that the ionization parameter could be constrained using a
combination of low ionization lines, such as \ion{Mg}{2}$\lambda 2800$
or \ion{N}{3}$\lambda 990$, which present decreasing optical depth as
a function of ionization parameter, and a high ionization line such as
\ion{P}{5}$\lambda 1118$, which has increasing optical depth as a
function of ionization parameter.  Density constraints are more
difficult; several candidate lines were identified, but because of
blending only \ion{C}{3}*$\lambda 1176$ seemed potentially useful.
Finally, we examine the broad band photometry and show that there is
probable evidence for reddening in FBQS~J1151$+$3822.  However, the
dramatic attenuation in the UV band sampled by {\it GALEX} is still a
factor of $> 2.5$ below that explained by the simulated spectrum and
the reddening correction.

In this section, we investigate the mass outflow rate and kinetic
luminosity inferred.  We also discuss plausible acceleration
mechanisms.  We discuss the potential that \ion{He}{1}* has for
detecting high-column-density BALQSOs and measuring their outflow
properties.  Finally, we discuss the future prospects for observing
additional \ion{He}{1}*$\lambda 10830$ BALQSOs, and the value of
constructing a low-redshift BALQSO sample.  

\subsection{Outflow Rate and Kinetic Luminosity}

Using the results presented in \S 3.1, we can compute physical
parameters of the outflow as a function of ionization parameter and
density.  The combination of the ionization parameter and the density
yield the radius of the outflow.  These are shown as a function of
ionization parameter and density in Fig.~\ref{fig14}.  

We can compute the mass outflow rate, given by e.g., \citet{dunn10}
$$\dot{M}=8 \pi \mu m_p \Omega R N_H v$$ 
where $\mu=1.4$ is the mean molecular weight, $m_p$ is the mass of the
proton, $\Omega$ is the global covering fraction, and $v$ is a
characteristic velocity.  For $v$, we use the column-weighted velocity
from the modeling.  This value depended slightly (range of $\sim
14$\%) on the form of the derivation of the \ion{He}{1}*$\lambda 3889$
optical depth, and whether we used a partial covering or powerlaw
model.  We used a central value of $-5400\rm\, km\, s^{-1}$.  The mass
outflow rate is also shown in Fig.~\ref{fig14}.

The kinetic luminosity is given by e.g., \citet{dunn10}
$$\dot{E}_k=\frac{\dot{M}v^2}{2}=4 \pi \mu m_p \Omega R N_H v^3$$ 
These are also shown in Fig.~\ref{fig14}.

We compute the bolometric luminosity by integrating over the {\it
  Cloudy} input flux density scaled to the dereddened flux density
  (Fig.~\ref{fig13}) and using a luminosity distance of 1748.7~Mpc.
For some of the higher ionization, higher column-density models, a
  considerable electron column density is  present which, by Thompson
  scattering, would attenuate the incident flux, making the observed
  flux smaller. The {\it Cloudy} simulations show that this attenuation
  can be up to $\sim$50\%.  However, the covering fraction of the
  electrons is unknown.  If the electrons are only present in the gas
  responsible for the UV absorption, then using a  typical value of
  the   covering fraction of 0.3, we find a  maximum attenuation of
  only 15\%.  On the other hand, the fraction of the source not
  absorbed by \ion{He}{1}* may be transmitted through completely
  ionized gas, yielding no absorption lines but resulting in
  attenuation due to Thompson scattering.  Given these geometrical
  complications, we ignore the effects of Thompson scattering. The
  bolometric luminosity then is $5.3 \times 10^{46}\rm \, erg\,
  s^{-1}$.  The log of the  ratio of the kinetic luminosity to the
  bolometric   luminosity is shown in Fig.~\ref{fig14}. 

Assuming an efficiency of 10\% of converting matter into radiation,
the inferred bolometric luminosity corresponds to a mass accretion
rate of 9.3 solar masses per year.  The outflow rate is larger than
this in every part of the allowed parameter space, ranging from being
almost equal at high densities and low ionization parameters, to being
more than 1000 times larger at the lowest densities.  The log of the
ratio of the outflow rate to the accretion rate is also shown in
Fig.\ 14.  

How do these results compare with others?  \citet{dunn10} compile
results from partial covering analyses of 8 quasars. Their radii
estimates range from 1 pc to 28 kpc.  Our possible solutions encompass
that range, with the smaller radii being inferred for higher density
or higher ionization parameter regions of parameter space.  Their mass
fluxes range from 0.2 to 590 $M_\odot \rm \, yr^{-1}$.  Our minimum mass
flux is about 10 $M_\odot \rm \, yr^{-1}$.  Their log kinetic luminosity
ranges from 41.1 to 45.7, while our minimum log kinetic luminosity is
$\sim 44$, and so is higher than at least half of their sample, and
can be very large in the low-density high-ionization region
of parameter space. Thus, it appears that FBQS~J1151$+$3822 is a
relatively powerful BALQSO.

\subsection{Acceleration Mechanisms}

It is not known how BALQSO winds are accelerated, and several theories
have been  proposed.  An attractive model supposes that the wind is
accelerated by the scattering of the photons that create the
absorption lines (radiative line driving).  This mechanism is also
believed to accelerate winds in other objects such as stars and CVs.
There is some observational evidence that this mechanism is at work in
at least some BALQSOs.  The ``ghost of Lyman $\alpha$'' is one piece
of evidence \citep{arav95, arav96, korista93, north06, cottis10}. This
is observed as a decrease in optical depth near $-5900 \rm km\, s^{-1}$.  It
occurs at this velocity offset because that is where Ly$\alpha$
broad-line region photons are resonantly scattered by \ion{N}{5} ions.
These extra photons, above the continuum, create additional
acceleration causing a decrease in optical depth in velocity space.  This
feature has not, by any means, been found in all objects, but it has
been confirmed in more than a few.

Another piece of evidence is the observed correlation of the maximum
velocity of the outflow, $\rm v_{max}$, with the UV luminosity
\citep{lb02, ganguly07}.  More precisely, an upper envelope has been
found so that low-luminosity objects have only low values of $\rm
v_{max}$, while high luminosity objects have a range of high to low
values.   Models of radiative acceleration predict that $v_{max}
\propto  L^\alpha$, where the power law index $\alpha$ lies in the
range 0.25--0.5 (see \citet{lb02,ganguly07} for details of these
arguments).  This correlation argues that radiative 
acceleration of some kind is generally active in BALQSOs because
the correlation is seen in  different large samples of
objects.  Note, though, that the driving force need not necessarily
be line driving; acceleration via dust-scattering opacity would also apply
\citep{lb02}.  Note also that this behavior has been seen to be
violated in at least one object \citep[WPVS~007;][]{leighly09}.  

Other mechanisms besides radiative line driving have been suggested
for the acceleration of the outflowing gas.  Radiative acceleration by 
dust-scattering opacity has been hypothesized
\citep[][e.g.,]{sn95}. This mechanism is promising since dust grains
have large opacity.  However, the outflow
needs to originate beyond the dust sublimation radius, though.  For
our object, with estimated bolometric luminosity of $5.3\times
10^{46}\rm\, erg\, s^{-1}$ (\S 4.1), the sublimation radius is estimated
to be about 1.2 pc \citep{sn95}.  As shown in Fig.~\ref{fig14}, that
radius is spanned by our {\it Cloudy} modeling solutions.  In
addition, we have found evidence for reddening in this object (\S
3.2.3).    

Another model appeals to a magnetocentrifugal wind \citep[e.g.,][and 
  references therein]{everett05}. In this model, gas is attached to
large-scale magnetic field lines, and then the rotation of the
accretion disk flings the matter away from the disk like beads on a
wire.  Therefore, in a pure magnetocentrifugal wind, the dynamics of
the outflow are divorced from the radiation field.  Many people
consider a hybrid model, assuming both magnetocentrifugal acceleration
and radiative line driving. Finally, thermally-driven winds have been
proposed \citep{kk01}, where the wind is essentially evaporated off
the inner edge of the torus.  This model has been applied principally
to warm absorbers rather than BALQSOs.  

As an interesting combination of the above, \citet{ge07} propose that
different models may be appropriate for different regions of the
outflow.  The X-ray absorbing region may be dominated by
magnetocentrifugal acceleration, the UV-absorbing region may be
dominated by radiative line driving, and the IR absorbing region may
be dominated by dust acceleration.   

In this section, we apply several of the radiative-driving models to
our results for FBQS~J1151$+$3822.

\citet{hamann98} presents a derivation of an equation that can be used
to determine, in an approximate way, if radiative-line driving is
sufficient to accelerate gas of a given column density to a
particular terminal velocity.
Basically, the equation of motion is solved making the assumption that
a fraction of the total luminosity $f_l$ is scattered or absorbed by
the outflowing gas, as follows:
$$ m_H N_H v \frac{dv}{dr} = \frac{f_l L}{4\pi R^2 c} - \frac{m_H N_H
    G M_{BH}}{R^2}$$ 
where $N_H$ is the column density of the outflowing gas.  
To put it another way, the momentum of that fraction of the
total radiation is converted to momentum of the wind, mediated
naturally by the gravitational attraction of the black hole.

To use this equation, we need to estimate the black hole mass; we
do that using standard methods from the SDSS
spectrum.   We measure a rest frame flux at 5100\AA\/ of $2.6 \times
10^{-15}\rm\, erg\,s^{-1}\,cm^{-2}$\AA\/$^{-1}$ from the dereddened
SDSS spectrum shown in Fig.~\ref{fig13}.  We compute the
broad-line-region radius using the regressions found by
\citet{bentz06} using the flux and a luminosity distance of $1773.7\rm
\, Mpc$ appropriate for their (inferred) cosmological parameters of $H_0=70\rm
\, km\, s^{-1}\, Mpc^{-1}$, $\Omega_M=27$ and $\Omega_\Lambda=0.73$.
That is found to be $\log(R_{BLR})=2.47$ in units of light-days.
Finally, we compute the  dispersion of the H$\beta$ line profile
obtained after subtracting the other fitted components, and referring
to \citet{collin06}, we use a scale factor of 1.5 to obtain a black
hole mass of $8.2 \times 10^{8}\rm \, M_\odot$.  For the bolometric
luminosity of $5.3 \times 10^{46}\rm \, erg\,
s^{-1}$, we find that the object is radiating at about half the
Eddington luminosity. 

The fraction of the bolometric luminosity that can be used to
accelerate the wind, $f_l$, can be estimated as the fraction of the incident
continuum that is absorbed.  This is easily obtained from
the {\it Cloudy} output by integrating over the transmitted continuum
and dividing the result by the integral over the incident continuum.
A difficulty is that for high column densities, the transmitted
continuum is attenuated significantly by Thompson scattering.
Thompson scattering will not transfer much momentum, so we correct the
incident continuum by multiplying it by the Thompson fraction
discussed in \S 4.1.   
We plot this fraction in Fig.~\ref{fig15}.   The fraction available is
relatively large, equal to or greater than 1/2 over the entire
parameter space.  It has a broad minimum near $\log U = -0.5$.  At
lower ionization parameters, more low-ionization species contribute to
the optical depth; at high ionization parameters, Compton scattering becomes
important.   

Next, we can compute the terminal velocity.  The bolometric
luminosity, obtained above, is estimated to be $5.3 \times 10^{46}\rm
\, erg\, s^{-1}$.  We need a value of the launch radius.  Above we
computed the radius of the wind from the density and the ionization
parameter used in the {\it Cloudy} modeling.  This is the radius at
which the wind has reached its observed velocity and is absorbing the
continuum, and so this value would be somewhat larger than the launch radius.
\citet{arav94} point out that a general characteristic of
radiation-driven winds is that most of the acceleration occurs on a
length scale comparable to the starting radius.  Therefore, for a
rough estimate, we use a launch radius that is half the radius of the
absorbing wind derived from the {\it Cloudy} simulations.   The column
density is that shown in Fig.~\ref{fig8}. The resulting predicted
terminal velocity is shown in Fig.~\ref{fig15}. A positive velocity 
(outflow) is attained over most of the parameter  space, with higher values
attained at higher densities and lower ionization parameters.  Higher
densities correspond to a smaller launch radius, where the flux is
more intense; lower ionization parameters correspond to smaller column
densities which are easier to accelerate.  Although most of parameter
space is characterized by an outflow,  the maximum velocity ($\sim
6400\rm \, km\, s^{-1}$) is almost a factor of two below the terminal
velocity observed  in this object.  Still, that we can attain high
positive velocities in this object is perhaps not surprising as it is
radiating at a large fraction of the Eddington limit. 

Could this outflow be a disk wind?  It does not seem likely that it
is.  The disk wind model predicts acceleration to $15,000\rm\, km\,
s^{-1}$ at about $10^{17}\rm \, cm$ from the central engine in a $10^8
\rm \, M_{\odot}$ black hole \citep{psk00}.  Our photoionization
modeling results show that the minimum radius predicted in our
parameter space is about 1 parsec for the models with the highest
density, out to hundreds of parsecs for lower density models.  The
distances then range from 10 to thousands of times larger than the
disk wind radius. 

The above estimate does not take into account the fact that the
opacity due to a single line is enhanced by the velocity gradient in an
optically-thin outflowing gas.  Thus, the momentum transfer for a
given column density is greater in an outflowing gas than in a static
gas, and higher velocities can be attained.  To take this into
account, we use the CAK formalism \citep{cak75}, originally applied to
stellar winds but adapted for use in quasars by e.g.,
\citet[][]{arav94} and \citet{alb94}.   

We compute the force multiplier due to line scattering as a function
of the equivalent electron optical depth $t$ in the standard
way according to Eq.~2.7 in \citet{arav94} using the ionic column
densities obtained from our {\it Cloudy} simulations, and line
opacities obtained from \citet{verner94}.  We ignore bound-free
transitions, which will of course also contribute to the acceleration,
but probably to a lesser degree than the lines
\citep[e.g.,][Fig.~2]{alb94}.  Fig.~\ref{fig16} shows the log of the
force multiplier for an example value of $t=10^{-6}$.  For this
particular value of $t$ (and noting that the force multiplier has a
strong $\sim 1/t$ dependence), we find that the force multiplier is
higher for larger ionization parameter.  This is because, as discussed
in \S 3.1, the column densities need to be larger at higher ionization
parameters to produce the observed \ion{He}{1}* column density.

Next, we assume a terminal velocity of $v_{term}=11,000\rm\, km\,
s^{-1}$ and a launch radius that is one-half of the radius derived
above for the absorbing gas, and solve the equation of motion for the 
force multiplier and equivalent optical depth $t_{needed}$ required to
reach that terminal velocity.  The equation of motion that we solve
is: 
$$v\frac{dv}{dr}=\frac{\sigma_T}{1.2 m_{H}c} \frac{L}{4\pi r^2} M_L(U,t) -
\frac{GM_{BH}}{r^2}$$
where we have assumed that $n_e=1.2 n_H$, i.e., the electron density
is approximately 1.2 times the hydrogen density, $L$ is the bolometric
luminosity, and $M_L(U,t)$ is the force multiplier.  The results are
shown in Fig.~\ref{fig16}.  They show almost the same dependence as the
radius of the absorbing gas, shown in Fig.~\ref{fig14}.  This makes
sense because the flux decreases rapidly with radius, so the optical
depth enhancement due to the velocity gradient has to be higher  at
larger radii to attain the same terminal velocity.
This enhancement is equivalent to a larger value of force
multiplier. Likewise, we can solve for the equivalent electron optical
depth $t$ needed to attain the observed terminal velocity.  That is
also shown in Fig.~\ref{fig16}. The equivalent electron optical depth
$t$ is smaller  for  larger values of the radius. 

The equivalent electron optical depth scale is defined as 
$$ t=\sigma_T \epsilon n_e v_{th} \left(\frac{dv}{dr}\right)^{-1}$$ 
where $\epsilon$ is the filling factor in the wind, and $v_{th}$ is
the thermal velocity of the gas.  We assume a temperature of $10^4\rm
\, K$, appropriate for a photoionized gas, and we again assume that
the electron density is $1.2 n_H$.  We make the zeroth-order assumption that $dv/dr$
is approximately equal to the terminal velocity divided by the
radius where the absorption occurs.  Then, we can solve for the
filling factor $\epsilon$.  That is also shown in Fig.~\ref{fig16}.  

The filling factor $\epsilon$ can be used to estimate the dimensions
of the outflow.  For a continuous flow, the length scale $l$ would be
$N_H/n_H$.  For a flow consisting of small clouds with filling factor
$\epsilon$, the length scale is increased by $1/\epsilon$, i.e.,
$N_H/\epsilon n_H$.  For the outflow to be physical, the length scale 
of the outflow must be less than the radius where the absorption
occurs; otherwise, there would not be sufficient length to accelerate
the outflow to the observed terminal velocity.  Working through the
approximations given above, we find:
$$\frac{l}{R_{abs}}=\frac{1.2 \sigma_T N_H v_{th}}{t_{needed} v_{term}}.$$  

This equation shows that larger column densities require larger region
sizes to accelerate all the gas to the terminal velocity.  The
terminal velocity appears on the bottom of the equation as it was
originally part of $dv/dr$, the differential expansion rate.  The
region size has to be larger for smaller differential expansion rates
in order to attain the same terminal velocity. 

The log of the ratio of the region size to the radius where the
absorption occurs is shown in Fig.~\ref{fig16}.   This shows that at
densities lower than $\sim 10^7 \rm \, cm^{-3}$, the flow is
unphysical since the region size is larger than the absorption radius.
This happens because low density flows occur at larger radii, and
at a large radius the flux is low, so a large equivalent electron
optical depth scale is needed to counteract the low flux and
accelerate the flow to the observed terminal velocity.  The equivalent
electron optical depth scale is linearly dependent on $\epsilon$.  

This argument, based on the flow dynamics, provides strong constraints
on the physical parameters of the outflow, excluding most of the
previously allowed parameter space.  Examining Fig.~\ref{fig8} and 
Fig.~\ref{fig14} and considering areas of parameter space where the
region size is smaller or equal to the radius size ($l/R_{abs} \leq
1$), we find that the 
log of the total hydrogen column density is constrained to be between
21.7 and 22.9, the radius is between 2 and 12 parsecs, the mass flux
is between 11 and 54 solar masses per year, the kinetic luminosity is
between 1 and 5$\times 10^{44}\rm \, erg\, s^{-1}$, and the kinetic
luminosity is between 0.2 and 0.9\% of the bolometric luminosity.  If
we now compare with previous results from \citet{dunn10}, we find that
FBQS~J1151$+$3822 has an absorbing region relatively close to the
central engine, a higher than average column density, a relatively
high kinetic luminosity, but a relatively low mass outflow rate.

It is also interesting to note that the allowed region lies in the
area where we expect \ion{C}{3}*$\lambda 1176$ to be a useful density
diagnostic.  We would also expect \ion{P}{5} to be weak in the UV
spectrum as the ionization parameter is now constrained to be between
$-1.4$ and $0.2$ (Fig.~\ref{fig12}; middle panel).  

This analysis assumes that radiative line driving is the acceleration
mechanism operating in this BALQSO; however, that is the favored
acceleration mechanism for the UV absorbing gas.  At any rate, it
implies that radiative line driving is feasible for higher densities
in our  parameter space.  Lower densities would require an additional
acceleration mechanism, perhaps hydromagnetic.  In addition, we have
assumed that the flow is radial.  It may be simply crossing our line
of sight; in that case, the terminal velocity would be larger than the 
observed $11,000\rm \, km\, s^{-1}$.  

\subsection{Size of the Absorbers}

Following \citet{hamann10}, we can compute the size of the absorbers.
Reviewing their argument, since partial covering is significant, the
upper limit on the size of the individual absorbing structure is the
size of the emitting region.  This is an upper limit since the
absorbers can be smaller, and there can be many of them.  

We compute the size of the emission region using classical relations
from accretion disk theory, namely, for a sum-of-blackbodies accretion
disk,
$$T_{disk}=\big(\frac{3G M\dot{M}}{8 \pi \sigma R^3}\big)^{1/4}.$$
We use the black hole mass and accretion rate derived in \S 4.2,
namely $M=8.2\times 10^{8}M_\odot$ and $\dot{M}=9.3\, \rm  M_\odot\,
yr^{-1}$.  This equation yields a temperature that decreases with
radius as  $R^{-3/4}$.  It has been argued that this radial dependence
is too steep (e.g., \citep{gaskell08}, who suggests that the
  index should be 0.57) yet may be acceptable for the
estimations made here.  Using this equation, we find that the
3888\AA\/ and 10830\AA\/ are emitted at 0.011 and 0.045 pc,
respectively.  Thus, the 10830\AA\/ emission region is a factor of
3.9 times larger than the 3888\AA\/ emission region; for a flatter 
temperature dependence of $-0.58$, that difference increases to a
factor of 5.8.  These sizes are comparable to those inferred by
\citet{hamann10}, despite the fact they are examining far UV lines.
The reason for this coincidence is that the black hole mass in their
object is much larger.

These numbers become more interested when compared with the much
smaller emission region size expected at for 1121\AA\/, where
\ion{P}{5} absorption should be seen in the continuum.  For
FBQS~J1151$+$3822 that is $0.0022\rm \, pc$, a factor of 21 times smaller than
the 10830\AA\/ emission region size (and a factor of 50 times smaller
for the flatter radial temperature dependence).  So, if partial
covering in \ion{P}{5} were observed in FBQS~J1151$+$3822, it would
imply that a swarm of at least 20--50 very small clouds cover the
10830\AA\/ emitting region.

These numbers become even more extreme when considering the low black
hole mass BALQSO WPVS~007 \citep{leighly09}.  That object has a black
hole mass of $4.1\times 10^6\rm \, M_\odot$ and an accretion rate
between 0.0086 and 0.0114 solar masses per year.  Yet, it also shows
evidence for partial covering in \ion{P}{5}.  Following the analysis
above, we obtain a continuum emission region for 1121\AA\/ of
3.6--4.8$\times 10^{-5}\rm pc$.  

A comparison of the covering fractions of \ion{He}{1}* and \ion{P}{5}
in the same object could be interesting due to the large differences
in size of the emitting region (a factor of 20--50, depending on the
radial dependence of the temperature).   Note that the covering
fractions of these lines can be directly compared because the gas
presents the same optical depth in these lines (Fig.~\ref{fig17}); as
discussed by \citet{hamann01}, lines that have higher optical depths
may have naturally higher covering fractions.   It is possible that
the swarm of small clouds may completely cover the 1121\AA\/ emitting
region, but not cover the entire 10830\AA\/ emitting region.   If this
were the case, the covering fraction would be different.

\subsection{\ion{He}{1}* as a High-Column-Density Diagnostic}

\subsubsection{Comparison with Other Lines}

High-column-density outflows are of great interest in BALQSOs.   For
radiative acceleration models,  the force multiplier is predicted
generally to decline with column density \citep[e.g.,][]{arav94}.
So high-column flows place the most stringent constraints on radiative
acceleration models.  In addition, the kinetic luminosity depends
linearly on the column density, so, depending on the velocity and
radius, these outflows may have the largest kinetic luminosities, thus
placing the most stringent constraints in general on
outflow-acceleration models.  

The study of BALQSOs underwent a paradigm shift about 15 years ago.
Previously, it was thought that the absorber completely covered the
continuum, and column densities could be measured directly from the
troughs.  But the observation of strong \ion{P}{5} in some objects was
puzzling, given that the solar abundance of phosphorus is about 765
times lower than that of carbon.  This conundrum was resolved with the
realization that the column densities were much larger than previously
thought, but the troughs are not black because the absorber only
partially covers the continuum source \citep{hamann98}.  

The low abundance of phosphorus means that  \ion{P}{5} lines will be
present only when the column densities are high.   Thus, \ion{P}{5} is
a good probe of high-column outflows.  But there is a practical
problem with using \ion{P}{5} to detect high-column outflows in
general: the resonance lines are in the far UV (at 1118 and 1128\AA\/) and
shortward of Ly$\alpha$.  So low-redshift objects must be observed
from space, while in high-redshift objects, when \ion{P}{5} is
shifted in to the optical bandpass (for redshifts larger than $\sim
2$), \ion{P}{5} can be obscured by Ly$\alpha$ forest absorption.

We propose that \ion{He}{1}* is an equally effective probe of high
column densities in BALQSOs.  We discuss the situation analytically
first, and then investigate using simulations.

As shown in \citet{ss91}, the optical depth is a function of $\lambda
f_{ik} N_{ion}$, where $\lambda$ and $f_{ik}$  are the wavelength and
the the oscillator strength of the transition, respectively, and
$N_{ion}$ is the column density of the relevant ion.  For \ion{P}{5},
$\lambda f_{ik}$ is 752 (for both of the resonance transitions for
this doublet).  For \ion{He}{1}*$\lambda 3889$, the product is 251,
while for \ion{He}{1}*$\lambda 10830$, the product is 5842.  

For an analytic estimation, the relevant metric is the $\lambda
f_{ik}$ times the fraction of the ion or metastable state relative to
hydrogen.  The lower the value, the higher the column density can be
before the line is saturated.  

What is the ratio of P$^{+4}$ to hydrogen, approximately?   For solar
abundances, there is one phosphorus atom for 
every 312,536 hydrogen atoms.  From \citet{hamann97b} we see the
fraction of phosphorus ionized three times is no larger than about
0.5--0.6.  Assuming 0.5, we then estimate the abundance of $P^{+4}$ to
be $1.6 \times 10^{-7}$.  Then the $\lambda f_{ik} N_{ion}$ value for
\ion{P}{5} is $1.2 \times 10^{-4}$.  

What is the ratio of \ion{He}{1}* to hydrogen, approximately? For
solar abundances, the helium abundance is 0.1.  The 
metastable level is populated by recombination of He$^+$.  As
discussed by \citet{clegg87} and \citet{ch89}, it is depopulated
principally by collisions.  \citet{clegg87} give an empirical formula
for the ratio of helium in the metastable state relative to the
number of He$^+$
$$\frac{N(2^3S)}{N(He^+)}=\frac{5.79\times 10^{-6} t_e^{1.18}}{1+3110
  t_e^{-0.51} N_e^{-1}}$$
where $t_e$ is the electron temperature in units of $10^4\rm\,K$ and
$N_e$ is the electron density in $\rm \, cm^{-3}$.  This function
shows that the number of helium in the metastable state is an
approximately constant fraction of the number of once-ionized helium.

In an ionized gas slab, only a fraction of helium will be He$^+$.  As
long as the slab is not too thick (i.e., optically thin to the
hydrogen continuum), the helium will be split between once and twice
ionized atoms.  Given that quasar photon continua fall steeply, we can
assume that the majority of the helium will be He$^+$.  Thus, the
abundance of metastable helium is approximately $5.8 \times 10^{-7}$.
Then $\lambda f_{ik} N_{ion}$ is $1.5 \times 10^{-4}$ for
\ion{He}{1}*$\lambda 3889$ and $3.4 \times 10^{-3}$ for
\ion{He}{1}*$\lambda 10830$.  Recalling that the $\lambda f_{ik}
N_{ion}$ value for \ion{P}{5} is $1.2 \times 10^{-4}$, we see that
\ion{He}{1}*$\lambda 3889$ component is equally sensitive as \ion{P}{5}, while
\ion{He}{1}*$\lambda 10830$ is a bit less sensitive.  In contrast, a
similar estimation for \ion{C}{4}$\lambda 1549$ (for an assumed
fraction of C$^{+3}$ of 0.65) is $7.0 \times 10^{-2}$.  So, for
example, \ion{C}{4} will become saturated at column densities a factor
of twenty lower than \ion{He}{1}*$\lambda 10830$, and a factor of
$\sim 470$ lower than \ion{He}{1}*$\lambda 3889$.  

The analytical analysis is limited because the fraction of these ions
in a slab depends on the ionization parameter and column density, as
well as the abundances.  So, we use the simulations discussed in \S
3.1 to investigate the sensitivity of these absorption lines to high
columns further.  As discussed in \S 3.1, we assume the {\it Cloudy}
{\tt AGN Kirk} continuum, solar abundances, and $\log(n)=5$.  Since all the
transitions discussed here are resonance transitions\footnote{Since \ion{He}1*
acts as a second ground state, it can also be regarded as a resonance
transition}, we expect negligible dependence on density .  We extract
ionic column densities as a function of ionization parameter and $\log
N_H - \log U$ from the {\it Cloudy} model results.  Contours of 
$\log \tau(v)=\log(2.654\times 10^{-15}\lambda f_{ik}
N_{ion})-\log(10,000)$ are show in Fig.~\ref{fig17}.  We divide  
by 10,000$\rm\, km\, s^{-1}$ to approximate the optical depth of a
square line profile with a width of 10,000$\rm\, km\, s^{-1}$ , 
consistent with the approximate width of the line observed in
FBQS~J1151$+$3822.  A smaller velocity width would yield a larger optical
depth for any combination of parameters.

These contour plots show that resonance-line absorption from  abundant
ions, such as \ion{O}{6} and \ion{C}{4}, are saturated at relatively
low column densities.  But lines such as \ion{P}{5} and
\ion{He}{1}*$\lambda 3889$, and to a slightly lesser extent,
\ion{He}{1}*$\lambda 10830$, remain optically thin over a wide range
of parameter space.   

These figures also show that the metal lines experience greater
ionization parameter dependence than the \ion{He}{1}* lines.  That is,
for example, \ion{Si}{4} optical depth is greater for any value of
$\log N_H - \log U$ at $\log U\sim -2.2$ than for other values of
ionization parameter.  This is of course due to the fact that at $\log
U\sim -2.2$, a greater fraction of silicon is Si$^{+3}$ than other
ionization states, while at lower and higher values of $\log U$, there
are relatively fewer Si$^{+3}$ ions.  Metastable \ion{He}{1} is more monotonic;
for a fixed radiation-bounded slab depth, the number of metastable
helium ions increases as the depth of the helium ionization front
increases.  

\subsubsection{Sensitivity Examined Using Simulated Spectra}

As discussed above, \ion{He}{1}* is a valuable probe of high column
density in quasars.  Moreover, using both the \ion{He}{1}*$\lambda
3889$ and \ion{He}{1}*$\lambda 10830$ lines, we can solve for the true
column density and covering fraction.  In this section, we discuss the
range of columns that can be probed using \ion{He}{1}*, and we discuss
the prospects of identifying low-reshift BALQSOs using \ion{He}{1}*.

Each quasar absorption line is sensitive to a certain range of column
density.  If the column is too low, the absorption line will not be
distinguishable from the continuum.  If the column is too high, the
absorption line will be saturated.  Obviously, the sensitivity will
depend on the properties of the transition, through the oscillator
strength, and it will depend on the abundance of the ion.  But it will
also depend on the properties of the outflow; in particular, it will
depend on the  the range
of velocity over which the line is spread, $\Delta v$.  A low column 
density is detectable in a narrow line, where $\Delta v$ is small,
while absorption spread over a large range of velocity would be
indistinguishable from the continuum.  The sensitivity depends on the
data properties, i.e., signal-to-noise ratio, obviously, with this
factor being more important for weak lines.  In addition, in some
cases, there is systematic uncertainty, such as the uncertainty in
defining the continuum.

We explore the sensitivity of the \ion{He}{1}* lines using
simulations.  We use the optical depth profile as a function of
velocity  obtained from the 
infrared spectrum.  Furthermore, we use this for both lines, that is,
we ignore velocity dependence of the optical depth and covering
fraction.   We smooth the profile slightly using the function 
mentioned in \S 2.1.   We investigate the  
region around the 3889\AA\/ and 10830\AA\/ lines in velocity space.  We
add gaussian noise to the initial ratio of observed to continuum
(equal to 1) so  that the resulting signal-to-noise ratio is 150.
Clearly, there will 
be sensitivity dependence on the signal-to-noise ratio, but since we
are more interested in the characteristics of the lines, we do not
explore that dependence here.   We investigate log \ion{He}{1}* column
densities between 14.0 and 17.0.  As discussed above, this
corresponds to approximate log hydrogen column densities of
20.25--23.25.  We emphasize again that the results depend on the absorption
profile assumed here to have $\Delta v \approx
11,000\rm\, km\, s^{-1}$.   We also investigate a range of covering
fractions between 0.1 and 0.9.   For each input column density of
\ion{He}{1}*, we scale the optical depth profile appropriately to
produce the desired column density.  Then, for each input covering
fraction, and using  equations 3 and 4 from \citet{sabra05}, we
simulate the observed ratio of absorbed  to continuum spectra for both the
3889\AA\/ region and the 10830\AA\/ region.   Finally, we fit the two 
resulting simulated ratios with the partial covering model using
the technique discussed in \S 2.8.   For each combination of parameters,
we run 10 simulations and plot the mean.

As discussed qualitatively above, metastable helium is a
valuable probe of high column densities because helium in this state
is relatively rare in the gas.  In that way, metastable helium
absorption is similar to P$^{+4}$ absorption.  But the
\ion{He}{1}*$\lambda 10830$ and \ion{He}{1}*$\lambda 3889$ lines have
another unique property: the high $\lambda f_{ik}$ ratio of 23.3.  In
contrast, the \ion{P}{5}$\lambda\lambda 1118, 1128$ lines, like other
lithium-like ions, have a $\lambda f_{ik}$ ratio of 2.  What effect
does the $\lambda f_{ik}$ ratio have on the sensitivity?  To explore
this question, we do another set of simulations in which $\lambda
f_{ik}$ for the 3889\AA\/ line is a factor of two lower than $\lambda
f_{ik}$ for the 10830\AA\/ line, instead of the normal case where the
ratio is 23.3.

The results are shown in Fig.~\ref{fig18}--\ref{fig21}.  We plot the
measured properties as a function of the input parameters.  We first
plot the number of points retained in Fig.~\ref{fig18}.  The optical
depth profile had 69 points, and as noted in \S 2.8, the software removes
unphysical points where $R_{3889} > R_{10830} > R_{3889}^{23.3}$ is
not obeyed.  We see that the number of points retained is lower for
the lowest and highest column densities.  This makes sense for the
lowest column densities, since in that case, the optical depth of both
lines is low, and it is more likely that a fluctuation due to noise
will cause a point to become unphysical.  At highest column densities,
the lines become saturated, and the ratio of the observed to continuum
looks the same for both lines.   Again, in this case, is it likely
that a fluctuation due to noise will cause a point to become
unphysical.  The number of points retained drops less rapidly for
$\lambda f_{ik}=23.3$ as the low value of the oscillator strength of
the 3889\AA\/ line means that it is not saturated until the column
density becomes very high.

Next, we plot the measured covering fraction and column density from
the partial covering model in Fig.~\ref{fig19}.  The measured covering
fraction and column density is consistent with the input for
intermediate $\log$  \ion{He}{1}* column densities for both sets of
simulations.  For lower column densities, the measured covering
fraction is consistently low, while the column density is consistently
high.  At lower column densities, the weaker line becomes
indistinguishable from the continuum, and unphysical points are
dropped.  Points are deemed physical if there is a downward
fluctuation in the 10830\AA\/ component, and an upward fluctuation in
3889\AA\/.  This situation looks the same as a higher column and a
lower covering fraction.   For the highest column densities, the
measured column is consistent with the input for the $\lambda
f_{ik}$ ratio $=23.3$ case, while it is consistently underestimated for the
$\lambda f_{ik}$ ratio $=2$ case. This is because at high column densities,
both lines become saturated when $\lambda f_{ik}$ ratio$=2$.  

How would we fare if we had only one of the two \ion{He}{1}* lines?
Fig.~\ref{fig20} shows the apparent column densities measured from the 
10830\AA\ line and the 3889\AA\/ lines separately.  We find that the
column density is uniformly less than the observed, as expected.  The
discrepancy is less for high covering fractions; the
presence of significant partial covering, i.e., small values of the
covering fraction, makes the optical depth look too low.   There is
also a saturation of column at high column densities for a given
covering fraction.  This is due to saturation of the lines.  The
3889\AA\/ line (with $\lambda f_{ik}$ ratio $=23.3$) performs the best
at high 
column densities because the low oscillator strength of
this line prevents it from becoming saturated.  The apparent
column density is lower than the input for small values of the
covering fraction, for the same reason as above: low values of the
covering fraction make the apparent optical depth low.

Finally, in Fig.~\ref{fig21} we plot the difference between the
log of the measured average column density and the log of the input
average column density. 
Recall that the average column density for the partial covering model
is the product of the covering fraction and the column density.  This
is the relevant parameter for the photoionization and dynamical
modeling.  For smaller column densities,  the accuracy is
better for larger covering fractions; for smaller covering fractions,
the observed column is overestimated but by less than an order of
magnitude. More importantly, this plot shows that the average column
is accurately reproduced for input log \ion{He}{1}* column densities
larger than $\sim 15$ when the $\lambda f_{ik}$ ratio $=23.3$.   For a 
$\lambda f_{ik}$ ratio $=2$, both lines become saturated at high
column densities, and the input column density is estimated to be
too low by an order of magnitude. 

In summary, these simulations show that while there is some bias in
measuring parameters at low column density (which is not surprising
since as the lines become indistinguishable from the continuum), the
large $\lambda f_{ik}$ ratio $=23.3$ for \ion{He}{1}$\lambda 10830$ and
\ion{He}{1}$\lambda 3889$ makes it superior for measuring high column
densities.  Basically, the large ratio gives these lines an increased
dynamic range for measuring high column densities.   An important
point, though, is that the values in 
these simulations apply for the optical depth profile derived for
\ion{He}{1}$\lambda f_{ik}$ for FBQS~J1151$+$3822; the columns
measurable will vary depending on the distribution of column over
velocity, $dN/dv$. 

\subsection{Are There More \ion{He}{1}*$\lambda 10830$ BALQSOs?}

In this paper, we report the discovery of the first
\ion{He}{1}*$\lambda 10830$ broad absorption line quasar.  Are there
any more of them?  In principle, there should be.  \ion{He}{1}* is a
high-ionization ion that appears in the He$^+$ region of the outflow,
coincident with C$^{+3}$ and Si$^{+3}$.  If the column densities are
generally large, then any high-ionization outflow should have
\ion{He}{1}*$\lambda 10830$.  

Actually, we have already found some.   An observing run in April
2010 using SPEX on the IRTF has netted four additional objects.  In
addition, observations using Lucifer on the LBT has
yielded two more (Leighly et al.\ in prep.). But not all BALQSOs have
\ion{He}{1}*$\lambda 10830$; we observed several well-known
low-redshift BALQSOs using SpeX on the IRTF and found that they do not
have significant  \ion{He}{1}*$\lambda 10830$ absorption.  These are
bright objects for the IRTF so the 
limits on \ion{He}{1}*$\lambda 10830$ will provide interesting upper
limits on the column density in these objects (Leighly et al.\ in
prep.).  

How can we find more candidates?  The Sloan Digital Sky Survey spectra
provide a useful starting place.  If we can identify
\ion{He}{1}*$\lambda 3889$ in SDSS spectra, then, because of the large
$\lambda f_{ik}$ ratio, \ion{He}{1}*$\lambda 10830$ will certainly be
seen.  How can we identify objects with \ion{He}{1}*$\lambda 3889$?
As discussed in \S 2.4, it can be difficult as it is generally a weak
line that can be confused with strong \ion{Fe}{2} emission.  One way
is to look at objects that have both \ion{Mg}{2} and
\ion{He}{1}*$\lambda 3889$ in the bandpass.  An informal examination of a
low-redshift subsample of BALQSOs cataloged by \citep{gibson09} 
shows that $\sim 1/3$ have something at 3889\AA\/ that may correspond
to the \ion{Mg}{2} absorption.  

Identifying \ion{He}{1}* BALQSOs using both \ion{Mg}{2} and the
3889\AA\/ component limits us to Lo-BALs.  How can we identify
high-ionization BALQSOs?  One way is to look for objects that do have 
\ion{He}{1}*$\lambda 3889$  but don't have \ion{Mg}{2}.  Another
possibility is to look at the broad band photometry.  The
FBQS~J1151$+$3822 photometry shown in Fig.~\ref{fig13} is rather
distinctive, as the decline in the UV is too  steep to be explained by
reddening.  Depending on the redshift, the usefulness of this
technique will depend on the availability of {\it GALEX} photometry.  

Observing \ion{He}{1}*$\lambda 10830$ limits us to low redshift
objects.  If we observe \ion{He}{1}*$\lambda 3889$ in the optical
band, we will observe the 10830 component in the 0.8--$2.4\mu$
short-wavelength band; then, we are limited to objects with redshifts
less than 1.2.    But these nearby
objects may be quite interesting.  Only a handful of low-redshift
BALQSOs are known \citep[e.g.,][]{sulentic06}.  Many of these were
discovered serendipitously; for example, they were observed because
they are PG quasars, or because they have weak [\ion{O}{3}] in their
optical spectra \citep{turnshek97}.   These low-redshift objects have lower
luminosities than the SDSS and LBQS BALQSOs \citep[e.g., Fig.\ 7
  in][]{ganguly07}, and so they  define the
low-$v_{max}$ end of the correlation between $v_{max}$ and
luminosity.  That region of the plot is rather sparsely populated, but
it may be important in our understanding of outflows in AGN as it
bridges the very low velocity outflows seen in Seyfert galaxies
\citep[e.g.,][]{crenshaw03} and the high velocity BALQSOs beyond
$z>1.5$.  FBQS~J1151$+$3822 is a luminous object; for
\citet{ganguly07}'s  cosmology, the dereddened 3000\AA\/ $\log \lambda
L_{\lambda}$ is 45.9 (without the reddening correction,  $\log \lambda
L_{\lambda}= 45.7$; Fig.~\ref{fig13}).  From \citet{ganguly07} we see that
lower luminosity objects of interest should have  $\log \lambda 
L_{\lambda} < \sim 45.1$, or greater than 2 magnitudes fainter than
FBQS~J1151$+$3822.  But FBQS~J1151$+$3822 is a relatively bright
object for SDSS, with an observed g magnitude of 15.8.  The DR7 SDSS
Quasar Catalog \citep{schneider10} has 37,430 quasars with redshift
smaller than 1.2, and the distribution of g-magnitude peaks at 19,
while there are only 35 quasars brighter than FBQS~J1151$+$3822.
Thus, it seems quite reasonable that an interesting low-redshift
low-luminosity sample could be defined and studied.

\section{Summary \& Conclusions}

The principal results and findings of this paper are as follows.

\begin{itemize}

\item We report the first 
  short wavelength (0.8--2.4$\mu$) infrared spectroscopic observation
  of FBQS~J1151$+$3822 using SpeX on the IRTF, as well as a new
  optical spectroscopic observation of  FBQS~J1151$+$3822 using  CCDS
  on the 2.4 meter Hiltner telescope at MDM observatory. In addition,
  we analyzed the SDSS spetrum.   We discovered  broad \ion{He}{1}*
  absorption lines,   both in the 3889\AA\/ transition and, for the
  first time in a quasar, in the  
  10830\AA\/ transition.  The terminal velocity $v_{max}$ is
  $11,000\,\rm km\, s^{-1}$, indicating that FBQS~J1151$+$3822 is a
  broad absorption line quasar.   

\item We extracted the apparent optical depths for the absorption
  lines.  The procedure was straightforward for the 10830\AA\/ line
  because the IR continuum is fairly simple.  But for the 3889\AA\/ line,
  this procedure was complicated by strong \ion{Fe}{2} emission, so a
  template-fitting  method had to be used.  An 
  upper limit on Balmer absorption was extracted from the SDSS
  spectrum. 

\item The \ion{He}{1}* column densities were first estimated by
  integrating   directly over the optical depth profiles.  The
  estimate based on the   3889\AA\/ line ($\log N_{HeI*} \sim 14.85$)
  was substantially higher than   the one based on the 10830\AA\/ line
  ($\sim 14.3$).  This result implies that the
  absorber does not fully cover the source.  Since the two transitions
  have the same lower (metastable) level, we can solve two-parameter
  inhomogeneous covering models as a function of velocity.  We use a
  pure partial covering model and a power law model.  Both models give
  the same average log \ion{He}{1}* column of 14.9.  The hydrogen
  $n=2$ limit depends on the details of the population of the $2p$ and
  the $2s$ states; assuming that the $2s$ metastable state was solely
  populated yielded an upper limit of log hydrogen $n=2$ column of
  13.5.

\item {\it Cloudy} modeling was used to constrain the range of parameter
  space permitted by the measurements.  The ionization parameter $\log 
  U$ must be greater than or equal to $-1.4$ in order to produce
  sufficient \ion{He}{1}*.  A log density less than $\sim 8$ was
  required, or too much Balmer absorption was predicted.  The log
  hydrogen column density was constrained to be greater or equal to
  21.6.   Comparison with other objects showed that FBQS~J1151$+$3822
  is a   relatively high-ionization, high-column-density BALQSO. 

\item The {\it Cloudy} simulations were used to see if UV spectra
  would be able to constrain parameters further.  It was found that
  ionization parameter would be reasonably easily constrained through
  observation of low ionization lines such as \ion{Mg}{2} and
  \ion{N}{3}, which decrease in optical depth as the ionization
  parameter increases, and observation of the high ionization line
  \ion{P}{5}, which increases in optical depth as the ionization
  parameter increases.  Several density diagnostic lines were
  examined; all were predicted to be too blended to be useful except
  for the high-velocity side of \ion{C}{3}*$\lambda 1176$.  

\item The broad band photometry was examined.  The observed decrease
  in the UV was too steep to be solely caused by reddening.
  Attenuation is expected also because of the absorption lines
  predicted in the UV.  However, a combination of correction for
  modest reddening  ($E(B-V)=0.1$ for an SMC reddening
  curve) and attenuation due to predicted high-ionization
  absorption lines left the observed UV still a factor of $\sim 2.5$
  too low.

\item The mass outflow rate, kinetic luminosity and ratio of kinetic
  to bolometric luminosity were estimated.  The range was large over
  allowed parameter space but generally the minimum values were higher
  than those estimated in other objects.  

\item Acceleration mechanisms were examined.  A force-multiplier
  analysis was done using the {\it Cloudy} results. It was found
  for this luminous object, radiating at about 1/2 the Eddington
  luminosity, the terminal velocity $v_{max}$ could be easily
  achieved.  Furthermore, a comparison of the size scale of the
  outflow with the radius of the outflow constrained the log of the
  density to be greater than $\sim 7$.  This strongly
  constrained parameter space, yielding an acceptable range log
  hydrogen column densities of 21.7--22.9, radii between 2--12
  parsecs, mass outflow rates between 11 and 54 solar masses per year,
  ratio of outflow to inflow rates between 1.2 and 5.8, kinetic
  luminosity between 1 and $5\times 10^{44}\rm \, erg\, s^{-1}$,  and
  kinetic luminosity between 0.2 and 0.9\% of the bolometric
  luminosity.  Compared with other objects, FBQS~J1151$+$3822 has a
  relatively powerful outflow originating relatively close to the
  central engine. 

\item We examine the potential for using the \ion{He}{1}* absorption
  lines to detect and measure the properties of high-column-density
  BALQSOs.  Objects with high column densities may require the highest
  kinetic energies and so may be important for testing models.  We
demonstrate that many prominent BAL lines become saturated at high
column densities.  We show that \ion{He}{1}*$\lambda 10830$ compares
favorably with \ion{P}{5}, a widely used probe of high column
densities, while \ion{He}{1}*$\lambda 3889$, because of its low
oscillator strength, is even more sensitive.  We use simulated spectra to
see how well we can reproduce input covering fractions and column 
densities using a partial covering model.  We find that the high
$\lambda f_{ik}$ ratio of 23.3 makes the \ion{He}{1}* lines more
sensitive to a wider range of column density than lines with a
$\lambda f_{ik}$ ratio equal to 2 (like \ion{P}{5}).  

\item We briefly discussed the prospects of finding other
  \ion{He}{1}*$\lambda 10830$ BALQSOs.  We have in fact already
  collected data on six additional objects; furthermore, we have
  discovered that several well-known, bright low-redshift BALQSOs have
  no \ion{He}{1}*$\lambda 10830$ absorption, a fact that will place
  upper limits on the column densities in those objects.  We discussed
  the utility of observing a sample of low-redshift BALQSOs identified
  via \ion{He}{1}* absorption.  Such objects may be valuable for
  understanding the relationship between luminosity and terminal
  velocity $v_{max}$ by allowing definition of a lower luminosity
  sample in a region currently sparsely populated.  

\end{itemize}

\acknowledgments
KML thanks Fred Hamann for suggesting the identification of the
absorption line in the IR spectrum.  KML \& MD thank the IRTF staff for
copious help during the observing runs. KML thanks Eddie Baron and Don
Terndrup for useful discussions.   KML also thanks the students in
the Spring 2010 ``Nebulae and AGN'' class for suffering through
discussions of \ion{He}{1}* absorption and associated homework.   This
research has made use 
of the NASA/IPAC Extragalactic Database (NED) which is operated by the
Jet Propulsion 
Laboratory, California Institute of Technology, under contract with
the National Aeronautics and Space Administration. Funding for the
SDSS and SDSS-II has been provided by the Alfred P. Sloan Foundation,
the Participating Institutions, the National Science Foundation, the
U.S. Department of Energy, the National Aeronautics and Space
Administration, the Japanese Monbukagakusho, the Max Planck Society,
and the Higher Education Funding Council for England. The SDSS Web
Site is http://www.sdss.org/.  
The SDSS is managed by the Astrophysical Research Consortium for the
Participating Institutions. The Participating Institutions are the
American Museum of Natural History, Astrophysical Institute Potsdam,
University of Basel, University of Cambridge, Case Western Reserve
University, University of Chicago, Drexel University, Fermilab, the
Institute for Advanced Study, the Japan Participation Group, Johns
Hopkins University, the Joint Institute for Nuclear Astrophysics, the
Kavli Institute for Particle Astrophysics and Cosmology, the Korean
Scientist Group, the Chinese Academy of Sciences (LAMOST), Los Alamos
National Laboratory, the Max-Planck-Institute for Astronomy (MPIA),
the Max-Planck-Institute for Astrophysics (MPA), New Mexico State
University, Ohio State University, University of Pittsburgh,
University of Portsmouth, Princeton University, the United States
Naval Observatory, and the University of Washington.  KML \& SB
acknowledge support by NSF AST-0707703. MB acknowledges support by NSF 
AST-0604066



{\it Facilities:} \facility{Infrared Telescope Facility},
\facility{Hiltner Telescope}.



\appendix

\section{Spectral Energy Distribution Dependence}

The incident spectral energy distribution influences the results.  We
examine this using an updated version of the semi-empirical spectral
energy distributions developed in \citet{casebeer06}.  They are
parameterized by the cutoff temperature of the UV bump measured in electron
volts.  The \citet{korista97} {\tt AGN Kirk} continuum is similar to a
spectral energy distribution with $T_{cut}=80 \rm \, eV$.  
Because the \ion{He}{1}* does not have significant density dependence,
we investigate the behavior as a function of ionization parameter and
$T_{cut}$ for a single value of the density $\log n = 5.5$.  

We first investigate the total hydrogen column density needed to
attain the required \ion{He}{1}* column.  The results are shown in
Fig.~\ref{fig9}.   We find that softer spectral energy distributions
(i.e., lower values of $T_{cut}$) require lower values of the hydrogen
column density for a particular ionization parameter, while harder
SEDs require larger values of the hydrogen column density.  This makes
sense because helium ionization front occurs at a smaller depth when
the SED is soft, and at a larger depth when the SED is hard.
Thus, the location of recombining He$^+$ shifts in the slab.

We also investigate the column density of P$^{+4}$ as a function of
ionization parameter and spectral energy distribution
(Fig.~\ref{fig9}).  As discussed in \citet{leighly09}, the P$^{+4}$
zone also occurs deeper in the slab when the spectral energy
distribution is hard than when it is soft.  But here we see it is
almost constant for a given ionization parameter (for higher values of
the ionization parameter where \ion{P}{5} absorption would be
detectable).  This simply shows that the depth of the P$^{+4}$ region
shifts in the gas in the same way as the He$^{+}$ region as a function
of SED.  Thus, there appears to be little relative dependence on
spectral energy distribution between these two lines. 

\section{Metallicity Dependence}

The metallicity of the gas will also influence the results.  We
briefly investigate the effect of metallicity by running a grid of
models with the \citet{korista97} {\tt AGN kirk} continuum and a
metallicity five times that of solar ($Z=5$).  As discussed in
\citet[][Table 2]{hamann02}, this means that each metal element is
five times more prevalent with respect to hydrogen than at solar
abundance, with the exception of nitrogen, which because of the
secondary production process is 25 times more abundant than at solar
metallicity.  In addition, helium is enhanced by a factor of 1.29.

Notable differences were seen between the $Z=5$ metallicity results
and the solar metallicity results presented in \S 3.1.  First, there is
a difference in the hydrogen column density required to produce the
observed metastable helium column density, namely, the allowable
solutions extend to lower ionization parameters, down to $\log U=-1.8$
versus $\log U=-1.4$ for solar metallicity.  At these lower ionization
parameters, we expect lower ionization species to be more prominent;
that could potentially be recognized in the UV spectrum (see below; \S
3.2.2).   

One might imagine that the column densities from the higher
metallicity grid might differ from the solar metallicity grid by a
constant factor reflecting the abundance differences.  However, the
presence of enhanced metals changes the cooling in the gas
\citep[e.g.,][]{hamann02}, and so the results are not so simple.  We
show the difference in the log column between the $Z=5$ and the $Z=1$
results in Fig.~\ref{fig10}, both for the total hydrogen column and the
P$^{+4}$ column density.  We include only regions of parameter space
were both solutions exist.   

For the total hydrogen density, we see that the difference in the log
of the column density is negative in all cases. This means that the
column required is lower in the $Z=5$ case, as expected.  As mentioned
above, in the $Z=5$ experiment, the abundance of helium is enhanced
by a factor of 1.29. This corresponds to a difference in log of 0.11.
We observe a much larger range of differences, with the largest
difference up to almost $-0.5$ for lower densities.  We suspect this
is a consequence of enhanced cooling at low densities due to the
higher metallicity.  As mentioned above,     the
metastable state is populated by recombination from He$^{+1}$ and
depopulated chiefly by collisions  \citep{clegg87, ch89}.  When the
temperature is lower, a 
larger fraction remains in the metastable state.  So a smaller total
hydrogen column is necessary to attain the required column of
\ion{He}{1}*.  The effect is enhanced at low densities where the gas
generally has more trouble cooling than it does at higher densities.    

For P$^{+4}$, we see that the difference is positive in all cases.
This means that there are more P$^{+4}$ at $Z=5$ than at
solar metallicity, as expected.  But at $Z=5$, the number of
phosphorus atoms is larger by a factor of five, while the enhancement in
the number of P$^{+4}$ reaches that value only at the highest
ionization parameters.  Part of the difference between the naive
expectation and the {\it Cloudy} results is caused by the lower
hydrogen column density.  The remainder of the difference may again be
due to enhanced cooling at higher metallicities; as discussed by
\citet{hamann02}, an increase in the metal abundance leads to enhanced
cooling and nearly constant flux ratios with respect to e.g.,
Ly$\alpha$.  This effect is most prominent at low ionization parameter
where there are plenty of species to cool the gas.  At high ionization
parameter, the number of lines accessible decreases, and the
difference in cooling between the $Z=1$ and $Z=5$ cases decreases.

\section{Differential Velocity Dependence}

In the {\it Cloudy} simulations presented in \S 3.1, we assumed that
the slab of gas was stationary.  This is probably not a realistic
model for the data given that we are modeling an outflow.
The presence of a differential velocity field can influence the
radiative transfer of Ly$\alpha$.  In a stationary slab and at high
densities, this line can be trapped.   That will increase the rate of
pumping of \ion{H}{1} into $n=2$, and it may increase the rate of
photoionization of \ion{He}{1}*.  In this section, we explore the
effects of a differential velocity field by including turbulence in
the {\it Cloudy} modeling.

\begin{figure}
\epsscale{1.0}
\plotone{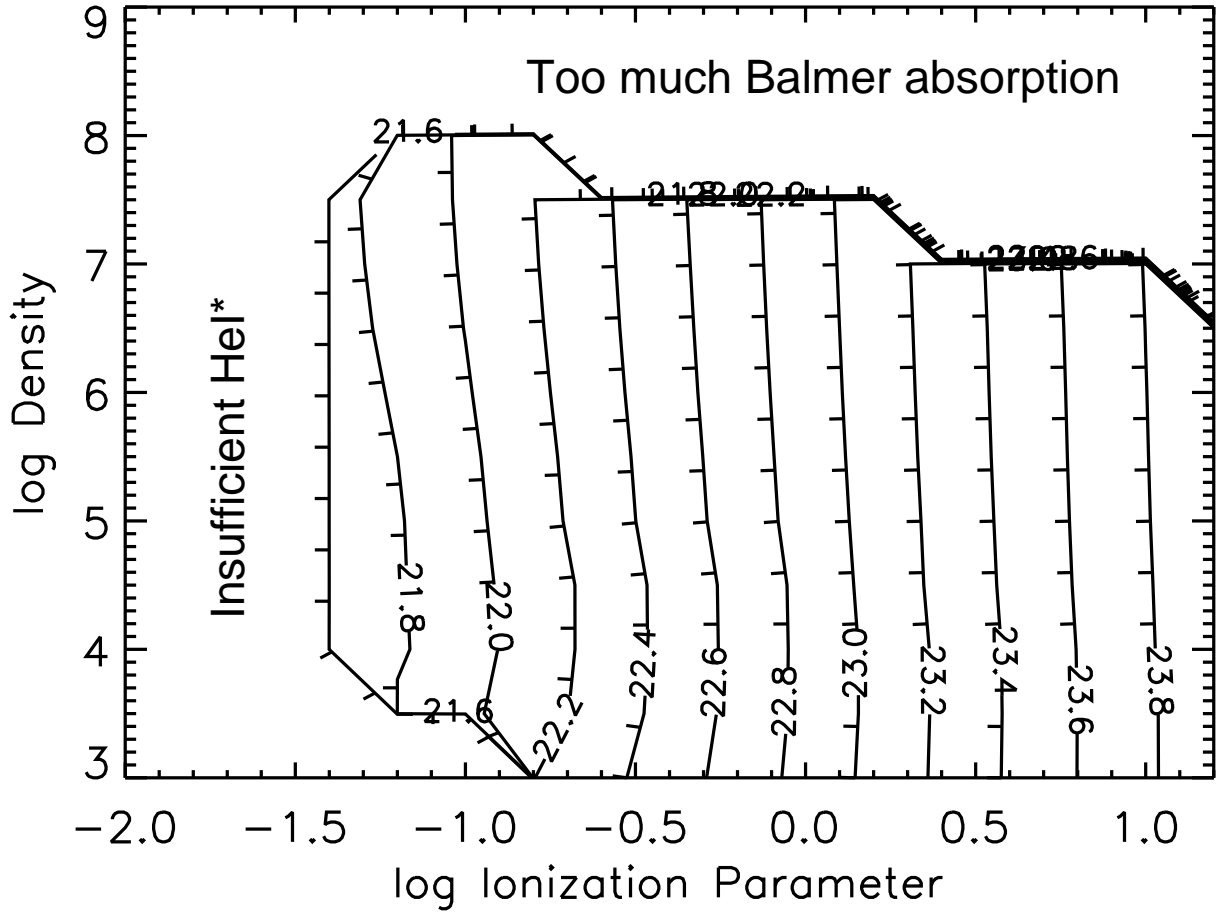}
\caption{Total hydrogen column density obtained from {\it Cloudy} modeling.
  At each point, the hydrogen column was adjusted until the
  \ion{He}{1}* average column was obtained.  Insufficient \ion{He}{1}*
  was produced for ionization parameters lower than $\sim -1.4$.  High
  densities predict Balmer absorption in excess of the lower limit
  derived from the spectra. \label{fig8}}  
\end{figure}

\begin{figure}
\epsscale{0.6}
\plotone{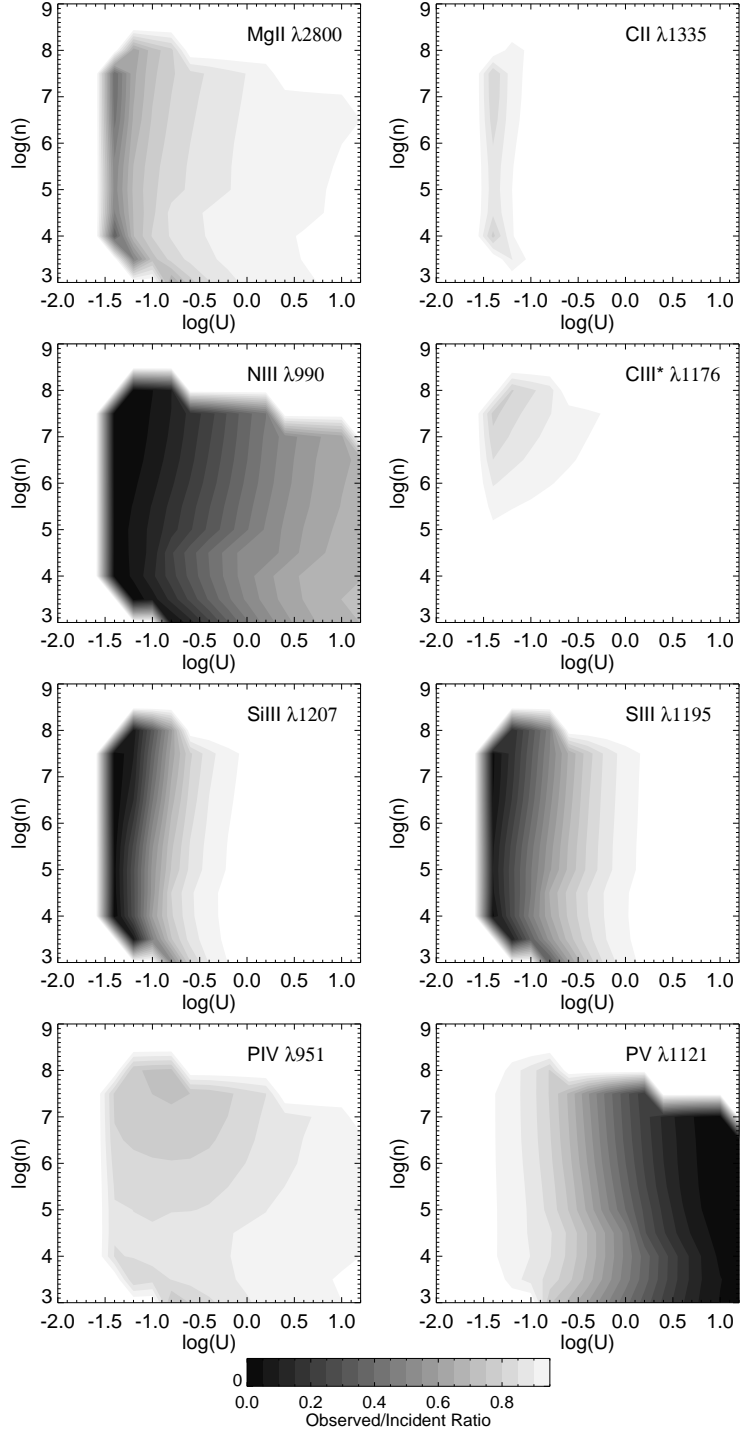}
\caption{Contours of the ratio of the predicted intensity divided by
  the continuum intensity for column densities obtained from the
  {\it Cloudy} modeling. The optical depth has been spread over a range of
  11,000$\rm\,  km\,s^{-1}$ appropriate for the line profile in this
  object.  Transitions that are shown are neither saturated nor too
  weak to be observed; they can in principle remove the
  degeneracy in the  modeling by constraining ionization parameter and
  density.\label{fig11}}   
\end{figure}

\begin{figure}
\epsscale{1.0}
\plotone{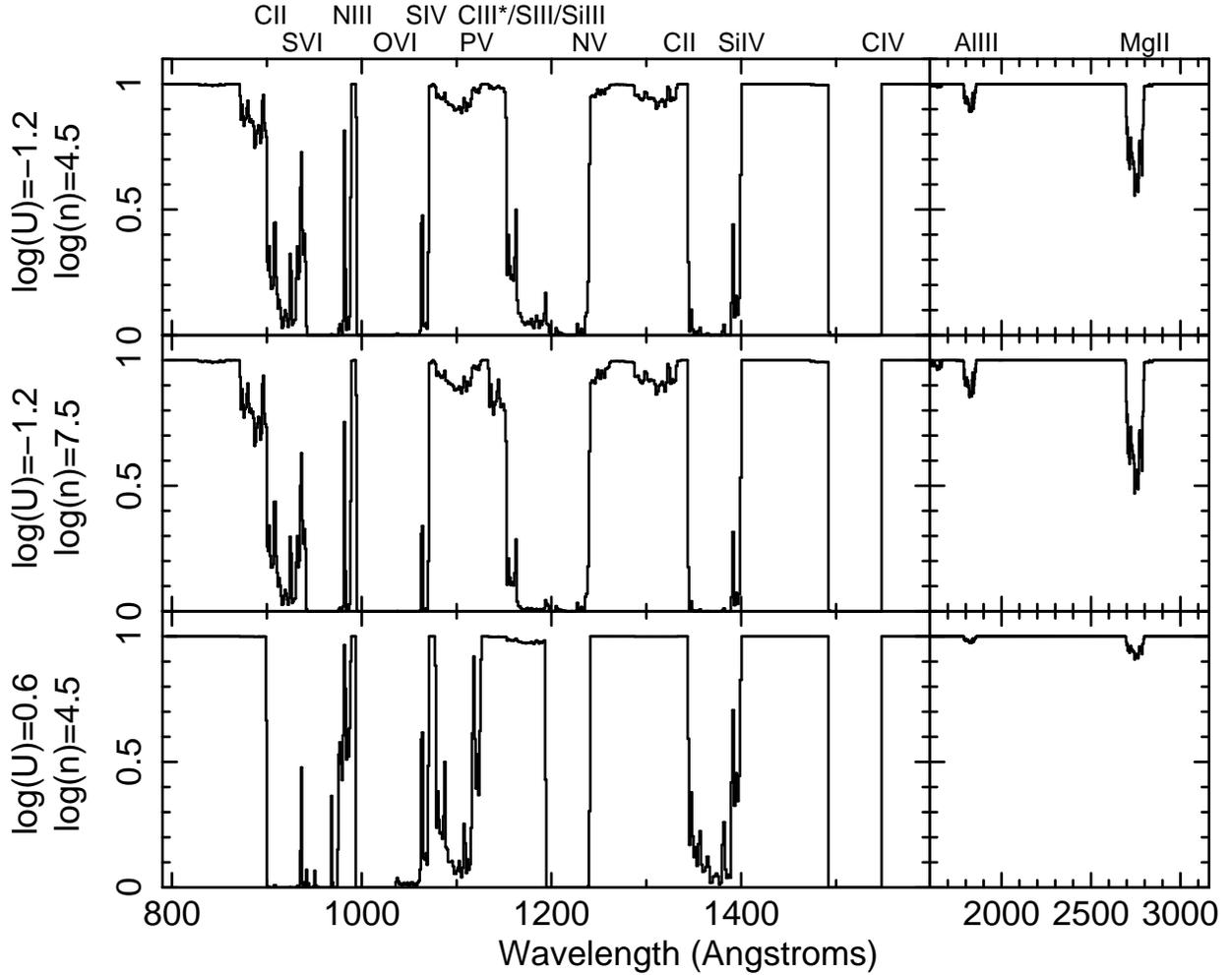}
\caption{Ratio of simulated spectrum to continuum for three
  combinations of density and ionization parameter from
  Fig.~\ref{fig10}; these are listed on the y axes.  The name of the
  absorption   line is listed above the spectra, centered at the rest
  wavelength.   Ionization parameter discrimination is
  straightforward:   \ion{P}{5}   is much stronger at high ionization
  parameters, while    low-ionization lines such as \ion{Mg}{2},
  \ion{C}{2}, and \ion{S}{3}   are stronger at low ionization
  parameters.  Blending makes density   discrimination more difficult,
  but   \ion{C}{3}*$\lambda 1176$ is present   at low ionization parameters
  and high densities.  \label{fig12}}    
\end{figure}

\begin{figure}
\epsscale{1.0}
\plotone{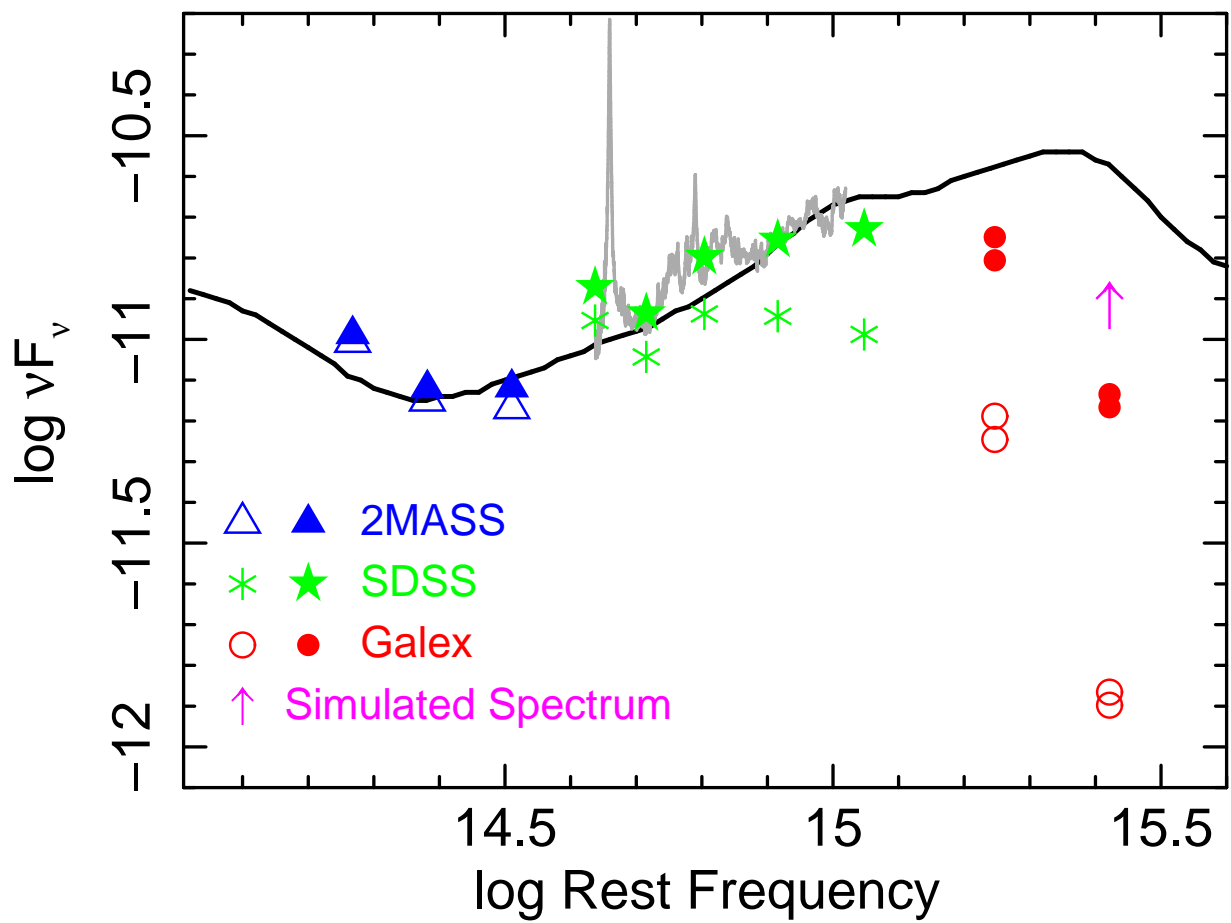}
\caption{The broadband spectral energy distribution constructed from
  non-simultaneous photometry data.  The solid line is the
  \citet{richards06} spectral energy distribution scaled to the 1
  micron break.  The open points show the observed data (corrected
  for Galactic reddening and redshift); the solid points show the
  observed data dereddened by $E(B-V)=0.1$ and an SMC reddening curve
  \citep{pei92}. The grey line shows the SDSS spectrum dereddened by
  the same amount.  The arrow shows the flux predicted by the simulated
  spectra.     The fact that the UV data lie below the SED shows
  that the UV is attenuated by   absorption lines as well as intrinsic
  reddening.  \label{fig13}}     
\end{figure}

\begin{figure}
\epsscale{0.8}
\plotone{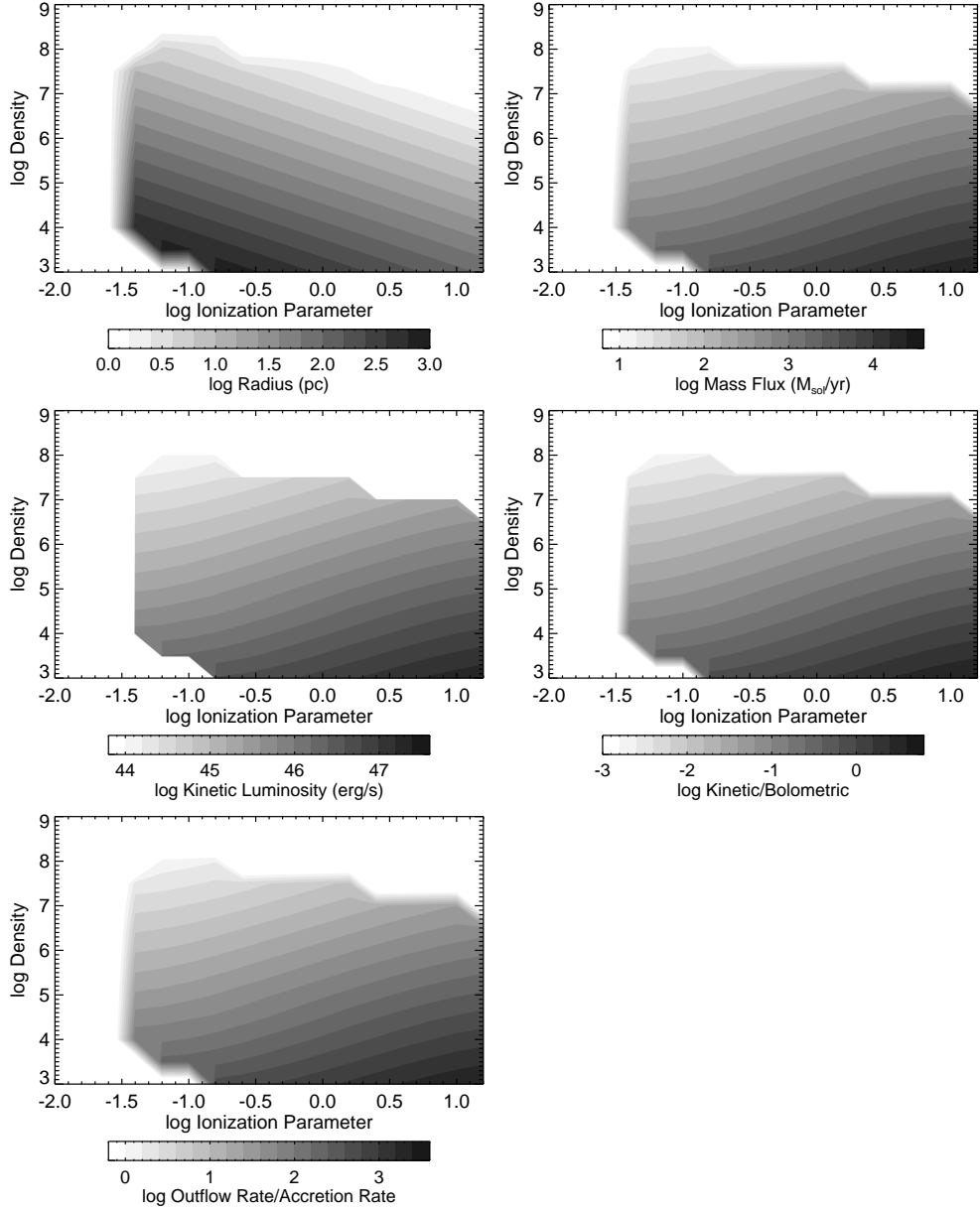}
\caption{{\it Top left:} The absorption-line radius inferred from the
  {\it Cloudy} modeling.  {\it Top right:}  The mass outflow rate.
  The minimum mass flux allowed by our data is about 10 $M_\odot \rm
  \, yr^{-1}$.    {\it Middle left:} The kinetic luminosity.  {\it Middle right:} The
  ratio of the kinetic luminosity and the bolometric luminosity.
  {\it Bottom left:} The ratio of the outflow rate to the accretion
  rate.    \label{fig14}}   
\end{figure}

\begin{figure}
\epsscale{1.0}
\plotone{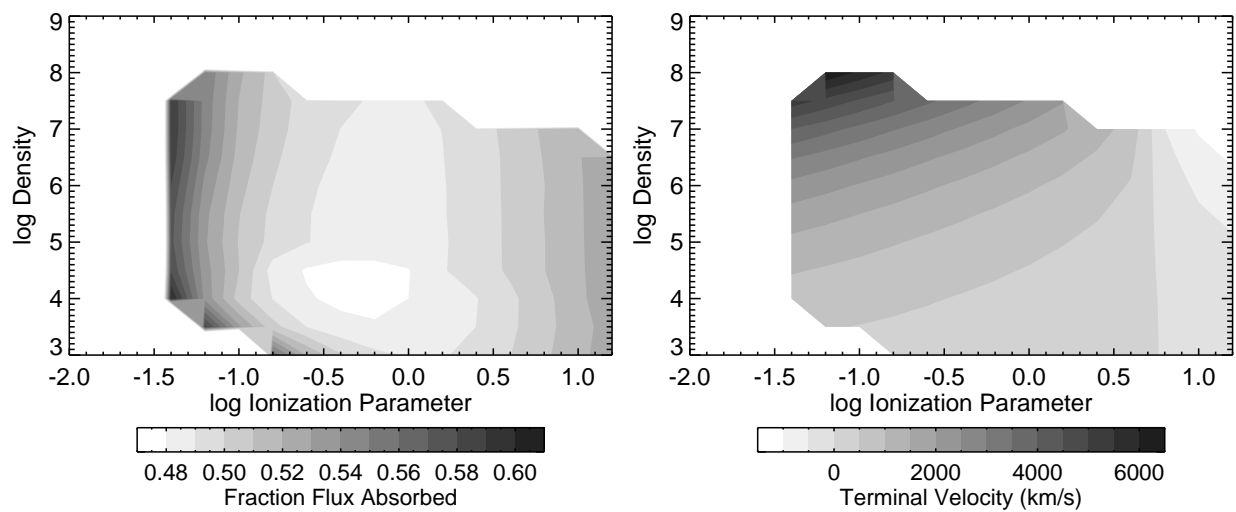}
\caption{{\it Left:} The fraction of the incident continuum absorbed.
  In principle, the photon momentum from this absorbed light can be
  converted to momentum of the outflow 
  \citep{hamann98}.  {\it Right:}  The terminal velocities predicted
  if the fraction of the incident continuum that is absorbed or
  scattered is converted to momentum of the outflow.  Positive
  velocities denote outflows.  Fairly large velocities can be
  attained; however, they are still lower than the observed terminal
  velocity of $11,000\rm \, km\, s^{-1}$ observed in
  FBQS~J1151$+$3822.  \label{fig15}}    
\end{figure}

\begin{figure}
\epsscale{0.8}
\plotone{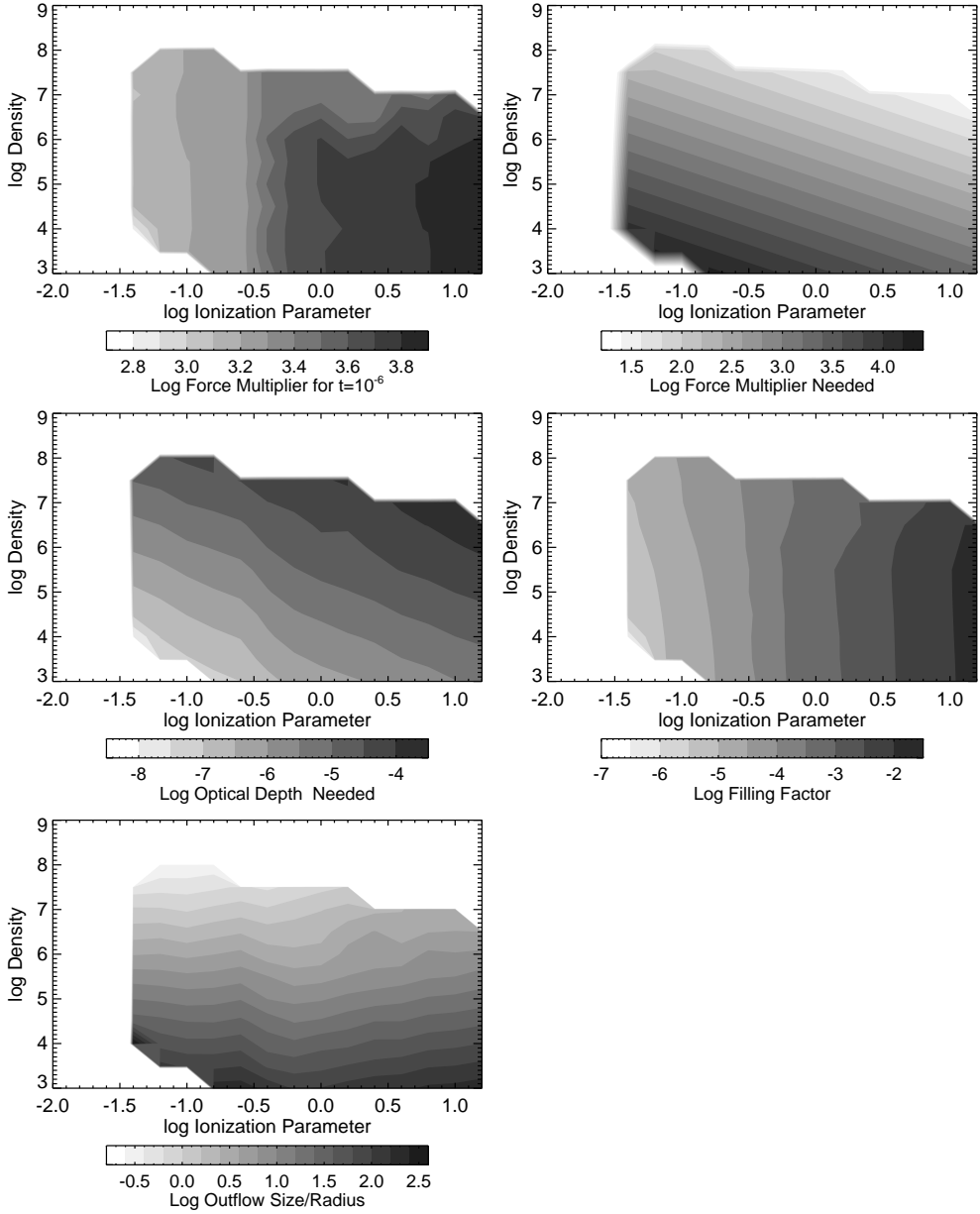}
\caption{{\it Top left:}  The line force multiplier estimated from the
  {\it Cloudy} output results for an example value of the equivalent
  electron optical depth of $10^{-6}$.  {\it Top right:}  The log of
  the force multiplier needed to attain the terminal velocity of
  $11,000 \rm\, km\, s^{-1}$ observed in FBQS~J1151$+$3822.  {\it
    Middle left:}  The log of the  equivalent electron optical depth
  required to attain the force multiplier needed to accelerate the
  outflow to the observed terminal velocity, given the force
  multiplier values computed from the {\it Cloudy} results.  {\it
    Middle right:}  The filling factor needed to accelerate the
  outflow to the observed terminal velocity.  We assume that $dv/dr$
  is approximately the  terminal velocity divided by the radius where
  the absorption occurs, obtained from the photoionization modeling
  results.  {\it Bottom left:}  The ratio of size of the outflow to
  the absorption radius.  Values much larger than zero (i.e., a ratio
  equal to one) are   unphysical.   \label{fig16}}   
\end{figure}

\begin{figure}
\epsscale{0.8}
\plotone{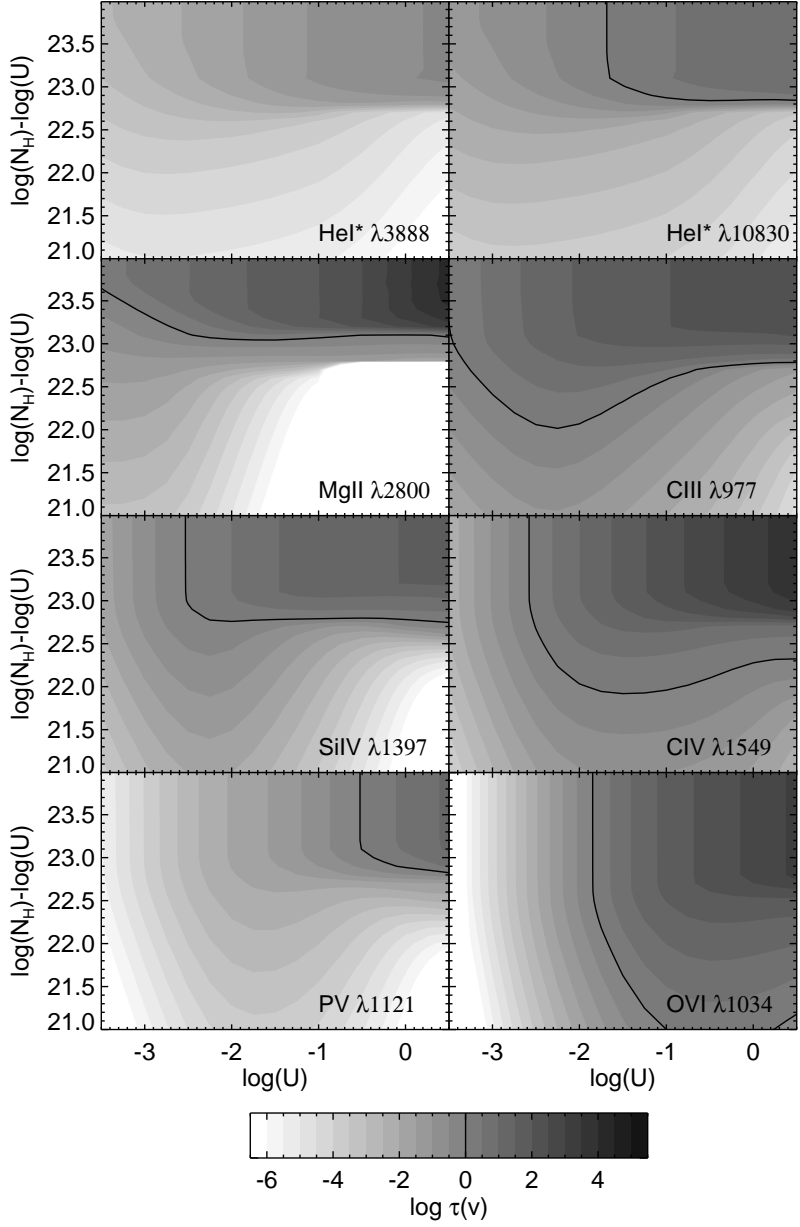}
\caption{Contours of $\log \tau(v)$ for a selection of lines as a function
  of ionization parameter and $\log(N_H)-\log(U)$.  We assumed a
  square absorption profile with  velocity width of $10,000\rm\, km\,
  s^{-1}$.   Lower velocity widths  would yield higher opacities for any
  combination of parameters.  $\tau(v)=1$ is shown by a contour line
  to guide the eye.  These contours show that resonance lines such as
  \ion{C}{3} and \ion{O}{6} are predicted to be saturated at
  relatively low column densities, while the high-column-density
  diagnostic lines including \ion{P}{5}, \ion{He}{1}*$\lambda 3889$,
  \ion{He}{1}*$\lambda 10830$ remain optically thin until high column
  densities are attained.   \label{fig17}}   
\end{figure}

\begin{figure}
\epsscale{1.0}
\plotone{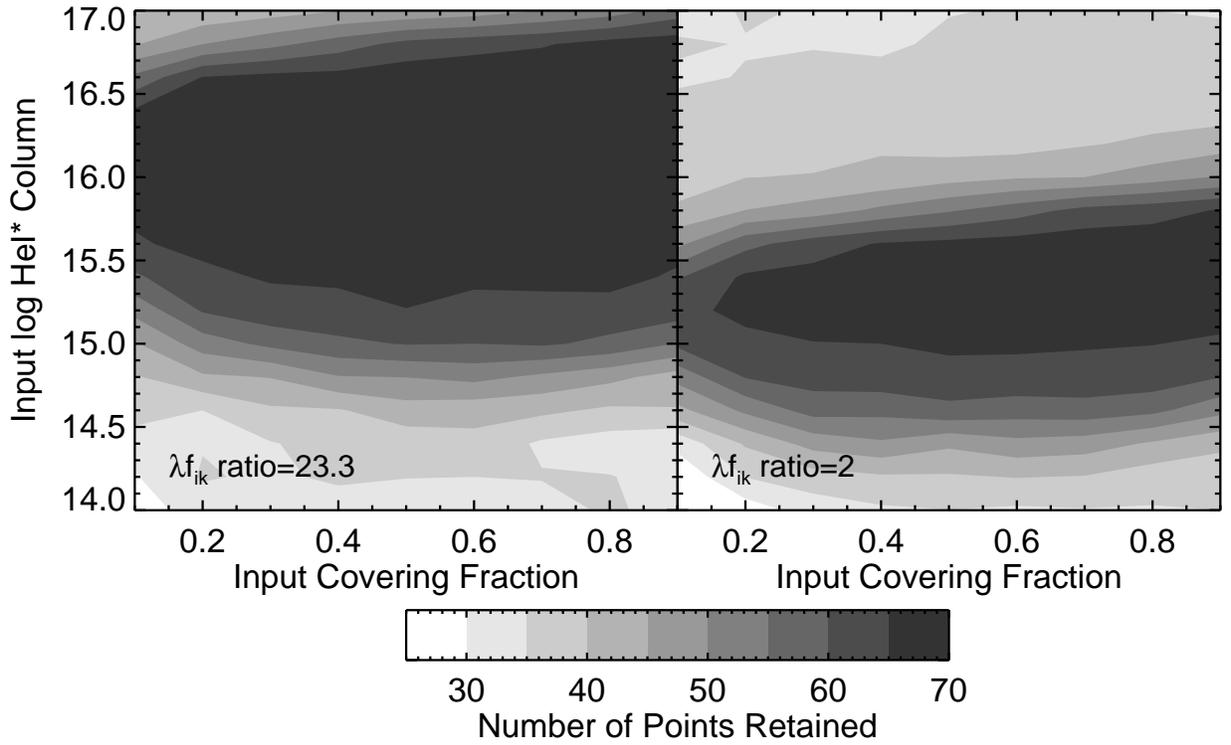}
\caption{There are 69 points in the IRTF velocity profile.  The
  fitting software excludes unphysical points where  $R_{3889} >
  R_{10830} > R_{3889}^{23.3}$ is not obeyed.  Contours show the
  number of points retained in the fitting.   \label{fig18}}   
\end{figure}

\begin{figure}
\epsscale{0.8}
\plotone{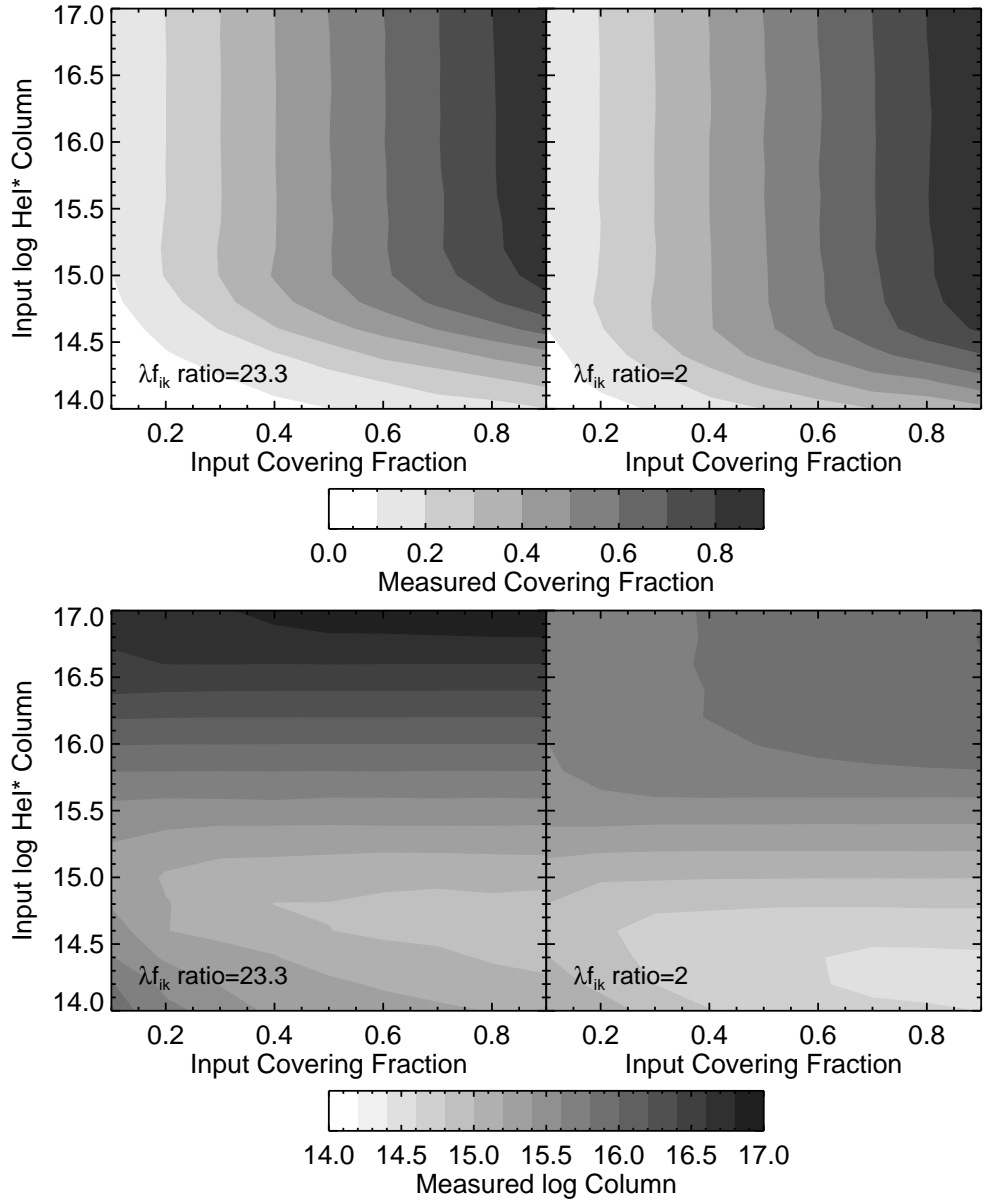}
\caption{The measured covering fraction (top) and column density
  (bottom) as a function of the input covering fraction and
  column.  \label{fig19}}   
\end{figure}

\begin{figure}
\epsscale{0.8}
\plotone{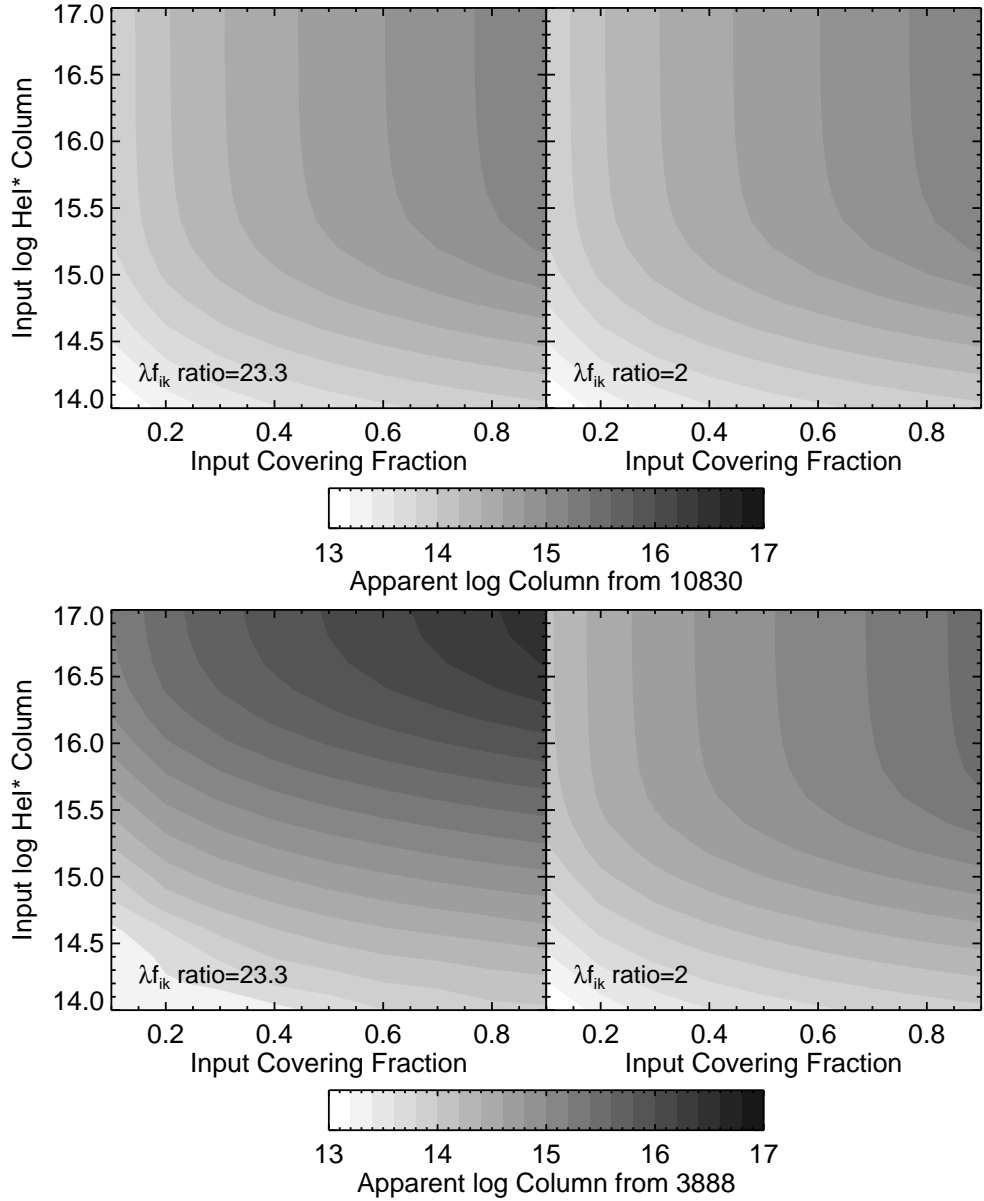}
\caption{Apparent column densities measured from the 10830\AA\/
  profile (top) and from the 3889\AA\/ profile (bottom).  \label{fig20}}   
\end{figure}

\begin{figure}
\epsscale{1.0}
\plotone{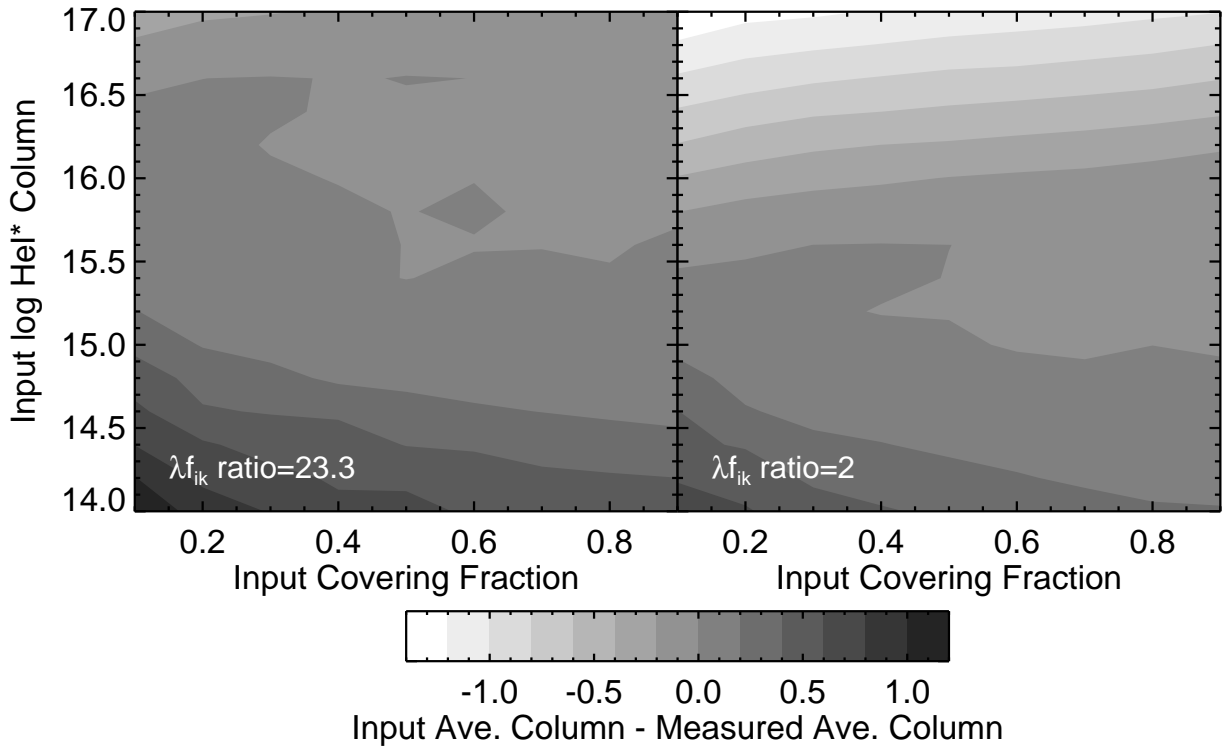}
\caption{Difference between the log of the input average column and
  the measured average column.  Values around zero show the regions
  of parameter space where the mean column (used in photoionization
  and dynamical models) is best reproduced.  \label{fig21}}   
\end{figure}

\begin{figure}
\epsscale{1.0}
\plotone{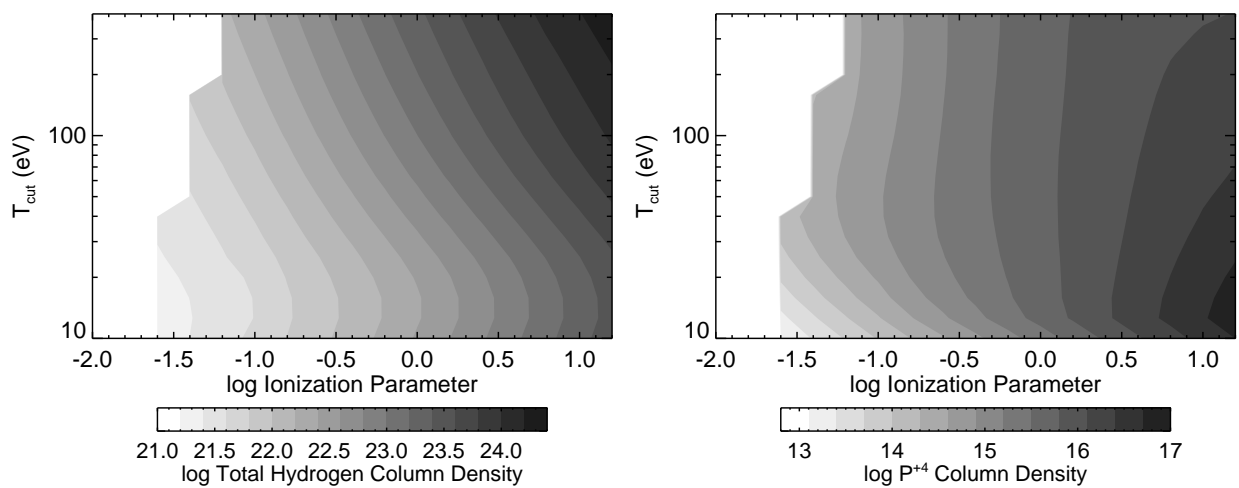}
\caption{The results depend on the spectral energy distribution.
  Semi-empirical spectral energy distributions from
  \citet{casebeer06} are parameterized by the cutoff temperature of
  the UV bump in eV.  {\it Left:} The total hydrogen column density
  required to produce the measured \ion{He}{1}* column density. Spectral
  energy dependence is seen, such that softer continua require lower
  column densities to produce the needed \ion{He}{1}* column. {\it
    Right:} The P$^{+4}$ column for 
  the simulations shown in the left   panel.  The fact that the
  contours are mostly vertical implies that there is little
  relative spectral energy distribution dependence between \ion{P}{5}
  and \ion{He}{1}*.  
 \label{fig9}}  
\end{figure}

\begin{figure}
\epsscale{1.0}
\plotone{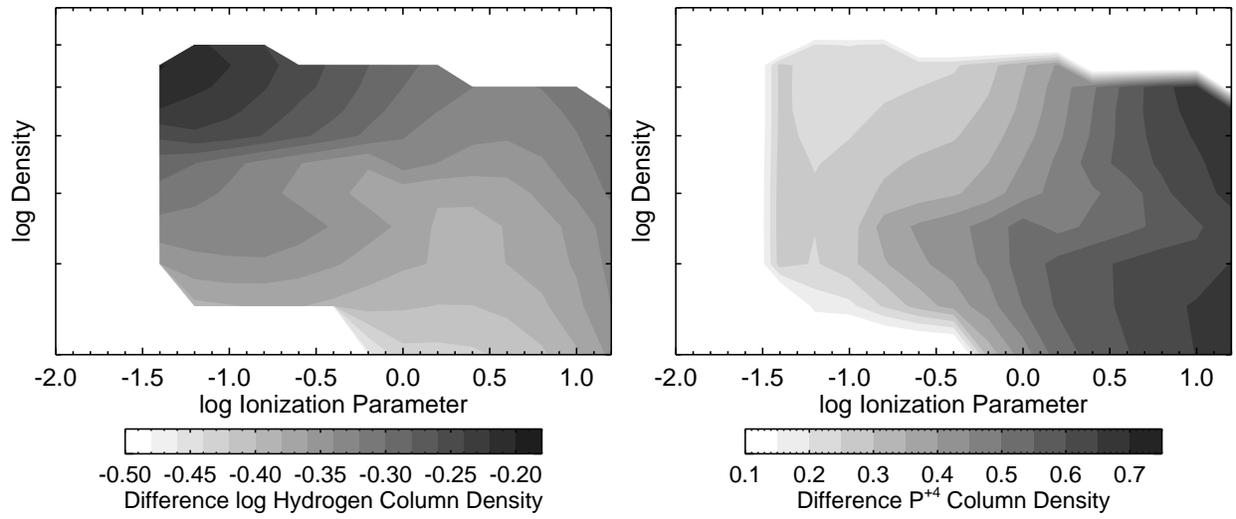}
\caption{The results depend on the metallicity of the gas.  We show
  the difference in the log of the column densities for the $Z=5$
  case and the solar metallicity case discussed in \S 3.1.  {\it
    Left:} 
  The difference in the log column density of the hydrogen column
  necessary to produce a sufficient metastable neutral helium
  column. {\it Right:} The difference in log column density of
  P$^{+4}$.    \label{fig10}}  
\end{figure}

\begin{figure}
\epsscale{1.0}
\plotone{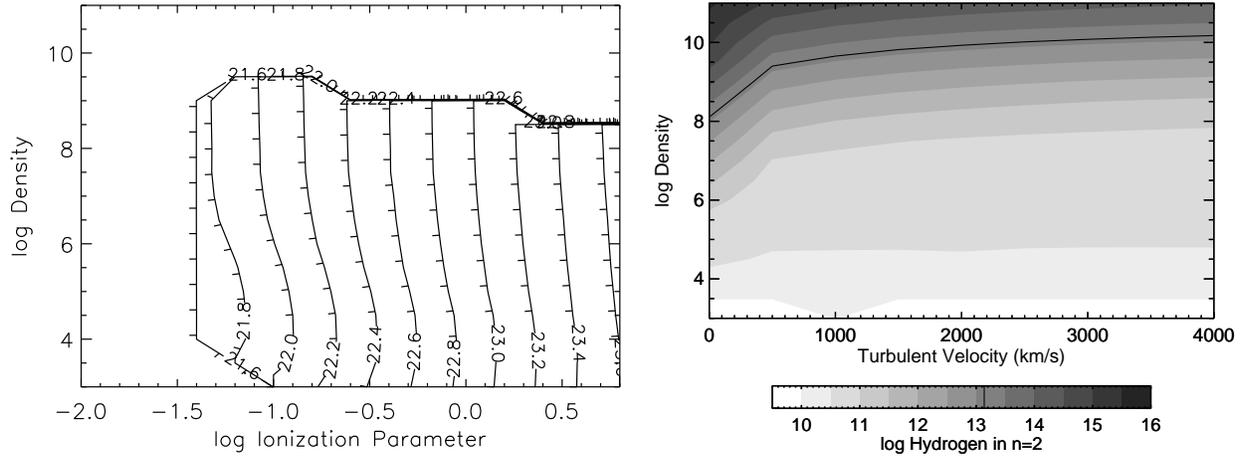}
\caption{The results depend on the differential velocity properties of
  the gas.  {\it Left:}  The total hydrogen column density required to
  produce the observed \ion{He}{1} column density for
  $v_{turb}=1000\rm \, km\, s^{-1}$.  The hydrogen column densities
  are nearly the same as those shown in Fig.~\ref{fig8}
  ($v_{turb}=0$), except that the allowed parameter space extends to
  higher densities.  {\it Right:} The column density of hydrogen in
  $n=2$ for $\log U=-1.0$  as a function of $v_{turb}$ and density for
  models where the hydrogen column  density is chosen to produce the
  observed \ion{He}{1}*.  Ly$\alpha$ trapping is more important at
 high densities, resulting in an increase in hydrogen in $n=2$.  The
 decrease in \ion{H}{1} in $n=2$ as a function of $v_{turb}$ is a
 consequence of the suppression of Ly$\alpha$ trapping when turbulence
 is  present.  The solid line shows the upper limit on \ion{H}{1}
 $n=2$ obtained from the lack of Balmer absorption.     \label{fig22}}  
\end{figure}

Initially, we include $v_{turb}=1000\rm\, km\, s^{-1}$ in the model.
We found that the execution time increased with $v_{turb}$, so we
chose this value to start with.  As discussed in \S 3.1, we ran
models using a large column density, used the output to determine the
column density required to produce the \ion{He}{1}* column density
measured from the data, then ran the models again. The grid sampled a
slightly smaller range of ionization parameter than in \S 3.1, as {\it
  Cloudy} crashed for the highest ionization parameters where $ \log U
> \sim 0.8$.   The 
results are shown in Fig.~\ref{fig22}.  Comparing this figure with
Fig.~\ref{fig8}, we see that the column densities are nearly the same,
as well as the ionization constraint on the allowed parameter space.
But higher densities are allowed, up to $\log n=9.5$ for $\log U \sim
-1.0$. This happens because as $v_{turb}$ increases, Ly$\alpha$ is less
trapped,  so the column density of \ion{H}{1}in $n=2$ decreases, and
so the upper limit on \ion{H}{1} in $n=2$ is attained at higher
densities. 

We note that we see no evidence for destruction of \ion{He}{1}* by 
Ly$\alpha$.  If that were important, we would expect an increase in
required hydrogen column at high densities, because we would have to
go deeper into the gas to attain sufficient \ion{He}{1}*.
Exploratory {\it Cloudy} modeling showed that such an increase in
total hydrogen column is only seen at $v_{turb}=0$ for very large
densities, i.e.,  $\log n > ~10$.  

Fig.~\ref{fig22} also shows the column density of  \ion{H}{1} in $n=2$
for $\log U=-1$ as a function of turbulent velocity between 0 and
4000$\,\rm km\, s^{-1}$.  Again, the total
hydrogen column density has been adjusted to provide the observed
\ion{He}{1}*, and that is nearly constant, lying between 21.91 and
22.06.  This graph shows that, for constant higher densities, after an
initial decrease in \ion{H}{1} $n=2$ column density as a function of
$v_{turb}$, the dependence levels off so that the column density in
\ion{H}{1} $n=2$ at $v_{turb}=3000\rm \, km\, s^{-1}$ is nearly the
same as that at $v_{turb}=4000\rm \, km\, s^{-1}$.  Therefore, our
upper limit on Balmer absorption lines seems to still rule out very
high densities, even when a differential velocity field is present.

\clearpage

\clearpage

\clearpage

\clearpage

\clearpage

\clearpage

\clearpage

\clearpage

\clearpage

\clearpage

\clearpage

\clearpage

\clearpage

\clearpage

\clearpage

\clearpage

\clearpage

\clearpage

\clearpage

\clearpage

\clearpage

\clearpage

\clearpage

\begin{deluxetable}{lccc}
\tablecaption{Inhomogeneous Absorber Model Fitting Results}
\tablewidth{0pt}
\tablehead{
\colhead{Spectrum\tablenotemark{a}} & \colhead{Points
  Retained\tablenotemark{b}} & \colhead{Partial Covering Model} & \colhead{Power Law Model}\\
& & \colhead{log \ion{He}{1}* Column Density} & \colhead{log \ion{He}{1}* Column Density}}
\startdata 
MDM -- 28 spectra & 63 & $14.9^{+0.6}_{-0.2}$ & $15.0^{+0.2}_{-0.1}$  \\
MDM -- excess variance & 61 &  $14.8^{+0.9}_{-0.2}$ & $14.8^{+0.2}_{-0.1}$   \\
SDSS -- 28 spectra & 64 & $15.0^{+0.8}_{-0.2}$ & $15.1^{-0.2}_{-0.2}$ \\
SDSS -- excess variance & 62 & $14.9^{+0.8}_{-0.2}$ & $15.0^{+0.2}_{-0.2}$\\
\enddata
\tablenotetext{a}{The origin of the optical spectrum and the type \ion{Fe}{2}
  model used to extract the 3889\AA\/ line. The IRTF spectrum was used
  throughout for the 10830\AA\/ line.}
\tablenotetext{b}{The number of points out of the original 69 that were
  retained after unphysical points were removed (see text for details).}
\end{deluxetable}


\begin{thebibliography}{}

\bibitem[Arav (1996)]{arav96} Arav, N.\ 1996, \apj, 465, 617

\bibitem[Arav et al.\ (1995)]{arav95} Arav, N., Korista, K.\ T.,
  Barlow, T.\ A., \&  Begelman, M.\ C.\ 1995, Nature, 376, 576

\bibitem[Arav \& Li (1994)]{arav94} Arav, N., \& Li, Z.-Y.\ 1994,
  \apj, 427, 700

\bibitem[Arav et al.\ (1994)]{alb94} Arav, N., Li, Z.-Y., \& Begelman,
  M.\ C.\ 1994, \apj, 432, 62

\bibitem[Arav et al.\ (2005)]{arav05} Arav, N., Kaastra, J., Kriss,
  G.\ A., Korista, K.\ T., Gabel, J., \& Proga, D.\ 2005, \apj, 620,
  665

\bibitem[Bentz et al.\ (2006)]{bentz06} Bentz, M.\ C., Peterson,
  B.\ M., Pogge, R.\ W., Vestergaard, M., \& Onken, C.\ A.\ 2006,
  \apj, 644, 133

\bibitem[Bethe \& Salpeter (1957)]{bs57} Bethe, H.\ A., \& Salpeter,
  E.\ E.\ 1957, ``Quantum Mechanics of One- and Two-Electron Atoms''
  (New York: Academic Press)

\bibitem[Boroson \& Green (1992)]{bg92} Boroson, T.\ A., \& Green, R.\
  F.\ 1992, \apjs, 80, 109

\bibitem[Boksenberg et al.\ (1997)]{boksenberg77} Boksenberg, A.,
  Carswell, R.\ F., Allen, D.\ A., Fosbury, R.\ A.\ E., Penston,
  M.\ V., \& Sargent, W.\ L.\ W.\ 1997, \mnras, 178, 451

\bibitem[Breit \& Teller (1940)]{bt40} Breit, G., \& Teller, E.\ 1940,
  \apj, 91, 215

\bibitem[Cardelli et al.\ (1989)]{cardelli89} Cardelli, J.\ A.,
  Clayton,
  G.\ C., \& Mathis, J.\ S.\ 1989, \apj, 345, 245

\bibitem[Casebeer et al.\ (2006)]{casebeer06} Casebeer, D.\ A.,
  Leighly, K.\ M., \& Baron, E.\ 2006, \apj, 637, 157

\bibitem[Castor et al.\ (1975)]{cak75} Castor, J. I., Abbott, D.\ C.,
  \& Klein, R.\ I.\ 1975, \apj, 195, 175

\bibitem[Cavaliere et al.\ (2002)]{cavaliere02} Cavaliere, A., Lapi,
  A., \& Menci, N.\ 2002, \apj, 581, L1

\bibitem[Clegg (1987)]{clegg87} Clegg, R.\ E.\ S.\ 1987, \mnras, 229,
  31

\bibitem[Clegg \& Harrington (1989)]{ch89} Clegg, R.\ E.\ S., \&
  Harrington, J.\ P.\ 1989, \mnras, 239, 869

\bibitem[Collin et al.\ (2006)]{collin06} Collin, S., Kawaguchi, T.,
  Peterson, B.\ M., \& Vestergaard, M.\ 2006, \aap, 456, 75

\bibitem[Cottis et al.\ (1010)]{cottis10} Cottis, C.\ E., Goad,
  M.\ R., Knigge, C., \& Scaringi, S.\ 2010, \mnras, accepted

\bibitem[Crenshaw et al. (2003)]{crenshaw03} Crenshaw, D.\ M.,
  Kraemer, S.\ B., \& George, I.\ M.\ 2003, ARA\&A, 41, 117

\bibitem[Cushing et al. (2004)]{cushing04} Cushing, M.\ C., Vacca, W.\
  D., \& Rayner, J.\ T.\ 2004, \pasp, 116, 362

\bibitem[Dai et al.\ (2008)]{dai08} Dai, X., Shankar, F., \& Sivakoff,
  G.\ R.\ 2008, \apj, 672, 108

\bibitem[De Kool et al.\ (2002)]{dekool02} De Kool, M., Korista, K.\
  T., \& Arav, N.\ 2002, \apj, 580, 54

\bibitem[Drake (1971)]{drake71} Drake, G.\ W.\ F., 1971,
  Phys.\ Rev.\ A, 3, 908

\bibitem[Drake \& Dalgarno (1968)]{dd68} Drake, G.\ W.\ F., \&
  Dalgarno, A.\ 1968, \apjl, 152, 121

\bibitem[Drake et al.\ (1969)]{dvd69} Drake, G.\ W.\ F., Victor,
  G.\ A., \& Dalgarno, A.\ 1969, Physical Review, 180, 25

\bibitem[Dunn et al.\ (2010)]{dunn10} Dunn, J.\ P., Bautista, M.,
  Arav, N., Moe, M., Korista, K., Constantini, E., Benn, C., Ellison,
  S., \& Edmonds, D.\ 2010, \apj, 709, 611

\bibitem[Edwards et al.\ (2003)]{edwards03} Edwards, S., Fischer, W.,
  Kwan, J., Hillenbrand, L., \& Dupree, A.\ K., 2003, \apjl, 599, 41

\bibitem[Everett (2005)]{everett05} Everett, J.\ E.\ 2005, \apj, 631,
  689 

\bibitem[Ferland et al.\ (1998)]{ferland98} Ferland, G.\ J., Korista,
  K.\ T., Verner, D.\ A., Ferguson, J.\ W., Kingdon, J.\ B. \& Verner, E.\
  M., 1998, PASP, 110, 761

\bibitem[Gabriel \& Jordon (1969)]{gj69} Gabriel, A.\ H., \& Jordon,
  C.\ 1969, Nature, 221, 947

\bibitem[Gallagher \& Everett (2007)]{ge07} Gallagher, S.\ C., \&
Everett, J.\ E.\ 2007, ASPC, 373, 305

\bibitem[Ganguly et al.\ (2007)]{ganguly07} Ganguly, R., Brotherton,
  M.\ S., Cales, S., Scoggins, B., Shang, Z., \& Vestergaard,
  M.\ 2007, \apj, 665, 990

\bibitem[Gaskell (2008)]{gaskell08} Gaskell, C.\ M., 2008, RMxAA, 32, 1

\bibitem[Gibson et al.\ (2009)]{gibson09} Gibson, R.\  R., et al.\ 2009, \apj,
  692, 758

\bibitem[Grevesse et al.\ (2007)]{grevesse07} Grevesse, N., Asplund, M., \&
  Sauval, A.\ J.\ 2007, Space Sci.\ Rev., 130, 105

\bibitem[Griem (1969)]{griem69} Griem, H.\ R.\ 1969, \apjl, 156, 103

\bibitem[Hall et al.\ (2003)]{hall03} Hall, P.\ B., Hutsem\'ekers, D.,
  Anderson, S.\ F., Brinkmann, J., Fan, X., Schneider, D.\ P., \&
  York, D.\ G.\ 2003, \apj, 593, 189

\bibitem[Hamann (1997)]{hamann97b} Hamann, F.\ 1997, \apjs, 109, 279

\bibitem[Hamann  (1998)]{hamann98} Hamann, F.\ 1997, \apj, 500, 798

\bibitem[Hamann et al. (1997)]{hamann97} Hamann, F., Barlow, T.\ A.,
  Junkkarinen, V., \& Burbidge, E.\ M.\ 1997, \apj, 478, 80

\bibitem[Hamann \& Ferland (1999)]{hf99} Hamann, F., \& Ferland,
  G.\ 1999, ARA\&A, 37, 487

\bibitem[Hamann et al.\ (2001)]{hamann01} Hamann, F.\ W., Barlow,
  T.\ A.., Chaffee, F.\ C., Foltz, C.\ B., \& Weymann, R.\ J., 2001,
  \apj, 550, 142

\bibitem[Hamann et al.\ (2002)]{hamann02} Hamann, F., Korista, K.\ T.,
  Ferland, G.\ J., Warner, C., \& Baldwin, J.\ 2002, \apj, 564, 592

\bibitem[Hamann et al.\ (2010)]{hamann10} Hamann, F., Kanekar, N.,
  J.\ X.\ Prochaska, Murphy, M.\ T., Ellison, S., Malec, A.\ L.,
  Milutinovic, N., \& Ubachs, W., 2010, MNRAS in press

\bibitem[Korista et al.\ (1997)]{korista97} Korista, K.\ T., Baldwin, J.,
  Ferland, G., \& Verner, D.\ 1997, ApJS, 108, 401

\bibitem[Korista et al.\ (1993)]{korista93} Korista, K.\ T., Voit,
  G.\ M., Morris, S.\ L., \& Weymann, R.\ J.\ 1993, \apjs, 88, 357

\bibitem[Kraemer et al.\ (2001)]{kraemer01} Kraemer, S.\ B., Crenshaw,
  D.\ M., Hutchings, J.\ B., George, I.\ M., Danks, A.\ C., Gull, T.\
  R., Kaiser, M.\ E., Nelson, C.\ H., Weistrop, D., \& Vieira, G.\ L.\
  2001, \apj, 551, 671

\bibitem[Kraemer et al.\ (2006)]{kraemer06} Kraemer, S.\ B., et al.\
  2006, \apjs, 167, 161

\bibitem[Kriss et al.\ (1992)]{kriss92} Kriss, G.\ A.,  et al.\ 1992,
  \apj, 392, 485

\bibitem[Krolik \& Kriss (2001)]{kk01} Krolik, J.\ H., \& Kriss,
  G.\ A.\ 2001, \apj, 561, 684

\bibitem[Landt et al. (2008)]{landt08} Landt, H., Bentz, M.\ C., Ward,
  M.\ J., Elvis, M., Peterson, B.\ M., Korista, K.\ T., \& Karovska,
  M.\ 2008, \apjs, 174, 282

\bibitem[Laor \& Brandt (2002)]{lb02} Laor, A., \& Brandt, W.\ N.,
  2002, \apj, 569, 641

\bibitem[Laor et al. (1997)]{laor97} Laor, A., Jannuzi, B.\ T., Green,
  R.\ F., \& Boroson, T.\ A., 1997, \apj, 489, 656

\bibitem[Leighly et al.\ (2009)]{leighly09} Leighly, K.\ M., Hamann,
  F., Casebeer, D.\ A., \& Grupe, D.\ 2009, \apj, 701, 176

\bibitem[Leighly et al. (2007)]{leighly07} Leighly, K.\ M., Halpern,
  J.\ P., Jenkins, E.\ B., \& Casebeer, D.\ 2007, \apjs, 173, 1

\bibitem[Mathis (1957)]{m57} Mathis, J.\ S.\ 1957, \apj, 125, 318

\bibitem[North et al.\ (2006)]{north06} North, M., Knigge, C., \&
  Goad, M.\ 2006, \mnras, 365, 1057

\bibitem[Ogle (1998)]{ogle98} Ogle, P.\ M., Ph.\ D.\ Thesis, California
  Institute of Technology 1998

\bibitem[Pei (1992)]{pei92} Pei, Y.\ C., 1992, \apj, 395, 130

\bibitem[Phillips (1978)]{phillips78} Phillips, M.\ M., 1978, 
\apj, 226, 736

\bibitem[Proga \& Kallman (2004)]{pk04} Proga, D., \& Kallman,
  T.\ R.\ 2004, \apj, 616, 688

\bibitem[Proga et al.\ (2000)]{psk00} Proga, D., Stone, J.\ M., \&
  Kallman, T.\ R.\ 2000, \apj, 543, 686

\bibitem[Rayner et al. (2003)]{rayner03} Rayner, J.\ T., Toomey, D.\
    W., Onaka, P.\ M., Denault, A.\ J., Stahlberber, W.\ E., Vacca,
    W.\ D., Cushing, M.\ C., \& Wang, S.\ 2003, \pasp, 115, 362

\bibitem[Richards et al.\ (2006)]{richards06} Richards, G.\ T., et
  al.\ 2006, \apjs, 166, 470

\bibitem[Sabra \& Hamann (2005)]{sabra05} Sabra, B.\ M., \& Hamann,
  F., 2005, arXiv:astro-ph/0509421

\bibitem[Saraph \& Storey (1999)]{ss99} Saraph, H.\ E., \& Storey,
  P.\ J.\ 1999, A\&AS, 143, 369

\bibitem[Savage \& Sembach (1991)]{ss91} Savage, B.\ D., \& Sembach,
  K.\ R.\ 1991, \apj, 379, 245

\bibitem[Scannapieco \& Oh (2004)]{so04} Scannapieco, E., \& Oh,
  S.\ P.\ 2004, \apj, 608, 62

\bibitem[Schlegel et al.\ (1998)]{schlegel98} Schlegel, D.\ J.,
  Finkbeiner, D.\ P., \& Davis, M.\ 1998, \apj, 500, 525

\bibitem[Schneider et al.\ (2010)]{schneider10} Schneider, D.\ et
  al.\ 2010, \aj, 139, 2360

\bibitem[Scoville \& Norman (1995)]{sn95} Scoville, N., \& Norman,
  C.\ 1995, \apj, 451, 510

\bibitem[Sulentic et al.\ (2006)]{sulentic06} Sulentic, J.\ W.,
  Dultzin-Hacyan, D, Marziani, P., Bongardo, C., Braito, V., Calvani,
  M., \& Zamanov, R.\ 2006, RMxAA, 42, 23

\bibitem[Turnshek et al.\ (1997)]{turnshek97} Turnshek, D.\ A.,
  Monier, E.\ M., Sirola, C.\ J., \& Espey, B.\ R.\ 1997, \apj, 476,
  40 

\bibitem[Vacca et al. (2003)]{vacca03} Vacca, W.\ D., Cushing, M.\ C.,
  \& Rayner, J.\ T.\ 2003, \pasp, 115, 389

\bibitem[Verner et al.\ (1994)]{verner94} Verner, D.\ A., Barthel, P.\
  D., \& Tytler, D.\ 1994, A\&AS, 108, 287

\bibitem[Verner et al.\ (1999)]{verner99} Verner, E.\ M., Verner, D.\
  A., Korista, K.\ T., Ferguson, J.\ W., Hamann, F., \& Ferland, G.\
  J.\ 1999 ApJS, 120, 101

\bibitem[Weymann et al.\ (1991)]{weymann91} Weymann, R.\ J., Morris,
  S.\ L., Foltz, C.\ B., \& Hewett, P.\ C.\ 1991, \apj, 373, 23

\bibitem[White et al. (2000)]{white00} White, R.\ L., et al.\ 2000,
  \apjs, 126, 133


\end{thebibliography}
\end{document}